\documentclass[aps, prd, floatfix, nofootinbib, superscriptaddress, twocolumn]{revtex4-1}

\usepackage{latexsym}
\usepackage{amsmath}
\usepackage{amssymb}
\usepackage{amsfonts}

\usepackage[mathscr,scaled=1.15]{urwchancal}
\DeclareFontFamily{OT1}{pzc}{}
\DeclareFontShape{OT1}{pzc}{m}{it}%
{<-> s * [1.15] pzcmi7t}{}
\DeclareMathAlphabet{\mathpzc}{OT1}{pzc}{m}{it}

\usepackage{CJKutf8}
\usepackage{color}
\usepackage{slashed}
\usepackage{dsfont}

\usepackage{supertabular}
\usepackage{placeins}
\usepackage{epsfig}
\usepackage{graphicx}

\definecolor{purple}{rgb}{0.5,0,0.5}
\definecolor{blue}{rgb}{0.0,0,0.9}
\definecolor{prdblue}{rgb}{0.133,0.118,0.498}
\usepackage[colorlinks=true, pdfstartview=FitV, linkcolor=prdblue, citecolor= prdblue, urlcolor=prdblue]{hyperref}

\begin{document}
\begin{CJK}{UTF8}{song}

\title{$\,$\\[-6ex]\hspace*{\fill}{\normalsize{\sf\emph{Preprint nos}.
NJU-INP 028/20, USTC-ICTS/PCFT-22-07}}\\[1ex]
Nucleon axial-vector and pseudoscalar form factors, and PCAC relations}

\author{Chen Chen} 
\email[]{chenchen1031@ustc.edu.cn}
\affiliation{Interdisciplinary Center for Theoretical Study, University of Science and Technology of China, Hefei, Anhui 230026, China}
\affiliation{Peng Huanwu Center for Fundamental Theory, Hefei, Anhui 230026, China}
\affiliation{Institut f\"ur Theoretische Physik, Justus-Liebig-Universit\"at Gie{\ss}en, D-35392 Gie{\ss}en, Germany}
\affiliation{Helmholtz Forschungsakademie Hessen f\"ur FAIR (HFHF),
GSI Helmholtzzentrum f\"ur Schwerionenforschung, Campus Gie{\ss}en, 35392 Gie{\ss}en, Germany}

\author{Christian S. Fischer}
\email[]{christian.fischer@theo.physik.uni-giessen.de}
\affiliation{Institut f\"ur Theoretische Physik, Justus-Liebig-Universit\"at Gie{\ss}en, D-35392 Gie{\ss}en, Germany}
\affiliation{Helmholtz Forschungsakademie Hessen f\"ur FAIR (HFHF),
GSI Helmholtzzentrum f\"ur Schwerionenforschung, Campus Gie{\ss}en, 35392 Gie{\ss}en, Germany}

\author{Craig D. Roberts}
\email[]{cdroberts@nju.edu.cn}
\affiliation{School of Physics, Nanjing University, Nanjing, Jiangsu 210093, China}
\affiliation{Institute for Nonperturbative Physics, Nanjing University, Nanjing, Jiangsu 210093, China}

\author{\mbox{Jorge Segovia}}
\email[]{jsegovia@upo.es}
\affiliation{Dpto.\ Sistemas F\'isicos, Qu\'imicos y Naturales, Univ.\ Pablo de Olavide, E-41013 Sevilla, Spain}
\affiliation{Institute for Nonperturbative Physics, Nanjing University, Nanjing, Jiangsu 210093, China}

\date{2022 April 04}

\begin{abstract}
We use a continuum quark+diquark approach to the nucleon bound-state problem in relativistic quantum field theory to deliver parameter-free predictions for the nucleon axial and induced pseudoscalar form factors, $G_A$ and $G_P$, and unify them with the pseudoscalar form factor $G_5$ or, equivalently, the pion-nucleon form factor $G_{\pi NN}$.
We explain how partial conservation of the axial-vector current and the associated Goldberger-Treiman relation are satisfied once all necessary couplings of the external current to the building blocks of the nucleon are constructed consistently; in particular, we fully resolve 
the seagull couplings to the diquark-quark vertices associated with the axial-vector and pseudoscalar currents.
Among the results we describe, the following are worth highlighting.
A dipole form factor defined by an axial charge $g_A=G_A(0)=1.25(3)$ and a mass-scale $M_A = 1.23(3) m_N$, where $m_N$ is the nucleon mass, can accurately describe the pointwise behaviour of $G_A$.
Concerning $G_P$, we obtain the pseudoscalar charge $g_p^\ast = 8.80(23)$, and find that the pion pole dominance approach delivers a reliable estimate of the directly computed result.
Our computed value of the pion-nucleon coupling constant, $g_{\pi NN}/m_N =14.02(33)/{\rm GeV}$ is consistent with
recent precision determinations.
%
%
\end{abstract}



\maketitle
\end{CJK}


\section{Introduction}
\label{secintro}

The properties of the proton and neutron (nucleons) are determined by the strong interaction; and a central aim of on-going experimental and theoretical efforts is to understand their structure as composite objects made of three valence light quarks \cite{Hofstadter:1955ae}.  Electron+nucleon scattering is a well developed experimental technique in such studies and it has delivered, for instance, precise measurements of nucleon electromagnetic form factors \cite{Jones:1999rz, Gayou:2001qd, Punjabi:2005wq, Puckett:2010ac, Puckett:2011xg, Cates:2011pz, Puckett:2017flj}.  While they are relatively well determined, less constrained are the axial and induced-pseudoscalar form factors derived from the isovector axial-vector current that describes the neutrino-nucleon scattering process. These nucleon axial form factors are important quantities for the understanding of weak interactions, neutrino-nucleus scattering and parity violation experiments.

The axial form factor, $G_A(Q^2)$, is experimentally determined from neutrino+proton scattering, $\nu_\mu + p \to \mu^+ + n$ \cite{Ahrens:1988rr, Bodek:2007vi, Mosel:2016cwa, Meyer:2016oeg}, while the induced pseudoscalar form factor, $G_P(Q^2)$, can be extracted from the longitudinal cross-section in pion electroproduction \cite{Choi:1993vt, Bernard:1994pk, Fuchs:2003vw} or, potentially, from single-spin asymmetries in $\nu(\bar\nu)$ charged-current quasielastic scattering on free nucleons \cite{Tomalak:2020zlv}.   At zero momentum transfer the axial form factor gives the axial charge $g_A := G_A(0)$, measured with high precision from $\beta$-decay experiments \cite{Mendenhall:2012tz, Mund:2012fq, Brown:2017mhw, Darius:2017arh}.  The induced pseudoscalar coupling $g_P^*$  can be determined via the muon capture process $\mu^- + p \rightarrow n + \nu_\mu$ from the singlet state of the muonic hydrogen atom at the muon capture point, $Q^2=0.88 m_\mu^2$ \cite{Castro:1977ep, Bernard:2000et, Andreev:2007wg, Andreev:2012fj, Andreev:2015evt}, where $m_\mu$ is the muon mass.

On the theory side, reliable extractions of the axial form factors can only be made within nonperturbative quantum chromodynamics (QCD). Prime contenders in this respect are lattice QCD (lQCD) and nonperturbative functional methods, like the Dyson-Schwinger equations (DSEs).

lQCD has made important progress in the past years.
For example, several groups have computed the axial form factors \cite{Capitani:2017qpc, Rajan:2017lxk, Alexandrou:2017hac, Shintani:2018ozy, Bali:2018qus, Ishikawa:2018rew, Jang:2019vkm, Bali:2019yiy, Alexandrou:2020okk}, some of them using simulations generated directly at the physical pion mass and with large volumes. The problem of excited state contamination \cite{Capitani:2010sg} has also been analysed in detail with chiral perturbation theory methods \cite{Bernard:2001rs, Bar:2018xyi, Bar:2019igf} and a variety of lattice improvements have been implemented \cite{Capitani:2012gj, Bali:2019yiy, Jang:2019vkm, Jang:2020ygs, Ottnad:2020qbw}.
The continuum limit and the physical pion mass are mandatory in lQCD if one wants to check important low-energy QCD relations, such as the partially conserved axial-vector current (PCAC) which, at the form factor level, connects $G_A(Q^2)$ and $G_P(Q^2)$ with the pseudoscalar form factor, $G_5(Q^2)$. Another important feature is that, at low $Q^2$, assuming pion pole dominance (PPD), $G_A(Q^2)$ and $G_P(Q^2)$ are directly connected, leading to the Goldberger-Treiman relation.

These observables can also be studied using the DSEs, see, e.g.\ Refs.\,\cite{Roberts:2000aa, Alkofer:2000wg, Maris:2003vk, Pawlowski:2005xe, Bashir:2012fs,  Eichmann:2016yit, Fischer:2018sdj, Qin:2020rad}. This theoretical framework ranges among different levels of complexity, from symmetry-preserving QCD-kindred model studies \cite{Chen:2017pse, Chen:2018nsg, Chen:2019fzn, Lu:2019bjs, Cui:2020rmu} to systematic analyses using a truncation derived from n-particle-irreducible effective interactions \cite{Fischer:2003rp, Fischer:2008wy, Eichmann:2008ae, Binosi:2009qm, Williams:2015cvx, Huber:2018ned}.

The three-valence-body problem in relativistic quantum  field  theory is a challenge, only solved in Refs.\,\cite{Eichmann:2009qa, Sanchis-Alepuz:2014sca, Sanchis-Alepuz:2014wea, Qin:2018dqp, Qin:2019hgk}.  Fortunately, the quark+diquark approximation \cite{Oettel:1998bk, Segovia:2015ufa} has delivered comparable results for baryon properties \cite{Eichmann:2016yit, Eichmann:2016hgl, Barabanov:2020jvn}. This is because any interaction capable of creating pseudo-Nambu-Goldstone modes as bound-states of a light dressed-quark and -antiquark, and reproducing the measured value of their leptonic decay constants, must also generate strong colour-antitriplet correlations between any two dressed quarks contained within a hadron.  This assertion is supported by evidence accumulated in two decades of studying two- and three-body bound-state problems in hadron physics, \emph{e.g}. Refs.\,\cite{Cahill:1987qr, Cahill:1988dx, Oettel:1998bk, Bloch:1999vk, Bender:2002as, Eichmann:2008ef, Roberts:2011cf, Eichmann:2016yit, Segovia:2015ufa}. No realistic counter examples are known; and the existence of such diquark correlations is also supported by simulations of lQCD \cite{Barabanov:2020jvn}.

In connection with the form factors of the nucleon axial current, Ref.\,\cite{Eichmann:2011pv} reported their consistent calculation using the Poincar\'e-covariant three-valence-body rainbow-ladder truncation \cite{Munczek:1994zz, Bender:1996bb}. However, the study underestimates the experimental value of the axial charge by 20-25\% and algorithmic inadequacies prevented computation of the axial form factors beyond $Q^2\sim2\,\text{GeV}^2$. The quark+diquark approximation can be used, \emph{inter alia}, to reach higher transferred momenta.

In the first computation within the quark+diquark picture, the interaction between the axial current and the
bystander quark \cite{Hellstern:1997pg} was the only one considered.  Subsequently, Refs.\,\cite{Bloch:1999rm, Oettel:2000jj, Roberts:2007jh} included the contributions from the interaction with diquarks and also from the so-called exchange-quark diagram. However, they still neglected the ``seagull terms'', which are essential in order to maintain, for instance, PCAC relations.

Herein, we construct all elements necessary for calculation of the nucleon's axial-vector current; in particular, we determine, for the first time, the seagull contributions.  We choose to approach the problem using the Poincar\'e-covariant quark+diquark QCD-kindred framework, which has successfully been employed in the description and unification of an array of properties of the nucleon, $\Delta$-baryon, and their low-lying excitations
\cite{Segovia:2014aza, Burkert:2017djo, Chen:2017pse, Chen:2018nsg, Chen:2019fzn, Lu:2019bjs, Cui:2020rmu, Liu:2022ndb}.  With its inputs tuned elsewhere via comparisons with different observables, many aspects of emergent hadron mass are implicitly expressed in the formulation. First results for the axial and induced pseudoscalar form factors are reported in Ref.\,\cite{Chen:2020wuq}.  This study expands significantly upon that analysis, including all mathematical details involved in proving PCAC on an analytic level, quantification of the range over which PCAC holds numerically, and a concise discussion of numerical results for the pion-nucleon form factor.

Section~\ref{secaffs} is devoted to the description of the matrix elements of the nucleon axial-vector and pseudoscalar currents, and their associated form factors. In Sec.\,\ref{secTheory}, we briefly survey our theoretical framework, introducing the elements necessary for a practical calculation of the axial and induced pseudoscalar form factors.  Section~\ref{secnume} discusses our numerical results and associated comparisons with available experimental data and results from lQCD. Section~\ref{secsum} supplies a summary and perspective.


\section{Nucleon axial and pseudoscalar currents}
\label{secaffs}

The axial form factor, $G_A(Q^2)$, and the induced-pseudoscalar form factor, $G_P(Q^2)$, are defined via the nucleon matrix element
\begin{subequations}
\label{jaxdq0}
\begin{align}
\label{jaxdq}
\hat J^j_{5\mu}(&K,Q)
:= \langle N(P_f)|{\mathpzc A}^j_{5\mu}(0)|N(P_i)\rangle \\
\label{jaxdqb}
=&\bar{u}(P_f)\frac{\tau^j}{2}\gamma_5
\bigg[ \gamma_\mu G_A(Q^2) +i\frac{Q_\mu}{2m_N}G_P(Q^2) \bigg]\,u(P_i)\,,
\end{align}
\end{subequations}
where $P_i$ and $P_f$ are, respectively, the initial and final momenta of the nucleon, defined such that the on-shell condition is fulfilled, $P_{i,f}^2=-m_N^2$, with $m_N$ the nucleon mass
and $u(P)$ the associated Euclidean spinor (here we have suppressed the spin label, see Ref.\,\cite[Appendix\,B]{Segovia:2014aza} for details).
Furthermore, we denote $K=(P_i+P_f)/2$ as the average momentum of the system and $Q=P_f-P_i$ the transferred momentum between initial and final states. Throughout this paper we work in the Euclidean space and use the SU$(2)_F$ isospin limit $m_u=m_d=:m_q$. The flavour structure is given by the Pauli matrices $\{\tau^j|j=1,2,3\}$, where $\tau^3$ represents the neutral current and $\tau^{1\pm i2}:=(\tau^1\pm i\tau^2)/2$ the charged currents. Moreover,
\begin{equation}
\label{jaxx}
{\mathpzc A}_{5\mu}^j(x) = \bar{\psi}(x) \frac{\tau^j}{2} \gamma_5 \gamma_\mu \psi(x), \,\,\,\,\, \psi=\left(\begin{array}{c} u \\ d \end{array}\right) \,,
\end{equation}
is the isovector axial current operator.

Similarly, the pseudoscalar form factor $G_5(Q^2)$ is defined by the kindred pseudoscalar current
\begin{subequations}
\label{jpsdq0}
\begin{align}
\label{jpsdq}
\hat J^j_{5}(K,Q) &:=
\langle N(P_f)|{\mathpzc P}^j_5(0)|N(P_i)\rangle \\
&=\bar{u}(P_f)\frac{\tau^j}{2}\gamma_5\,G_5(Q^2)\,u(P_i)\,,
\end{align}
\end{subequations}
where
\begin{equation}
\label{jpsx}
{\mathpzc P}^j_5(x) = \bar{\psi}(x)\frac{\tau^j}{2}\gamma_5\psi(x),	
\,\,\,\,\, \psi=\left(\begin{array}{c} u \\ d \end{array}\right) \,,
\end{equation}
is the isovector pseudoscalar current operator.

The form factors $G_A$, $G_P$ and $G_5$ can be obtained directly from these currents using any set of sensible projection operators.  For example, with
\begin{equation}
J^j_{5\mu}(K,Q) := \sum_{{\rm spins}\,(i,f)} \!\!\! u(P_f) \hat J^j_{5\mu}(K,Q) \bar u(P_i)
\end{equation}
and obvious analogues, then
{\allowdisplaybreaks
\begin{subequations}
\begin{align}
\label{gaproj}
G_A &= -\frac{1}{4(1+\tau)}{\rm tr}_{\rm D}[ J_{5\mu}\gamma_5\gamma_\mu^T ]\,,\\
\label{gpproj}
G_P &= \frac{1}{\tau}\bigg(G_A - \frac{Q_\mu}{4im_N\tau}	
{\rm tr}_{\rm D}[ J_{5\mu}\gamma_5]\bigg)\,,\\
\label{g5proj}
G_5 &= -\frac{1}{2\tau}{\rm tr}_{\rm D}[ J_{5}\gamma_5]\,,
\end{align}
\end{subequations}
where the trace is over spinor indices; $\tau=Q^2/4m_N^2$; $\gamma_\mu^T = \gamma_\mu-(\gamma\cdot Q)Q_\mu/Q^2$; $J_{5\mu}$ and $J_5$ represent the Dirac parts of the corresponding currents, \emph{viz.} $J_{5\mu}^j=:(\tau^j/2)J_{5\mu}$ and $J_{5}^j=:(\tau^j/2)J_{5}$; and the flavour matrix $\tau^j/2$ must be projected onto the isospinors of the proton, ${\mathrm p}\equiv(1,0)^{\rm T}$, or neutron, ${\mathrm n}\equiv(0,1)^{\rm T}$.
}

A detailed analysis of the consequences of PCAC within functional methods can be found in Ref.\,\cite{Eichmann:2011pv}.  We briefly recapitulate here. By using standard Ward-Green-Takahashi identities one can immediately infer the PCAC relation between the current operators:
\begin{equation}
\label{pcacx}
\partial_\mu {\mathpzc A}_{5\mu}^j(x)+2m_q {\mathpzc P}_5^j(x)	= 0\,.
\end{equation}
Using the nucleon matrix element expression of the identity above, one then obtains the PCAC relation at the nucleon level
\begin{equation}
\label{pcacn}
Q_\mu J^j_{5\mu}(K,Q)+2im_q	J^j_{5}(K,Q)=0 \,,
\end{equation}
and this entails
\begin{equation}
\label{pcacp}
G_A(Q^2)-\frac{Q^2}{4m_N^2}G_P(Q^2)=\frac{m_q}{m_N}G_5(Q^2)\,.
\end{equation}
It should be pointed out that PCAC is an operator relation and thus any realistic results for $G_A$, $G_P$ and $G_5$ should precisely satisfy Eq.\,\eqref{pcacp}.

At the pion mass pole: $Q^2=-m_\pi^2$, the residue of $G_5$ is the pion-nucleon coupling constant $g_{\pi NN}$. Thus, one defines the pion-nucleon
form factor $G_{\pi NN}(Q^2)$ via
\begin{align}
\label{gpinn}
G_5(Q^2) =: \frac{m_\pi^2}{Q^2+m_\pi^2}\frac{f_\pi}{m_q}G_{\pi NN}(Q^2)\,,
\end{align}
with $G_{\pi NN}(-m_\pi^2)=g_{\pi NN}$ and $f_\pi$ the pion's leptonic decay constant. Since $G_P(Q^2)$ is analytic at $Q^2=0$, contraction
of Eq.\,\eqref{jaxdqb} with $Q_\mu$ and using Eq.\,\eqref{gpinn} together with the PCAC relation \eqref{pcacp} leads to the original Goldberger-Treiman relation,
\begin{align}
\label{gtr}
G_A(0)=\frac{f_\pi}{m_N}G_{\pi NN}(0)\,.	
\end{align}
This relation is valid for all current quark masses.


\section{Theoretical framework}
\label{secTheory}

\subsection{Diquark correlations}
\label{secdiquarks}

The bulk of QCD's particular features and nonperturbative phenomena can be traced to the evolution of the strong running coupling. Its unique characteristics are primarily determined by the three-gluon vertex: the four-gluon vertex does not contribute dynamically at leading order in perturbative analyses of matrix elements; and nonperturbative continuum analyses of QCD's gauge sector indicate that satisfactory agreement with gluon propagator results from lQCD simulations is typically obtained without reference to dynamical contributions from the four-gluon vertex, \emph{e.g}.\ Refs.\,\cite{Aguilar:2008xm, Aguilar:2009ke, Aguilar:2009nf, Maas:2011se, Boucaud:2011ug, Pennington:2011xs, Binosi:2012sj, Strauss:2012dg, Meyers:2014iwa, Huber:2018ned, Huber:2020keu}.

Analyses of the three valence-quark scattering problem reveals that the binding energy contribution of the diagram in which each leg of the three-gluon vertex is attached to one quark vanishes when projected onto the baryon's colour-singlet wave function, \emph{i.e.}
\begin{equation}
\label{primafacie}
\varepsilon_{f_1 f_2 f_3} \otimes
\big(f^{abc} [\lambda^a]_{f_1 i_1} \, [\lambda^b]_{f_2 i_2} \,
[\lambda^c]_{f_3 i_3}\big) \otimes
\varepsilon_{i_1 i_2 i_3}  = 0\,.
\end{equation}
Here $\varepsilon_{i_1 i_2 i_3}$ and $\varepsilon_{f_1 f_2 f_3}$ are the Levi-Civita tensors representing the colour structure of initial
and final baryon states, $\{\lambda^a\}$ are $SU(3)_c$ Gell-Mann matrices and $f^{abc}$ is the structure tensor of $SU(3)_c$.

Consequently, the three-gluon vertex is the primary factor in generating the class of renormalisation-group-invariant running interactions \cite{Binosi:2014aea} that produce strong attraction between two quarks, creating tight diquark correlations in analyses of the three valence-quark scattering problem.

Further details on diquark physics can be found elsewhere \cite{Barabanov:2020jvn}; however, some technical details are needed here. The quark+quark correlation can be described by the 4-point Green function
\begin{align}
\label{4qgreen}
\nonumber
G(x_1,x_2;y_1,y_2) = &\langle0| T\big\{
\psi(x_1)\psi(x_2)\bar{\psi}(y_1)\bar{\psi}(y_2)\big\}|0\rangle\\
\equiv & \langle T\big\{\psi(x_1)\psi(x_2)\bar{\psi}(y_1)\bar{\psi}(y_2)\big\}\rangle\,,
\end{align}
and its Fourier transformation, $G(k_1,k_2,q_1,q_2)$, defined in momentum space is
\begin{align}
\nonumber
&(2\pi)^4\delta^{(4)}(k_1+k_2+q_1+q_2) G(k_1,k_2,q_1,q_2) \\
\nonumber
=& \int d^4{x_1}\, d^4{x_2}\, d^4{y_1}\, d^4{y_2}\, e^{-ik_1\,x_1} e^{-ik_2\,x_2} e^{-iq_1\,y_1} e^{-iq_2\,y_2} \\
\times &\langle T\big\{\psi(x_1)\psi(x_2)\bar{\psi}(y_1)\bar{\psi}(y_2)\big\}\rangle\,.
\end{align}

\begin{figure}[!t]
\centerline{%
\includegraphics[clip, height=0.10\textwidth, width=0.45\textwidth]{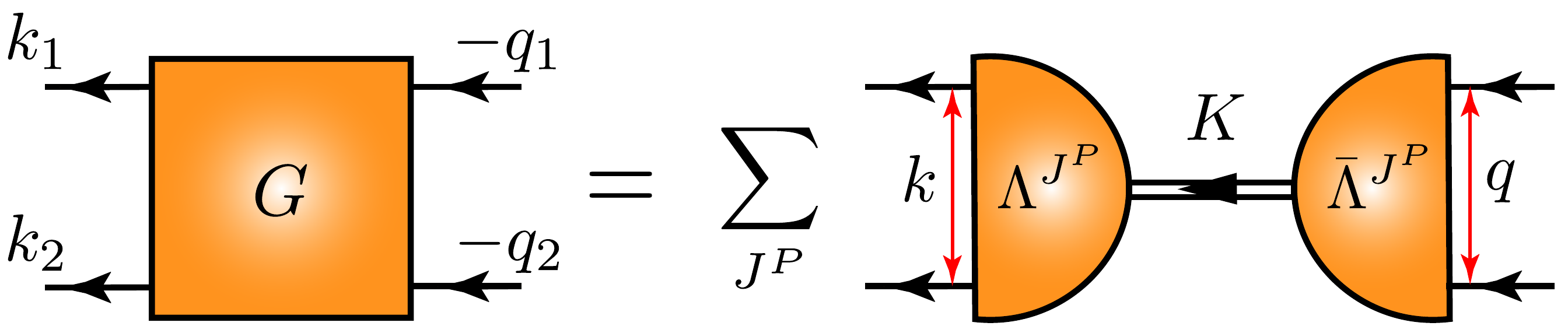}}
\caption{\label{figdqaz} The diquark \emph{Ansatz} for the 4-point Green function of the quark+quark correlations.}
\end{figure}

One can study the phenomenological consequences of having diquark correlations inside baryons using the following \emph{Ansatz} (depicted in Fig.\,\ref{figdqaz}):
\begin{align}
\label{sqansatz}
\nonumber
&G(k_1,k_2,q_1,q_2) \\
= &\sum_{J^P=0^+,1^+,\dots}
\Lambda^{J^P}(k;K)\Delta^{J^P}(K)\bar{\Lambda}^{J^P}(q;-K)\,.
\end{align}
where $\{\Lambda^{J^P}\}$ ($\{\bar{\Lambda}^{J^P}\}$) is the (charge conjugated) Bethe-Salpeter (BS) wave function describing the diquark correlation characterised by the quantum numbers $J^P$. The diquark propagator is $\Delta^{J^P}(K)$ and $K$ is the total momentum of the quark+quark pair. Furthermore, $q=(q_1-q_2)/2$ and $k=(k_1-k_2)/2$ are the relative momenta between quarks in the initial and final diquark states.

It is worth remarking here that in a dynamical theory based on SU$(2)$-colour, diquarks are colour-singlets. They would thus exist as asymptotic states and form mass-degenerate multiplets with mesons composed from like-flavoured quarks. (These properties are a manifestation of Pauli-G\"ursey symmetry \cite{Pauli:1957ot, Guersey:1958ro}.) Consequently, the isoscalar-scalar, $[ud]_{J^P=0^+}$, diquark would be massless in the presence of dynamical chiral symmetry breaking (DCSB), matching the pion, and the isovector-pseudovector, $\{ud\}_{1^+}$, diquark would be degenerate with the theory's $\rho$-meson.  Such identities are lost in changing the gauge group to SU$(3)$-colour; but clear and instructive similarities between mesons and diquarks nevertheless remain, such as: (\emph{i}) isoscalar-scalar and isovector-pseudovector diquark correlations are the strongest, but others could appear inside a hadron so long as their quantum numbers are allowed by Fermi-Dirac statistics; (\emph{ii}) the associated diquark mass-scales express the strength and range of the correlation and are each bounded below by the partnered meson's mass; and (\emph{iii}) realistic diquark correlations are soft, \emph{i.e}.\ they possess an electromagnetic size that is bounded below by that of the analogous mesonic system.


\subsection{Faddeev quark+diquark amplitude}
\label{secfaddeev}

\begin{figure}[t]
\centerline{%
\includegraphics[clip, height=0.14\textwidth, width=0.45\textwidth]{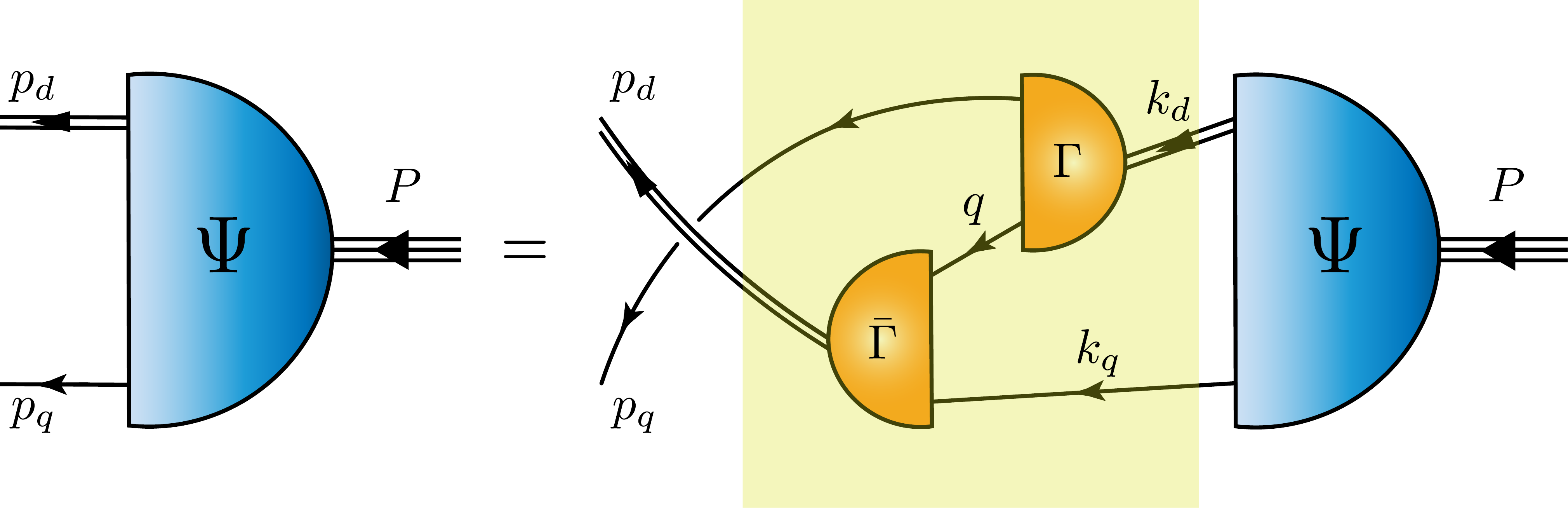}}
\caption{\label{figFaddeev}
Quark+diquark Faddeev equation, a linear integral equation for the Poincar\'e-covariant matrix-valued function $\Psi$, the Faddeev amplitude for a nucleon with total momentum $P=p_q+p_d=k_q+k_d$. $\Psi$ describes the relative momentum correlation between the dressed-quarks and -diquarks. Legend. \emph{Shaded rectangle} -- Faddeev kernel; \emph{single line} -- dressed-quark propagator; $\Gamma$ -- diquark correlation amplitude; and \emph{double line} -- diquark propagator. Ground-state nucleons ($n$ - neutron, $p$ - proton) contain both isoscalar-scalar diquarks, $[ud]\in(n,p)$, and isovector--axial-vector diquarks $\{dd\}\in n$, $\{ud\}\in (n,p)$, $\{uu\}\in p$. Other possible correlations play no measurable role in nucleons \cite{Barabanov:2020jvn}.}
\end{figure}

The existence of tight diquark correlations simplifies analyses of the three valence-quark scattering problem, hence, baryon bound states, because it reduces that task to solving a Poincar\'e covariant Faddeev equation \cite{Cahill:1988dx, Reinhardt:1989rw, Efimov:1990uz, Oettel:1999gc}, depicted in Fig.\,\ref{figFaddeev}.

In the quark+diquark picture, the binding of a nucleon (and kindred baryons) has two contributions. One part is expressed in the formation of tight (but not point-like) diquark correlations. That is augmented, however, by the attraction between quark and diquark generated through the quark exchange depicted in the shaded area of Fig.\,\ref{figFaddeev}. This exchange ensures that diquark correlations within the baryon are fully dynamical: no quark holds a special place because each one participates in all diquarks to the fullest extent allowed by its quantum numbers. The continual rearrangement of the quarks guarantees, amongst other things, that the nucleon's dressed-quark wave function complies with Pauli statistics.

We now specify the elements of the Faddeev equation kernel: the dressed-quark propagator $S(p)$; and the diquark masses, propagators $\Delta^{J^P}(P)$ and BS wave functions $\Lambda^{J^P}(p;P)$. It is convenient to define the diquark BS amplitudes $\Gamma^{J^P}(p;P)$ by amputating the external quark legs from their BS wave functions:
\begin{equation}
\Gamma^{J^P}(p;P):=S^{-1}(p_+)	\Lambda^{J^P}(p;P)(S^{-1})^{\rm T}(-p_-)\,,
\end{equation}
where $p_\pm = p\pm P/2$ and ``${\rm T}$'' denotes matrix transpose.

In working towards realistic QCD-connected predictions, one can adapt the pattern used for mesons, \emph{viz.} solve the DSE for the dressed-quark propagator and BS equations for the diquark correlation amplitudes; therewith, build the Faddeev kernel and determine baryon masses and corresponding Faddeev amplitudes. This \emph{ab initio} approach has delivered many successful results, but it is computationally intensive and limited in application by existing algorithms. An alternative \cite{Oettel:1999gc, Bloch:1999rm, Hecht:2002ej} is to construct a QCD-kindred framework, in which all elements of the Faddeev kernel and interaction currents are momentum dependent, and consistent with QCD scaling laws.

A QCD-kindred framework has recently provided a large range of predictions for various baryons, including their spectrum \cite{Chen:2017pse,Chen:2019fzn} and form factors \cite{Segovia:2014aza, Segovia:2015hra, Segovia:2015ufa, Segovia:2016zyc, Chen:2018nsg, Lu:2019bjs, Cui:2020rmu}. This approach uses an efficacious algebraic parametrisation for the dressed light-quark propagator which is consistent with contemporary numerical results \cite{Chen:2017pse}, expresses confinement and dynamical chiral symmetry breaking, retains the leading diquark amplitudes and describes diquark propagation in a manner consistent with colour confinement and asymptotic freedom.

Our QCD-kindred framework is sketched in Appendix~\ref{appendixQCD}.  Here, we just mention some key points: accounting for Fermi-Dirac statistics, five types of diquark correlations are possible in a $J=1/2$ bound-state: isoscalar-scalar, isovector-pseudo\-vector, isoscalar-pseudo\-scalar, iso\-scalar-vector, and iso\-vector-vector. However, only the first two are quantitatively important in positive-parity systems \cite{Barabanov:2020jvn}; thus, the only ones considered herein. For the scalar and axial-vector diquark masses we use the following values
\begin{subequations}
\label{dqmasses}
\begin{align}
m_{[ud]_{0^+}} &= 0.80\,\mbox{GeV}\,,\\
m_{\{uu\}_{1^+}} = m_{\{ud\}_{1^+}} &= m_{\{dd\}_{1^+}} = 0.89\,\mbox{GeV}\,,
\end{align}
\end{subequations}
which are drawn from Refs.~\cite{Segovia:2014aza, Segovia:2015hra}. Together with a Euclidean constituent mass $M_q^E = 0.33$ GeV for light quarks, they provide a good description of many dynamical properties of the nucleon, $\Delta$-baryon and Roper resonance.

By solving the nucleon's Faddeev equation, we obtain the mass $m_N = 1.18$ GeV. This value is intentionally large because Fig.\,\ref{figFaddeev} describes the so-called nucleon's dressed-quark core. The complete nucleon is obtained by including resonant (meson-cloud) contributions to the Faddeev kernel. Such effects are known to produce a reduction of about $0.2\,\text{GeV}$ in the physical nucleon mass \cite{Hecht:2002ej, Sanchis-Alepuz:2014wea}. For this reason, $G_A$ and $G_P$ shall be expressed in terms of $x=Q^2/m_N^2$. In addition, since the computation of $G_5$ involves explicitly the current quark mass $m_q$, Eqs.\,\eqref{pcacp} and \eqref{psvx2}, it is appropriate to study the normalised combination $(m_q/m_N)G_5(Q^2)$.


\begin{figure}[!t]
\centerline{\includegraphics[clip, height=0.33\textwidth, width=0.45\textwidth]{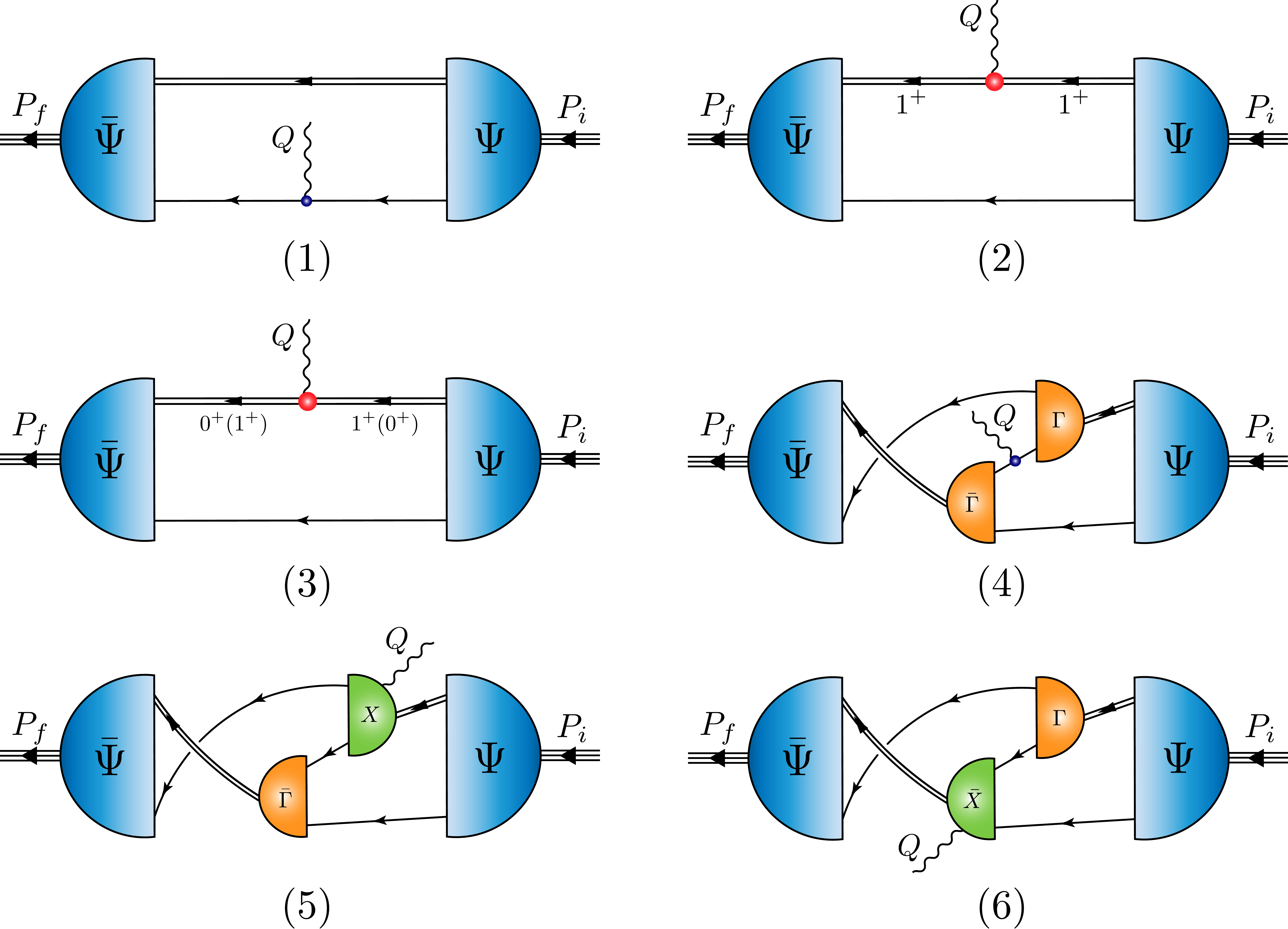}}
\caption{\label{figcurrent} The axial or pseudoscalar currents that ensures PCAC for on-shell baryons that are described by the Faddeev amplitudes produced by the equation depicted in Fig.\,\ref{figFaddeev}: \emph{single line}, dressed-quark propagator; \emph{undulating line}, the axial or pseudoscalar current; $\Gamma$,  diquark correlation amplitude; \emph{double line}, diquark propagator; and $\chi$, seagull terms. Diagram~1 is the top-left image; the top-right is Diagram~2;
and so on, with Diagram~6 being the bottom-right image.}
\end{figure}

\subsection{Microscopic calculation of axial and pseudoscalar currents}
\label{secpcac}

\subsubsection{General decompositions}
\label{subsecdec}

Within the quark+diquark picture of the nucleon, the axial and pseudoscalar currents, Eqs.~\eqref{jaxdq0} and~\eqref{jpsdq0}, can both be decomposed into a sum of six terms, depicted in Fig.~\ref{figcurrent}. One can see therein that the current interacts with the quarks and diquarks in various ways.  In order to perform practical computations, we need to specify three more building blocks, \emph{viz}.\ the coupling of the external current to: the quark lines; the diquark lines; and the diquark-quark wave functions, \emph{i.e.} the seagull terms.


\subsubsection{Axial-vector and pseudoscalar vertices}
\label{subsecaxpsvx}

We begin with the dressed-quarks' axial-vector vertex, $\Gamma_{5\mu}^j(k_+,k_-)$, and the pseudoscalar vertex, $\Gamma_{5}^j(k_+,k_-)$. They are related by the axial-vector Ward-Takahashi identity (AXWTI)
\begin{align}
\label{axwti}
\nonumber
&Q_\mu \Gamma^j_{5\mu}(k_+,k_-)+2im_q\,\Gamma^j_5(k_+,k_-)\\
= & S^{-1}(k_+)i\gamma_5\frac{\tau^j}{2}+
\frac{\tau^j}{2}i\gamma_5S^{-1}(k_-)\,,	
\end{align}
where $Q$ is the incoming momentum of the weak boson, $k_-$ and $k_+$ are the quark's incoming and outgoing momenta, $k_\pm=k\pm Q/2$.

The general expressions for $\Gamma_{5\mu}^j(k_+,k_-)$ and $\Gamma_{5}^j(k_+,k_-)$ can be found in Ref.~\cite{Maris:1997hd} and are given by
\begin{align}
\label{axvx}
\nonumber
\Gamma^j_{5\mu}(k_+,k_-) &= \frac{\tau^j}{2}\gamma_5
\bigg[ \gamma_\mu F_R(k,Q)+\gamma\cdot k k_\mu G_R(k,Q)\\
\nonumber
&- \sigma_{\mu\nu}k_\nu H_R(k,Q)\bigg]
+ \tilde{\Gamma}^j_{5\mu}(k,Q)\\
&+ \frac{f_\pi \,Q_\mu}{Q^2+m_\pi^2}\Gamma^j_\pi(k,Q)\,,
\end{align}
and
\begin{align}
\label{psvx}
\nonumber
i\Gamma^j_{5}(k_+,k_-) &= \frac{\tau^j}{2}\gamma_5
\bigg[ iE_R^P(k,Q) + \gamma\cdot Q F_R^P(k,Q) \\
\nonumber
& +\gamma\cdot k \,k\cdot Q G_R^P(k,Q)
+ \sigma_{\mu\nu}k_\mu Q_\nu H_R^P(k,Q)\bigg] \\
&+ \frac{r_P}{Q^2+m_\pi^2}\Gamma^j_\pi(k,Q)\,,
\end{align}
respectively, where $F_R$, $G_R$, $H_R$, $E_R^P$, $F_R^P$, $G_R^P$, $H_R^P$ and $\tilde{\Gamma}^j_{5\mu}$ are regular functions when $Q^2\to -m_\pi^2$, with $Q_\mu \tilde{\Gamma}^j_{5\mu}(k,Q)\sim {\rm O}(Q^2)$, and
the residue $r_P$ at the pion pole is given by $r_P = f_\pi m_\pi^2/{2m_q}$. The pion mass and decay constant are given by
$m_\pi = 0.14$ GeV and $f_\pi = 0.092$ GeV \cite{Bloch:1999rm}.

{\allowdisplaybreaks
The Bethe-Salpeter amplitude of the pion, $\Gamma^j_\pi$, has the general form\,\cite{Maris:1997hd}
\begin{align}
\label{pibsa}
\nonumber
\Gamma^j_\pi& (k,Q)= \tau^j\gamma_5
\bigg[ iE_\pi(k,Q) +\gamma\cdot Q F_\pi(k,Q) \\
& +\gamma\cdot k \, k\cdot Q G_\pi(k,Q) +
\sigma_{\mu\nu}k_\mu Q_\nu H_\pi(k,Q)\bigg]\,.
\end{align}

Substituting Eqs.\,\eqref{axvx}\,--\,\eqref{pibsa} into Eq.\,\eqref{axwti}, and equating both sides, one obtains
\begin{align}
\frac{\tau^j}{2}\gamma_5 & \bigg[
\gamma\cdot Q \big(F_R(k,Q)+2m_q F_R^P(k,Q)\big) \nonumber \\
&
+\gamma\cdot k\, k\cdot Q \big(G_R(k,Q)+2m_q G_R^P(k,Q)\big) \nonumber \\
&
+\sigma_{\mu\nu}k_\mu Q_\nu \big(H_R(k,Q)+2m_q H_R^P(k,Q)\big) \nonumber\\
&
+2im_q E_R^P(k,Q)\bigg] + f_\pi\Gamma_\pi^j(k,Q) \nonumber \\
&
\hspace*{-0.90cm} = \frac{\tau^j}{2}\gamma_5\bigg[\gamma\cdot Q\Sigma_A(k_+^2,k_-^2) +
2\gamma\cdot k \,k\cdot Q\Delta_A(k_+^2,k_-^2) \nonumber \\
&
+ 2i\Sigma_B(k_+^2,k_-^2)\bigg]\,,
\end{align}
where
\begin{subequations}
\begin{align}
\Sigma_F(\ell_1^2,\ell_2^2)=\frac{1}{2}[F(\ell_1^2)+F(\ell_2^2)]\,,\\
\Delta_F(\ell_1^2,\ell_2^2)=\frac{F(\ell_1^2)-F(\ell_2^2)}
{\ell_1^2-\ell_2^2}\,,
\end{align}
\end{subequations}
with $F \in \{A,B\}$, where $A$ and $B$ are the dressing functions of the quark propagator (see Eq.\,\eqref{SpAB} in Appendix~\ref{appendixQCD}).
}

For the BS amplitude of the pion, $\Gamma_\pi^j$, we use the following \emph{Ansatz} \cite{Hellstern:1997pg, Bloch:1999rm, Roberts:2007jh}
\begin{subequations}
\begin{align}
E_\pi(k,Q) &= \frac{1}{2f_\pi}\big(B(k_+^2)+B(k_-^2)\big)\\
&\equiv \frac{1}{f_\pi}\Sigma_B(k_+^2,k_-^2)\,,
\end{align}
\end{subequations}
and
\begin{align}
F_\pi(k,Q) = G_\pi(k,Q) = H_\pi(k,Q) &= 0\,.
\end{align}
This retains the leading tensor component of the pion.  The subleading components typically provide corrections at the level of 10\% in pion-related observables. We then obtain
\begin{subequations}
\begin{align}
E_R^P(k,Q) &= 0\,,\\
F_R(k,Q) +2m_q\,F_R^P(k,Q) &= \Sigma_A(k_+^2,k_-^2)\,,\\
G_R(k,Q) +2m_q\,G_R^P(k,Q) &= 2\Delta_A(k_+^2,k_-^2)\,,\\
H_R(k,Q) +2m_q\,H_R^P(k,Q) &= 0\,.
\end{align}
\end{subequations}
Considering now that $F_R$, $G_R$, $H_R$, $F_R^P$, $G_R^P$ and $H_R^P$ are functions which do not depend explicitly on $m_q$, the set of equations is solved by
\begin{subequations}
\begin{align}
F_R(k,Q) &= \Sigma_A(k_+^2,k_-^2)\,,\\
G_R(k,Q) &= 2\Delta_A(k_+^2,k_-^2)\,,\\
H_R(k,Q) &= 0\,,
\end{align}
\end{subequations}
and
\begin{align}
E_R^P = F_R^P = G_R^P = H_R^P = 0\,.
\end{align}
Therefore, we finally arrive at
\begin{align}
\nonumber
\Gamma^j_{5\mu}(k_+,k_-) &= \frac{\tau^j}{2}\gamma_5
\bigg[ \gamma_\mu \Sigma_A(k_+^2,k_-^2)  \nonumber \\
& \quad + 2\gamma\cdot k k_\mu \Delta_A(k_+^2,k_-^2) \nonumber \\
&
\quad + 2i\frac{Q_\mu}{Q^2+m_\pi^2}\Sigma_B(k_+^2,k_-^2)\bigg]\,,
\label{axvx2}
\end{align}
and
\begin{align}
\nonumber
i\Gamma^j_{5}(k_+,k_-) &= \frac{m_\pi^2}{Q^2+m_\pi^2}
\frac{f_\pi}{2m_q}\,\Gamma_\pi^j(k,Q)\\
&
\equiv \frac{\tau^j}{2} \frac{m_\pi^2}{Q^2+m_\pi^2} \frac{1}{m_q}i\gamma_5\Sigma_B(k_+^2,k_-^2) \,,
\label{psvx2}
\end{align}
as the expressions for the axial-vector and pseudoscalar vertices, which are completely determined by the dressed-quark propagator.

Note that, in the chiral limit, Eq.\,\eqref{axvx2} becomes
\begin{align}
\label{axvx20}
\nonumber
& \Gamma^{j,m_q=0}_{5\mu}(k_+,k_-) = \frac{\tau^j}{2}\gamma_5
\bigg[ \gamma_\mu \Sigma_A(k_+^2,k_-^2) \\
&
\quad +2\gamma\cdot k k_\mu \Delta_A(k_+^2,k_-^2)
+ 2i\frac{Q_\mu}{Q^2}\Sigma_B(k_+^2,k_-^2)\bigg]\,,
\end{align}
which is precisely the solution of the chiral-limit AXWTI,
\begin{align}
\label{axwti0}
Q_\mu \Gamma^{j,m_q=0}_{5\mu}&(k_+,k_-) \nonumber \\
& \quad=  S^{-1}(k_+)i\gamma_5\frac{\tau^j}{2}+
\frac{\tau^j}{2}i\gamma_5S^{-1}(k_-)\,.
\end{align}
Eq.\,\eqref{axvx20} has been employed previously \cite{Hellstern:1997pg, Bloch:1999rm, Roberts:2007jh}.

It is worth emphasising here that Eqs.\,\eqref{axvx2}, \eqref{psvx2} are minimal \emph{Ans\"atze}, deliberately constructed to be consistent with all other elements in our QCD-kindred framework.  Since there are no associated parameters, these \emph{Ans\"atze} could lead to poor results for the form factors considered herein.  As will become evident, they do not.  On the contrary, they yield sound predictions for the entire array of studied form factors.  If one were to add additional terms, then other elements of the framework would require compensating modifications.  That would serve no useful purpose herein.  We therefore leave such considerations to the future instant when \emph{ab initio} treatments are available for every Schwinger function needed to deliver predictions for all nucleon form factors on the entire domain of empirically accessible $Q^2$.


\subsubsection{Seagull terms}
\label{subsecsg}

Since diquark correlations are nonpointlike, any considered current must couple to the diquark BS amplitudes, leading to the seagull terms. Within the nucleon's quark+diquark picture, the seagull diagrams for the electromagnetic current were first derived in Ref.\,\cite{Oettel:1999gc}: they are essential to guarantee current conservation in any realistic study. However, the contribution coming from the seagull diagrams when considering the axial and pseudoscalar currents has eluded understanding for more than twenty years, partly because there is no trivial relation between the vertices and propagators of the diquarks for these currents.

We derive the axial-vector and pseudoscalar seagull terms here by using a strategy similar to the one followed in Ref.\,\cite{Oettel:1999gc}. First, the equal-time commutators of the axial current operator can be calculated straightforwardly from Eq.\,\eqref{jaxx}:
\begin{subequations}
\begin{align}
\label{jcomm}
[{\mathpzc A}_{5\mu=4}^j(x),\psi(y)]_{x_4=y_4} &= \frac{\tau^j}{2}\gamma_5\psi(x)\delta^{(4)}(x-y),\\
[{\mathpzc A}_{5\mu=4}^j(x),\bar{\psi}(y)]_{x_4=y_4} &= \bar{\psi}(x)\gamma_5\frac{\tau^j}{2}\delta^{(4)}(x-y).
\end{align}
\end{subequations}

Second, we define the five-point Schwinger function $G^{j}_{5\mu}$ as the coupling of the quark+quark scattering kernel with the axial current,
\begin{align}
\label{sg0a}
\nonumber
 &(2\pi)^4\delta^{(4)}(k_1+k_2+q_1+q_2-Q)G^{j}_{5\mu}(k,K';q,K)\\
 \nonumber
:= &\int d^4x_1d^4x_2d^4y_1d^4y_2d^4z\,
e^{-ik_1x_1}e^{-ik_2x_2}e^{-iq_1y_1}e^{-iq_2y_2}\\
\times & e^{iQz}\big\langle T\big\{\psi(x_1)\psi(x_2)\bar{\psi}(y_1)\bar{\psi}(y_2){\mathpzc A}_{5\mu}^{j}(z)
\big\}\big\rangle \,,
\end{align}
and, correspondingly, $G^{j}_5$ is the coupling of the same quark+quark scattering kernel with the pseudoscalar current,
\begin{align}
\label{sg0b}
\nonumber
 &(2\pi)^4\delta^{(4)}(k_1+k_2+q_1+q_2-Q)G^{j}_5(k,K';q,K)\\
 \nonumber
:= &\int d^4x_1d^4x_2d^4y_1d^4y_2d^4z\,
e^{-ik_1x_1}e^{-ik_2x_2}e^{-iq_1y_1}e^{-iq_2y_2}\\
\times & e^{iQz}\big\langle T\big\{\psi(x_1)\psi(x_2)\bar{\psi}(y_1)\bar{\psi}(y_2){\mathpzc P}_5^j(z)
\big\}\big\rangle \,,
\end{align}
where $k=(k_1-k_2)/2$, $q=(q_1-q_2)/2$, $k_1+k_2=K'$, $-q_1-q_2=K$ and $K'=K+Q$.

Now, taking the partial derivative of the five-point correlation function $\big\langle T\big\{ \psi(x_1) \psi(x_2) \bar{\psi}(y_1) \bar{\psi}(y_2){\mathpzc A}_{5\mu}^{j}(z) \big\}\big\rangle$, and using the PCAC relation, Eq.~\eqref{pcacx}, we get
\begin{align}
\label{sg1}
\nonumber
&\frac{\partial}{\partial z_\mu}
\big\langle T\big\{\psi(x_1)\psi(x_2)\bar{\psi}(y_1)\bar{\psi}(y_2){\mathpzc A}_{5\mu}^{j}(z)
\big\}\big\rangle\\
\nonumber
+& 2m_q\big\langle T\big\{\psi(x_1)\psi(x_2)\bar{\psi}(y_1)\bar{\psi}(y_2){\mathpzc P}_5^j(z)
\big\}\big\rangle \\
\nonumber
=&\delta^{(4)}(z-x_1)\big\langle T\big\{\big(\frac{\tau^j}{2}\gamma_5\psi(x_1)\big)\psi(x_2)\bar{\psi}(y_1)\bar{\psi}(y_2)\big\}\big\rangle\\
\nonumber
+& \delta^{(4)}(z-x_2)\big\langle T\big\{\psi(x_1)\big(\frac{\tau^j}{2}\gamma_5\psi(x_2)\big)\bar{\psi}(y_1)\bar{\psi}(y_2)\big\}\big\rangle\\
\nonumber
+& \delta^{(4)}(z-y_1)\big\langle T\big\{\psi(x_1)\psi(x_2)\big(\bar{\psi}(y_1)\gamma_5\frac{\tau^j}{2}\big)\bar{\psi}(y_2)\big\}\big\rangle\\
+& \delta^{(4)}(z-y_2)\big\langle T\big\{\psi(x_1)\psi(x_2)\bar{\psi}(y_1)\big(\bar{\psi}(y_2)\gamma_5\frac{\tau^j}{2}\big)\big\}\big\rangle \,.
\end{align}
Performing a Fourier transform of both sides and introducing a compact notation for the sum over diquark contributions in the resulting Fourier amplitudes, where $j$ expresses the evident isospin structure:
\begin{subequations}
\begin{align}
G_{5\mu}^j(k,K^\prime;q,K) & = \sum_{J^P} G_{5\mu}^{j,J^P}(k,K^\prime;q,K)\,, \\
G_{5}^j(k,K^\prime;q,K) & = \sum_{J^P} G_{5\mu}^{j,J^P}(k,K^\prime;q,K)\,,
\end{align}
\end{subequations}
one can use Eqs.\,\eqref{sqansatz}, \eqref{sg0a}, \eqref{sg0b}, to arrive at the momentum space form of Eq.\,\eqref{sg1}, \emph{viz}.\
\begin{subequations}
\begin{align}
\nonumber
& Q_\mu G^{j}_{5\mu}(k,K';q,K)+2im_q G_5^{j}(k,K';q,K)\\
=& \sum_{J^P}\left(Q_\mu G^{j,J^P}_{5\mu}+2im_q G^{j,J^P}_{5}\right)(k,K';q,K)\\
= & \sum_{J^P}\sum_{i=1}^{4} \tilde{G}^{j,J^P}_{(i)}(k,K';q,K) \label{NewOne} \\
=: & \sum_{J^P}\sum_{i=1}^{4} \left(Q_\mu G^{j,J^P}_{5\mu,(i)}+2im_q G^{j,J^P}_{5,(i)}\right)(k,K';q,K)\,.
\label{NewOneTwo}
\end{align}
\end{subequations}
Here, the identification of terms is straightforward: the first sum is over the separate diquark contributions; the second sum expresses that over the Fourier amplitudes of the four terms on the right-hand-side in Eq.\,\eqref{sg1}; and Eq.\,\eqref{NewOneTwo} serves as a formal expedient for subsequent use.  In Eq.\,\eqref{NewOne}, writing diquark wave functions as $\Lambda^{J^P}$:
\begin{subequations}
\begin{align}
\label{sg2a}
\nonumber
\tilde{G}^{j,J^P}_{(1)} =& \frac{\tau^j}{2}i\gamma_5\Lambda^{J^P}(k-Q/2;K)\,\\
&\hspace*{2cm}\times\Delta^{J^P}(K)\,
\bar{\Lambda}^{J^P}(q;-K) \,,\\
\nonumber
\label{sg2b}
\tilde{G}^{j,J^P}_{(2)} =& \Lambda^{J^P}(k+Q/2;K)(i\gamma_5\frac{\tau^j}{2})^{\rm T}\,\\
&\hspace*{2cm}\times\Delta^{J^P}(K)\,\bar{\Lambda}^{J^P}(q;-K) \,,\\
\nonumber
\label{sg2c}
\tilde{G}^{j,J^P}_{(3)} =& \Lambda^{J^P}(k;K')\,\Delta^{J^P}(K')\,\\
&\hspace*{1cm}\times(i\gamma_5\frac{\tau^j}{2})^{\rm T} \bar{\Lambda}^{J^P}(q-Q/2;-K') \,,\\
\label{sg2d}
\nonumber
\tilde{G}^{j,J^P}_{(4)} =& \Lambda^{J^P}(k;K')\,\Delta^{J^P}(K')\,\\
&\hspace*{1cm}\times \bar{\Lambda}^{J^P}(q+Q/2;-K')\frac{\tau^j}{2}i\gamma_5\,.
\end{align}
\end{subequations}

Continuing toward expressions for the one-particle-irreducible 4-point seagull couplings of the axial-vector and pseudoscalar currents to the quarks and the diquarks, we exploit Eq.\,\eqref{NewOneTwo} and define $\chi^{j,J^P}_{5\mu,(i)}$ and $\chi^{j,J^P}_{5,(i)}$ via
\begin{align}
\nonumber
&\big( S(k_1)\chi^{j,J^P}_{5\mu,(i)}(k_1,k_2,K)S^{\rm T}(k_2)\big)\Delta^{J^P}(K)\\
:=& ({\cal Z}^{J^P})^{-1}\int\frac{d^4q}{(2\pi)^4}\,G^{j,J^P}_{5\mu,(i)}(k,K';q,K)\,\Gamma^{J^P}(q;K)\,,
\end{align}
and
\begin{align}
\nonumber
&\big( S(k_1)\chi^{j,J^P}_{5,(i)}(k_1,k_2,K)S^{\rm T}(k_2)\big)\Delta^{J^P}(K)\\
:=& ({\cal Z}^{J^P})^{-1}\int\frac{d^4q}{(2\pi)^4}\,G^{j,J^P}_{5,(i)}(k,K';q,K)\,\Gamma^{J^P}(q;K)\,.
\end{align}
Here
\begin{equation}
{\cal Z}^{J^P} :=\int \frac{d^4q}{(2\pi)^4}\,{\rm tr_{CDF}}\big[	\bar{\Lambda}^{J^P}(q;-K)\,
\Gamma^{J^P}(q;K)\big]\,,
\end{equation}
and the trace is over colour, flavour and Dirac matrices. Then, Eq.\,\eqref{sg2a} entails
\begin{align}
\nonumber
&\big( S(k_1)\big\{Q_\mu\,\chi^{j,J^P}_{5\mu,(1)}(k_1,k_2,K)\\
\nonumber
+&2im_q\,\chi^{j,J^P}_{5,(1)}(k_1,k_2,K)\big\}S^{\rm T}(k_2)\big)\Delta^{J^P}(K)\\
\nonumber
=& ({\cal Z}^{J^P})^{-1}\int\frac{d^4q}{(2\pi)^4}\,
\tilde{G}^{j,J^P}_{(1)}(k,K';q,K)\Gamma^{J^P}(q;K)\\
\nonumber
=& ({\cal Z}^{J^P})^{-1}\int\frac{d^4q}{(2\pi)^4}\,
\big(\frac{\tau^j}{2}i\gamma_5\Lambda^{J^P}(k- Q/2;K)\big)\,\,\\
\nonumber
\times &\Delta^{J^P}(K){\rm tr_{CDF}}\big[	\bar{\Lambda}^{J^P}(q;-K)\,
\Gamma^{J^P}(q;K)\big]\\
=& \frac{\tau^j}{2}i\gamma_5\Lambda^{J^P}(k-Q/2;K)\,\Delta^{J^P}(K)\,,
\end{align}
and using the AXWTI, Eq.\,\eqref{axwti}, we arrive at
\begin{align}
\nonumber
& Q_\mu\,\chi^{j,J^P}_{5\mu,(1)}(k_1,k_2,K)
+2im_q\,\chi^{j,J^P}_{5,(1)}(k_1,k_2,K) \\
\nonumber
=& \big(Q_\mu\Gamma^{j}_{5\mu}(k_1,k_1-Q)+2im_q\,\Gamma_5^{j}(k_1,k_1-Q)\big)
\\ \times & S(k_1-Q)\Gamma^{J^P}(k- Q/2;K) \nonumber \\
& \quad - \frac{\tau^j}{2}i\gamma_5\Gamma^{J^P}(k- Q/2;K)\,.
\end{align}
Proceeding similarly, we arrive at
\begin{align}
\nonumber
& Q_\mu\,\chi^{j,J^P}_{5\mu,(2)}(k_1,k_2,K)
+2m_q\,\chi^{j,J^P}_{5,(2)}(k_1,k_2,K) \\
\nonumber
=& \Gamma^{J^P}(k+Q/2;K)S^{\rm T}(k_2-Q)
\big(Q_\mu\Gamma_{5\mu}^{j}(k_2,k_2-Q) \\
+&2im_q\,\Gamma^{j}_5(k_2,k_2-Q)\big)^{\rm T} \nonumber \\
& \quad -\Gamma^{J^P}(k+Q/2;K)(i\gamma_5\frac{\tau^j}{2})^{\rm T}\,,
\end{align}
for the second contribution, Eq.\,\eqref{sg2b}. Besides, the third and fourth contributions are
\begin{align}
\nonumber
&\big( S(k_1)\big\{Q_\mu\,\chi^{j,J^P}_{5\mu,(3)}(k_1,k_2,K)\\
\nonumber
+&2im_q\,\chi^{j,J^P}_{5,(3)}(k_1,k_2,K)\big\}S^{\rm T}(k_2)\big)\Delta^{J^P}(K)\\
\nonumber
=& ({\cal Z}^{J^P})^{-1}\int\frac{d^4q}{(2\pi)^4}\,
\Lambda^{J^P}(k;K')\,\Delta^{J^P}(K')\times\\
&{\rm tr_{CDF}}\big[\big((i\gamma_5\frac{\tau^j}{2})^{\rm T}\bar{\Lambda}^{J^P}(q-Q/2;-K')\big)
\Gamma^{J^P}(q;K)\big]\,,
\end{align}
for Eq.~\eqref{sg2c}, and
\begin{align}
\nonumber
&\big( S(k_1)\big\{Q_\mu\,\chi^{j,J^P}_{5\mu,(4)}(k_1,k_2,K)\\
\nonumber
+&2im_q\,\chi^{j,J^P}_{5,(4)}(k_1,k_2,K)\big\}S^{\rm T}(k_2)\big)\Delta^{J^P}(K)\\
\nonumber
=& ({\cal Z}^{J^P})^{-1}\int\frac{d^4q}{(2\pi)^4}\,
\Lambda^{J^P}(k;K')\,\Delta^{J^P}(K')\times\\
&{\rm tr_{CDF}}\big[\big(\bar{\Lambda}^{J^P}(q+Q/2;-K')\frac{\tau^j}{2}i\gamma_5\big)
\Gamma^{J^P}(q;K)\big] \,,
\end{align}
for Eq.~\eqref{sg2d}, respectively.

Let us first study the third and fourth contributions in more detail. Their flavour structure involves four different diquark configurations: scalar diquark, $[ud]_{0^+}$, and axial-vector diquarks, $\{ud\}_{1^+}$, $\{uu\}_{1^+}$, $\{dd\}_{1^+}$. When the correlations $[ud]_{0^+}$ and $\{ud\}_{1^+}$ are involved, we obtain trivial zeros because the traces of the flavour parts vanish (the diquarks' flavour matrices are listed in Appendix\,\ref{subappendixdqbsa}):
\begin{align}
{\rm tr_F}\big[(t^0_f)^\dagger(\frac{\tau^j}{2})(t^0_f)\big] &=
{\rm tr_F}\big[(\frac{\tau^j}{2})^\dagger(t^0_f)^\dagger(t^0_f)\big] = 0,\\
{\rm tr_F}\big[(t^2_f)^\dagger(\frac{\tau^j}{2})(t^2_f)\big] &=
{\rm tr_F}\big[(\frac{\tau^j}{2})^\dagger(t^2_f)^\dagger(t^2_f)\big] = 0\,.
\end{align}

However, when the correlations $\{uu\}_{1^+}$ and $\{dd\}_{1^+}$ are involved, the flavour traces are not zero and thus we need to investigate them more closely. For $\{uu\}_{1^+}$ we have
\begin{align}
\label{sg3}
&
\big( S(k_1)\big\{Q_\mu\,\chi^{j,1^+}_{5\mu,\alpha,(3)}(k_1,k_2,K) \nonumber \\
&
+ 2im_q\,\chi^{j,1^+}_{5,\alpha,(3)}(k_1,k_2,K)\big\}S^{\rm T}(k_2)\big)\Delta^{1^+}_{\alpha\delta}(K) \nonumber \\
&
= ({\cal Z}_{\gamma\delta}^{1^+})^{-1}
\Lambda_\alpha^{1^+}(k;K^\prime)\Delta^{1^+}_{\alpha\beta}(K') \nonumber \\
&
\times \int\frac{d^4q}{(2\pi)^4}{\rm tr_{CDF}}\bigg[(i\gamma_5\frac{\tau^j}{2})^{\rm T}
\bar{\Lambda}_\beta^{1^+}(q-Q/2;-K^\prime)\Gamma_\gamma^{1^+}(q)\bigg] \nonumber \\
&
= ({\cal Z}_{\gamma\delta}^{1^+})^{-1} \Lambda_\alpha^{1^+}(k;K^\prime)\Delta^{1^+}_{\alpha\beta}(K')\times
g_{1^+}^2\times\delta_{j3} \nonumber \\
&
\times \int\frac{d^4q}{(2\pi)^4}
{\mathpzc F}((q-Q/2)^2/\omega_{1^+}^2)
{\mathpzc F}(q^2/\omega_{1^+}^2) \nonumber \\
&
\times 4\sigma_V((q-K/2-Q)^2)\sigma_V((q+K/2)^2) \nonumber \\
&
\times \big[\epsilon_{\rho\sigma\beta\gamma}q_\rho (K_\sigma+Q_\sigma)
-\epsilon_{\rho\sigma\beta\gamma}Q_\rho K_\sigma/2\big]\,,
\end{align}
where $\sigma_V$ is a quark-propagator dressing function, Eq.\,\eqref{Spsigma}, and
\begin{equation}
{\cal Z}^{1^+}_{\gamma\delta} :=\int \frac{d^4q}{(2\pi)^4}\,{\rm tr_{CDF}}\big[\bar{\Lambda}_\gamma^{1^+}(q;-K)\,
\Gamma^{1^+}_\delta(q)\big]\,.
\end{equation}
In this derivation, we used the explicit form of the axial-vector BS amplitude, Eq.\,\eqref{axBSAs}, which only depends on the relative momentum. Similarly, we obtain
{\allowdisplaybreaks
\begin{align}
\label{sg4}
&
\big( S(k_1)\big\{Q_\mu\,\chi^{j,1^+}_{5\mu,\alpha,(4)}(k_1,k_2,K) \nonumber \\
&
+ 2im_q\chi^{j,1^+}_{5,\alpha,(4)}(k_1,k_2,K)\big\}S^{\rm T}(k_2)\big)\Delta^{1^+}_{\alpha\delta}(K) \nonumber \\
&
= ({\cal Z}_{\gamma\delta}^{1^+})^{-1}
\Lambda_\alpha^{1^+}(k;K^\prime)\Delta^{1^+}_{\alpha\beta}(K')\times
g_{1^+}^2\times\delta_{j3} \nonumber \\
&
\times \int\frac{d^4q}{(2\pi)^4}
{\mathpzc F}((q+Q/2)^2/\omega_{1^+}^2)
{\mathpzc F}(q^2/\omega_{1^+}^2) \nonumber \\
&
\times (-4)\sigma_V((q+K/2+Q)^2)\sigma_V((q-K/2)^2) \nonumber \\
&
\times \big[\epsilon_{\rho\sigma\beta\gamma}q_\rho (K_\sigma+Q_\sigma)
+\epsilon_{\rho\sigma\beta\gamma}Q_\rho K_\sigma/2\big]\,.
\end{align}
}

Obviously, neither Eq.\,\eqref{sg3} nor Eq.\,\eqref{sg4} vanishes if $j=3$. Nevertheless, if we add them together and subsequently take the limit $Q\to0$, we find
\begin{align}
\nonumber
0& =  Q_\mu\big\{\chi^{j,J^P}_{5\mu,(3)}(k_1,k_2,K) +
\chi^{j,J^P}_{5\mu,(4)}(k_1,k_2,K)\big\} \\
+& 2im_q\big\{\chi^{j,J^P}_{5,(3)}(k_1,k_2,K) +
\chi^{j,J^P}_{5,(4)}(k_1,k_2,K)\big\}\,. \label{OnTheWay}
\end{align}
The same result can be shown for $\{dd\}_{1^+}$.

Therefore, for all kinds of diquark correlations we can write a unified result\footnote{Here the diquark amplitudes only depend on the relative momenta, since the derivation for $\{uu\}_{1^+}$ or $\{dd\}_{1^+}$ depends on its explicit form in our framework, the result \emph{may} be more complicated for its full amplitude.}
{\allowdisplaybreaks
\begin{align}
\nonumber
& Q_\mu\,\chi^{j,J^P}_{5\mu}(k_1,k_2,K)+2im_q\,\chi^{j,J^P}_5(k_1,k_2,K)\\
\nonumber
:=& \sum_{i=1}^{4}
 Q_\mu\,\chi^{j,J^P}_{5\mu,(i)}(k_1,k_2,K)+2im_q\,\chi^{j,J^P}_{5,(i)}(k_1,k_2,K)\\
\nonumber
\approx \; & \big(Q_{\mu}\Gamma^{j}_{5\mu}+2im_q\,\Gamma^{j}_5\big)S(k_1-Q)
\Gamma^{J^P}(k- Q/2) \\
\nonumber
+&\Gamma^{J^P}(k+Q/2)S^{\rm T}(k_2-Q)
\big(Q_\mu\Gamma^{j}_{5\mu}+2im_q\,\Gamma^{j}_5\big)^{\rm T}\\
-&\frac{\tau^j}{2}i\gamma_5\Gamma^{J^P}(k-Q/2)
-\Gamma^{J^P}(k+Q/2)(i\gamma_5\frac{\tau^j}{2})^{\rm T}\,. \label{TotalWGTI}
\end{align}
In arriving at Eq.\,\eqref{TotalWGTI}, we used Eq.\,\eqref{OnTheWay} and supposed it to be valid outside the neighbourhood $Q^2\simeq 0$.

The approximation expressed in Eq.\,\eqref{TotalWGTI} is kindred to that employed in deriving the seagull terms for the electromagnetic current \cite{Oettel:1999gc}; nevertheless, one might ask after its accuracy.
That is best judged by explicitly computing the impact of the neglected terms, \emph{viz}.\ the sum of Eqs.\,\eqref{sg3}, \eqref{sg4}, on the form factors calculated herein.

In view of this aim, we first observe that the axial form factor, $G_A$, is determined by the $Q$-transverse piece of the nucleon axial-current, see Eq.\,\eqref{gaproj}.  Hence, $G_A$ receives no contribution from any part of the seagull construction, which leaves only $G_P$, $G_5$.

The complete nucleon axial current, Fig.\,\ref{figcurrent}, contains the seagull contribution expressed in the sum of diagrams $(5)$ and $(6)$.
Considering this part, using the approximation in Eq.\,\eqref{TotalWGTI}, one is required to evaluate an eight dimensional (two loop) integral.  However, with the inclusion of Eqs.\,\eqref{sg3}, \eqref{sg4}, the additional contribution is seemingly a sixteen dimensional (four loop) integral.  Fortunately, using the Faddeev equation, it can be reduced to a three-loop calculation.  All such multi-loop integrals must be evaluated using Monte Carlo techniques.

Following this procedure, we compared the results obtained for $G_P$, $G_5$ by first neglecting Eqs.\,\eqref{sg3}, \eqref{sg4} and then after their inclusion.
%
In both cases, the approximation and the complete calculation produce results that are indistinguishable within the linewidth of any legible curves.  Stated quantitatively: when evaluated on the physically relevant domain, the relative ${\mathpzc L}_1$ difference between comparable curves is $\lesssim 1$\%.

Evidently, then, Eq.\,\eqref{TotalWGTI} is a sound approximation.  Its merits are amplified by the fact that it enables us to both avoid the need for calculation of three-loop diagrams and, as we shall subsequently see, obtain closed algebraic expressions for the seagull terms and diquark vertices.

%
%
%

As in the electromagnetic case~\cite{Oettel:1999gc}, $\chi^{j,J^P}_{5\mu}$ and $\chi^{j,J^P}_{5}$ can be both separated into two parts
\begin{subequations}
\begin{align}
\chi^{j,J^P}_{5\mu} &= \chi^{j,J^P}_{5\mu,[{\rm legs}]} + \chi^{j,J^P}_{5\mu,[{\rm sg}]}\,,\\
\chi^{j,J^P}_{5} &= \chi^{j,J^P}_{5,[{\rm legs}]} + \chi^{j,J^P}_{5,[{\rm sg}]}\,,
\end{align}
\end{subequations}
where: $\chi^{j,J^P}_{5\mu,[{\rm legs}]}$ and $\chi^{j,J^P}_{5,[{\rm legs}]}$ describe the coupling of the axial and pseudoscalar currents to the amputated quark legs separately;
and $\chi^{j,J^P}_{5\mu,[{\rm sg}]}$ and $\chi^{j,J^P}_{5,[{\rm sg}]}$ describe the axial-vector and pseudoscalar one-particle irreducible seagull couplings.

The ``legs'' and ``sg'' terms separately satisfy the following Ward-Green-Takahashi identities:
\begin{align}
\label{chilegsaxwti}
\nonumber
& Q_\mu\,\chi^{j,J^P}_{5\mu,[{\rm legs}]}(k_1,k_2,K)+2im_q\,\chi^{j,J^P}_{5,[{\rm legs}]}(k_1,k_2,K)\\
\nonumber
=& \big(Q_{\mu}\Gamma^{j}_{5\mu}+2im_q\,\Gamma^{j}_5\big)S(k_1-Q)
\Gamma^{J^P}(k- Q/2) \\
+&\Gamma^{J^P}(k+Q/2)S^{\rm T}(k_2-Q)
\big(Q_\mu\Gamma^{j}_{5\mu}+2im_q\,\Gamma^{j}_5\big)^{\rm T}\,,
\end{align}
\begin{align}
\label{sgwti}
&
Q_\mu\,\chi^{j,J^P}_{5\mu,[{\rm sg}]}(k,Q)+2im_q\,\chi^{j,J^P}_{5,[{\rm sg}]}(k,Q)  \nonumber \\
&
= - \frac{\tau^j}{2}i\gamma_5\Gamma^{J^P}(k-Q/2) - \Gamma^{J^P}(k+Q/2)(i\gamma_5\frac{\tau^j}{2})^{\rm T}\,,
\end{align}
whereby Eq.\,\eqref{TotalWGTI} is saturated.  It is worth highlighting that Eqs.\,\eqref{chilegsaxwti}, \eqref{sgwti} are complete because axial-vector/pseudoscalar probes do not couple to scalar targets and the axial-vector diquark propagator is symmetric.  Hence, in contrast to the electromagnetic case, the seagull identity, Eq.\,\eqref{sgwti}, does not have a diquark part.  As a consequence, AXWTIs of the relevant current-diquark vertices are both equal to zero, see Section \,\ref{subsecdqvx}.


In this work, we therefore employ the following \emph{Ans\"atze}
\begin{align}
\nonumber
\chi^{j,J^P}_{5\mu,[{\rm sg}]}(k,Q)=
&- \frac{Q_\mu}{Q^2+m_\pi^2}
\bigg[\frac{\tau^j}{2}i\gamma_5\Gamma^{J^P}(k-Q/2)
+\\
&\Gamma^{J^P}(k+Q/2)(i\gamma_5\frac{\tau^j}{2})^{\rm T}\bigg]\,, \label{axsg}
\end{align}
and
\begin{align}
\label{pssg}
\nonumber
 i\chi^{j,J^P}_{5,[{\rm sg}]}(k,Q)
= &
- \frac{1}{2m_q}\frac{m_\pi^2}{Q^2+m_\pi^2}
\bigg[\frac{\tau^j}{2}i\gamma_5\Gamma^{J^P}(k- Q/2)
+\\
&\Gamma^{J^P}(k+Q/2)(i\gamma_5\frac{\tau^j}{2})^{\rm T}\bigg] \,.
\end{align}

In a similar way, we can obtain the seagulls' charge conjugations:
\begin{align}
\nonumber
\bar{\chi}^{j,J^P}_{5\mu,[{\rm sg}]}(k,Q)=
&- \frac{Q_\mu}{Q^2+m_\pi^2}
\bigg[\bar{\Gamma}^{J^P}(k+Q/2)\frac{\tau^j}{2}i\gamma_5
+\\
&(i\gamma_5\frac{\tau^j}{2})^{\rm T}\bar{\Gamma}^{J^P}(k-Q/2)\bigg]\,, \label{axsgcg}
\end{align}
and
\begin{align}
\label{pssgcg}
\nonumber
 i\bar{\chi}^{j,J^P}_{5,[{\rm sg}]}(k,Q)
= &
- \frac{1}{2m_q}\frac{m_\pi^2}{Q^2+m_\pi^2}
\bigg[\bar{\Gamma}^{J^P}(k+Q/2)\frac{\tau^j}{2}i\gamma_5
+\\
&(i\gamma_5\frac{\tau^j}{2})^{\rm T}\bar{\Gamma}^{J^P}(k-Q/2)\bigg]\,.
\end{align}
Equations~(\ref{axsg})-(\ref{pssgcg}) are novel and the most important results in this section.

The axial-vector Ward-Takahashi identity in Eq~\eqref{sgwti}, cannot completely determine the seagull terms, e.g.\ their transverse parts are neglected herein. It is possible to use more complicated \emph{Ans\"atze}; however, Eqs.\,(\ref{axsg})\,--\,(\ref{pssgcg}) are simple expressions which ensure that the computed $G_A$, $G_P$ and $G_5$ form factors satisfy the PCAC and the Goldberger-Treiman relations.

It should also be stressed that the starting point for the construction of the electromagnetic seagull terms was to keep them regular~\cite{Oettel:1999gc}, which is not a suitable strategy here. The axial-vector and pseudoscalar seagulls could be singular, just like the analogous quark vertices. Consequently, Eqs.\,(\ref{axsg})\,--\,(\ref{pssgcg}) have a pole at the pion mass.


With the dressed quark vertices and the seagull terms in hand, we can construct the remaining elements: the axial-vector and pseudoscalar current-diquark vertices. Here, we use the same strategy as in Ref.~\cite{Oettel:2000jj}.
In short, we focus on the dominant tensor structure and compute its strength in the on-shell nucleon by: inserting Fig.\,\ref{figdqvx} into diagrams (2) and (3) of Fig.\,\ref{figcurrent};
evaluating the result in each case;
then equating the values with those obtained by computing the original one-loop diagrams from Fig.\,\ref{figcurrent} with the probe-diquark vertex written as the appropriate coupling times the dominant associated tensor structure.
Plainly, each current-diquark vertex receives four contributions, \emph{viz}.\ those depicted in Fig.\,\ref{figdqvx}. Two of them are generated by the coupling of the current to the upper and lower quark lines of the resolved diquark. The other two are the current couplings to the diquark amplitudes, which are again seagull terms. In Ref.~\cite{Oettel:2000jj}, the authors only took into account the first two contributions; all four are included in this work.

\begin{figure}[!t]
\centerline{%
\includegraphics[clip, width=0.45\textwidth]{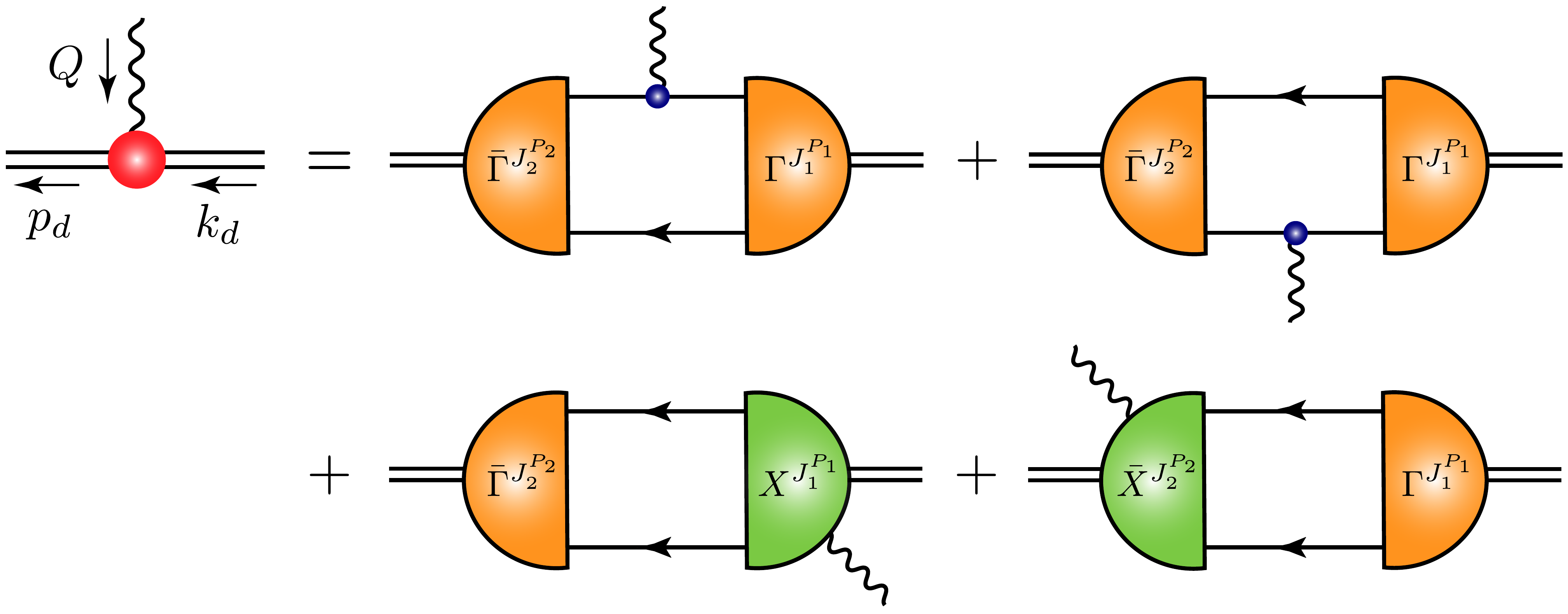}}
\caption{\label{figdqvx} Interaction vertex for the $J_1^{P_1}\to J_2^{P_2}$ diquark transition ($p_d=k_d+Q$): \emph{single line}, quark propagator; \emph{undulating  line}, the axial or pseudoscalar current; $\Gamma$, diquark correlation amplitude; \emph{double line}, diquark propagator; and $\chi$, seagull interaction.}
\end{figure}

\subsubsection{Current-diquark vertices}
\label{subsecdqvx}

Furthermore, we need to study the flavour structures of the currents and diquark correlations. Given that the scalar diquark is also isoscalar, associated with $\tau^2$, then there is no $[ud]_{0^+}\to [ud]_{0^+}$ contribution. On the other hand, transitions between scalar and axial-vector diquarks are possible in the form of: charged currents, $\{dd\}_{1^+}\to [ud]_{0^+}$ and $\{uu\}_{1^+}\to [ud]_{0^+}$; and neutral currents, $\{ud\}_{1^+}\to [ud]_{0^+}$. Note too that, in the isospin limit, the coupling strengths are identical in magnitude. Moreover, the constructed vertices must properly express the momentum-dependence of the diquark amplitudes, which is a natural consequence of the nonpointlike character of diquark correlations.

Taking the above arguments into account, we obtain the following expressions:
{\allowdisplaybreaks
\begin{itemize}
\item[i)] $\{qq\}_{1^+}$--pseudoscalar-current vertex
\begin{align}
\label{psdqvxAA}
&
\Gamma^{aa}_{5,\alpha\beta}(p_d,k_d) =  -\frac{1}{2m_q}\frac{m_\pi^2}{Q^2+m_\pi^2} \nonumber \\
&
\times \bigg[\kappa_{\rm ps}^{aa}
\frac{M_q^E}{m_N}\epsilon_{\alpha\beta\gamma\delta}(p_d+k_d)_\gamma Q_\delta\bigg]\,d(\tau^{aa})\,,
\end{align}
\item[ii)] $\{qq\}_{1^+}$--axial-current vertex
\begin{align}
\label{axdqvxAA}
i\Gamma^{aa}_{5\mu,\alpha\beta}&(p_d,k_d)=\bigg[\frac{\kappa_{\rm ax}^{aa}}{2}
\epsilon_{\mu\alpha\beta\nu}(p_d+k_d)_\nu + \frac{Q_\mu}{Q^2+m_\pi^2} \nonumber \\
&
\times  \kappa_{\rm ps}^{aa}
\frac{M_q^E}{m_N}\epsilon_{\alpha\beta\gamma\delta}(p_d+k_d)_\gamma Q_\delta \bigg]\,d(\tau^{aa})\,,
\end{align}
\item[iii)] Pseudoscalar-current induced ${0^+}\leftrightarrow{1^+}$ transition vertex
{\allowbreak
\begin{align}
\label{psdqvxSA}
&
\Gamma^{sa}_{5,\beta}(p_d,k_d) = \nonumber \\
&
=
\frac{1}{2m_q}\frac{m_\pi^2}{Q^2+m_\pi^2} \bigg[2i\kappa_{\rm ps}^{sa}M_q^E Q_\beta\bigg]\,d(\tau^{sa})\,,
\end{align}}
\item [iv)] Axial-current induced ${0^+}\leftrightarrow{1^+}$ transition vertex
\begin{align}
\label{axdqvxSA}
&
\Gamma^{sa}_{5\mu,\beta}(p_d,k_d) = \bigg[m_N\kappa_{\rm ax}^{sa}\delta_{\mu\beta} + \nonumber \\
&
+ \frac{Q_\mu}{Q^2+m_\pi^2}\big(-2\kappa_{\rm ps}^{sa}M_q^EQ_\beta\big)\bigg]\,d(\tau^{sa}) \,.
\end{align}
\end{itemize}}

For the ${1^+}\leftrightarrow{0^+}$ transition, we have $\Gamma^{as}_{5,\beta}=-\Gamma^{sa}_{5,\beta}$ and $\Gamma^{as}_{5\mu,\beta}=-\Gamma^{sa}_{5\mu,\beta}$.
In the expressions above, $k_d$ and $p_d$ are the diquark's incoming and outgoing momenta, with $p_d-k_d=Q$;
$M_q^E = 0.33\,$GeV is the Euclidean constituent-quark mass, defined as the solution of $p^2=M^2(p^2)$ (see Appendix\,\ref{subappendixqprop});
$\tau^{aa}_{\rm ax}=Q^2/[4m_{1^+}^2]$, $\tau^{sa}_{\rm ax}=Q^2/[4m_{0^+}m_{1^+}]$;
and $\kappa_{\rm ax}^{aa}$, $\kappa_{\rm ps}^{aa}$, $\kappa_{\rm ax}^{sa}$ and $\kappa_{\rm ps}^{sa}$ are the \emph{computed} $Q^2=0$ values of these couplings.
Further, emulating the electromagnetic current construction \cite{Cloet:2008re}, $d(x)=1/(1+x)$ is introduced to express diquark compositeness via form factor suppression on $Q^2>0$.  It is noteworthy that using dipole suppression instead, \emph{viz}.\ $d(x)^2$, no prediction in any image drawn herein changes by more than the linewidth because, in all cases, diagram (1) in Fig.\,\ref{figcurrent} is both dominant and hard, whereas all weak-boson--diquark interactions are soft and subdominant.

On the other hand, using Eqs.~\eqref{axwti} and~\eqref{sgwti}, we can get the AXWTIs of the current-diquark vertices:
\begin{subequations}
\label{dqwti}
\begin{align}
\label{dqwtiaa}
0 & = Q_\mu\,\Gamma^{aa}_{5\mu,\alpha\beta}(p_d,k_d)+2im_q\,	
\Gamma^{aa}_{5,\alpha\beta}(p_d,k_d) \,,\\
\label{dqwtisa}
0 & = Q_\mu\,\Gamma^{sa}_{5\mu,\beta}(p_d,k_d)+2im_q\,\Gamma^{sa}_{5,\beta}(p_d,k_d) \,.
\end{align}
\end{subequations}
It should be emphasised that Eqs.\,\eqref{dqwti} are independent of the explicit forms of the current-diquark vertices involved; hence, can be used as constraints for the vertices' algebraic forms. It is easy to see that Eqs.\,\eqref{psdqvxAA}\,--\,\eqref{axdqvxSA} satisfy the above AXWTIs precisely, from which we can get
\begin{subequations}
\begin{align}
\kappa_{\rm ps}^{aa} &= \frac{m_N}{2M_q^E}\kappa_{\rm ax}^{aa},\\
\kappa_{\rm ps}^{sa} &= \frac{m_N}{2M_q^E}\kappa_{\rm ax}^{sa}.
\end{align}
\end{subequations}
We verified these constraints numerically
\begin{subequations}
\label{kappav}
\begin{align}
\kappa_{\rm ax}^{aa}  &= 0.73\,,\\
\kappa_{\rm ps}^{aa}  &= 1.31\,,\\
\kappa_{\rm ax}^{sa}  &= 0.75\,,\\
\kappa_{\rm ps}^{sa}  &= 1.34\,,
\end{align}
\end{subequations}
from the direct computation of the four diagrams in Fig.\,\ref{figdqvx}.

As anticipated following Eq.\,\eqref{sgwti}, Eqs.\,\eqref{dqwti} both equate to zero because, unlike the electromagnetic case, the corresponding seagull terms have no contribution from the diquarks' leg.

We have now specified all the necessary building blocks to construct the diagrams for $J_{5\mu}^j(K,Q)$ and $J_{5}^j(K,Q)$ drawn in Fig.\,\ref{figcurrent}, with the corresponding expressions given in Appendix~\ref{appcurrentdia}.  We are therefore in a position to analytically verify the PCAC relation, Eq.\,\eqref{pcacn}.  This is completed in Appendix~\ref{subsecpcac1}.


\section{Numerical results}
\label{secnume}
Our predictions for the nucleon's axial and induced pseudoscalar form factors were sketched elsewhere \cite{Chen:2020wuq}.  Herein, so as to provide a complete, self-contained report, we recapitulate those results.  We subsequently describe predictions for the pseudoscalar (pion-nucleon) form factor and quantify the domain on which PCAC is satisfied.


The axial form factor, $G_A(Q^2)$, can be expanded at small momenta as
\begin{equation}
G_A(Q^2) = g_A \Big(\,1\,-\,\frac{\langle r_A^2\rangle}{6}Q^2 \,+\, \dots\, \Big) \,,	
\end{equation}
where $g_A:=G_A(0)$ is the nucleon axial charge, and $\langle r_A^2\rangle^{1/2}$ is the axial mean square radius with
\begin{align}
\label{DipoleGA}
\langle r_A^2\rangle &=
-6 \frac{d}{dQ^2} \bigg(\frac{G_A(Q^2)}{G_A(0)}\bigg)\bigg|_{Q^2=0} \,.
\end{align}
Sometimes, when comparing with other theoretical estimations, it is convenient to use a dipole \emph{Ansatz} for the axial form factor:
\begin{equation}
\label{ma}
G_A(Q^2) = \frac{g_A}{\big(1+Q^2/m_A^2\big)^2}\,,
\end{equation}
where $m_A$ is the so-called axial mass. It should be stressed that we use Eq.\,\eqref{ma} to interpolate the \emph{global} $Q^2$-behaviour of $G_A$, instead of relating it to $\langle r_A^2\rangle$ via $m_A=\sqrt{12/\langle r_A^2\rangle}$.


\begin{figure}[t]
\leftline{\hspace*{0.5em}{\large{\textsf{A}}}}
\vspace*{-5ex}
\includegraphics[clip, width=0.42\textwidth]{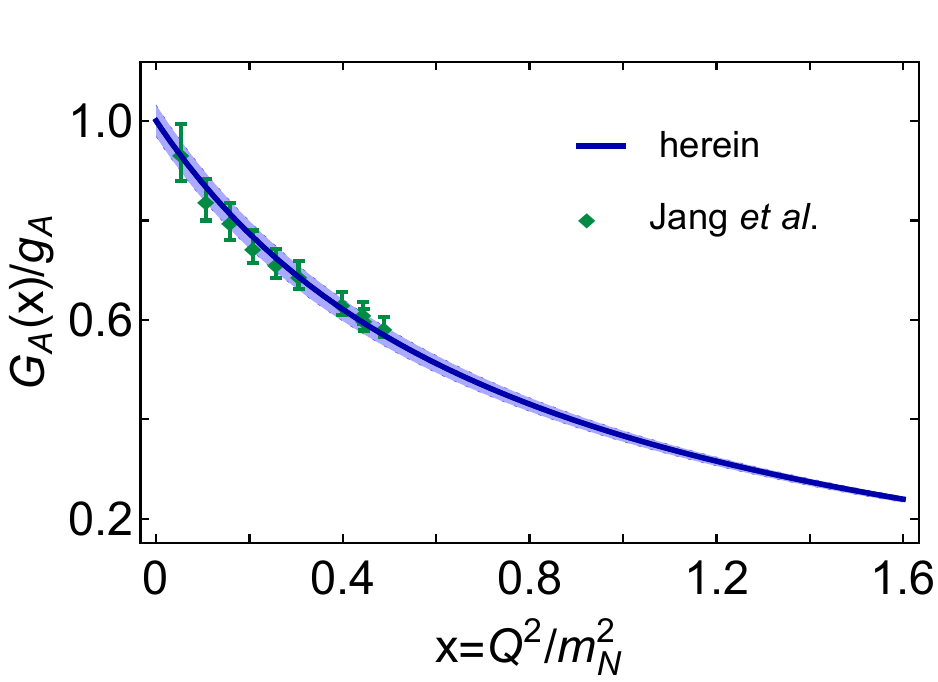}
\vspace*{1ex}

\leftline{\hspace*{0.5em}{\large{\textsf{B}}}}
\vspace*{-5ex}
\includegraphics[clip, width=0.42\textwidth]{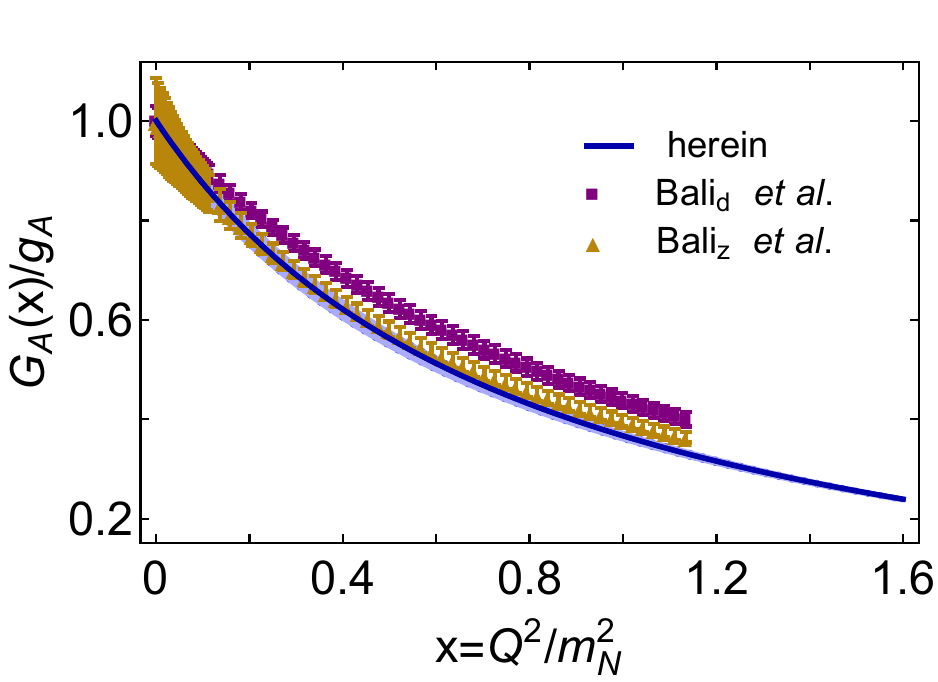}
\caption{\label{FigGAx}
\emph{Upper panel}\,--\,{\sf A}.  $G_A(x)/g_A$ calculated herein -- blue curve within lighter blue uncertainty band, compared with lQCD results from Ref.\,\cite{Jang:2019vkm} -- green diamonds. With respect to our central results, this comparison may be quantified by reporting the mean-$\chi^2$ value, which is 0.27.
\emph{Lower panel}\,--\,{\sf B}.  $G_A(x)/g_A$ calculated herein -- blue curve within lighter blue uncertainty band, compared with lQCD results from Ref.\,\cite{Bali:2019yiy} -- purple boxes [dipole] and golden triangles [$z$ expansion]. The mean-$\chi^2$ values are 10.86 [dipole] and 1.88 [$z$ expansion].
}
\end{figure}

\subsection{The axial and induced pseudoscalar form factors}
\label{subsecga}

Our Faddeev equation result for $G_A(x)/g_A$ is depicted in Fig.\,\ref{FigGAx},
together with results from lQCD \cite{Jang:2019vkm, Bali:2019yiy}. The lighter blue band expresses the impact of $\pm 5\%$ variations in our diquark masses. Notably, scalar and axial-vector diquark variations interfere destructively, e.g.\ reducing $m_{[ud]}$ increases $g_A$, whereas $g_A$ decreases with the same change in the axial-vector mass. It is clear from Fig.\,\ref{FigGAx} that our prediction agrees with the lQCD results in Refs.\,\cite{Jang:2019vkm} and is similar to that in Ref.\,\cite{Bali:2019yiy} obtained from $z$ expansion. We will return to this point when we discuss PCAC below.
The extracted values for $g_A$, the charge radius and the axial mass are
\begin{subequations}
\begin{align}
g_A &= 1.25(3)\,, \\
\langle r^2_A\rangle^{1/2} m_N &= 3.25(4)\,, \\
m_A/m_N &= 1.23(3)\,.
\end{align}
\end{subequations}
As signalled above, the listed uncertainties in our values express the impact of varying the diquark masses. 

\begin{table}[!t]
\caption{\label{tablegr}
Referring to Fig.\,\ref{figcurrent}, separation of $G_A(0)$, $G_P(0)$ and $G_5(0)$ into contributions from various diagrams, listed as a fraction of the total $Q^2=0$ value.
Diagram (1): $\langle J \rangle^{S}_{\rm q}$ -- weak-boson strikes dressed-quark with scalar diquark spectator; and $\langle J \rangle^{A}_{\rm q}$ -- weak-boson strikes dressed-quark with axial-vector diquark spectator.
Diagram (2): $\langle J \rangle^{AA}_{\rm qq}$ -- weak-boson interacts strikes axial-vector diquark with dressed-quark spectator.
Diagram (3): $\langle J \rangle^{SA+AS}_{\rm dq}$ -- weak-boson mediates transition between scalar and axial-vector diquarks, with dressed-quark spectator.
Diagram (4): $\langle J \rangle_{\rm ex}$ -- weak-boson strikes dressed-quark ``in-flight'' between one diquark correlation and another.
Diagrams (5) and (6): $\langle J \rangle_{\rm sg}$ -- weak-boson couples inside the diquark correlation amplitude.
The listed uncertainty in these results reflects the impact of $\pm 5$\% variations in the diquark masses in Eq.\,\eqref{dqmasses}, \emph{e.g}.\ $0.71_{4_\mp} \Rightarrow 0.71 \mp 0.04$.
The extension of these
}
\begin{center}
\begin{tabular*}
{\hsize}
{
l@{\extracolsep{0ptplus1fil}}
|l@{\extracolsep{0ptplus1fil}}
l@{\extracolsep{0ptplus1fil}}
l@{\extracolsep{0ptplus1fil}}
l@{\extracolsep{0ptplus1fil}}
l@{\extracolsep{0ptplus1fil}}
l@{\extracolsep{0ptplus1fil}}}\hline
 & $\langle J \rangle^{S}_{\rm q}$  & $\langle J \rangle^{A}_{\rm q}$ &$\langle J \rangle^{AA}_{\rm qq}$ & $\langle J \rangle^{SA+AS}_{\rm qq}$ & $\langle J \rangle_{\rm ex}$ & $\langle J \rangle_{\rm sg}$ \\\hline
 $G_A(0)\ $ & $0.71_{4_\mp}$ & $0.064_{2_\pm} $ & $0.025_{5_\pm}$ & $0.13_{0_\mp}$ & $0.072_{32_\pm}$ & $\phantom{-}0$\\
 $G_P(0)\ $ & $0.74_{4_\mp}$ & $0.070_{5_\pm} $ & $0.025_{5_\pm}$ & $0.13_{0_\mp}$ & $0.22_{4_\pm}$ & $-0.19_{1_\mp}$\\
 $G_5(0)\ $ & $0.74_{4_\mp}$ & $0.069_{5_\pm} $ & $0.025_{5_\pm}$ & $0.13_{0_\mp}$ & $0.22_{4_\pm}$ & $-0.19_{1_\mp}$\\  \hline
\end{tabular*}
\end{center}
\end{table}

In Table~\ref{tablegr}, referring to Fig.\,\ref{figcurrent}, we list the relative strengths of each diagram's contribution to the nucleon's axial charge. Diagram~(1), with the weak-boson striking the dressed-quark in association with a spectator scalar diquark, is dominant. On the other hand, the contribution from diagrams~(5) and~(6) are zero in this case because the seagull terms, Eqs.\,\eqref{axsg} and \eqref{axsgcg}, are proportional to $Q_\mu$ and thus they do not contribute to $G_A$, which is determined by the $Q$-transverse piece of the nucleon axial current, see $G_A$'s projection, Eq.\,\eqref{gaproj}.
(The impacts on all form factors of the pion pole terms in our current construction are detailed in Appendix~\ref{Apppionpole}.)


\begin{figure}[t]
\leftline{\hspace*{0.5em}{\large{\textsf{A}}}}
\vspace*{-5ex}
\includegraphics[clip, width=0.42\textwidth]{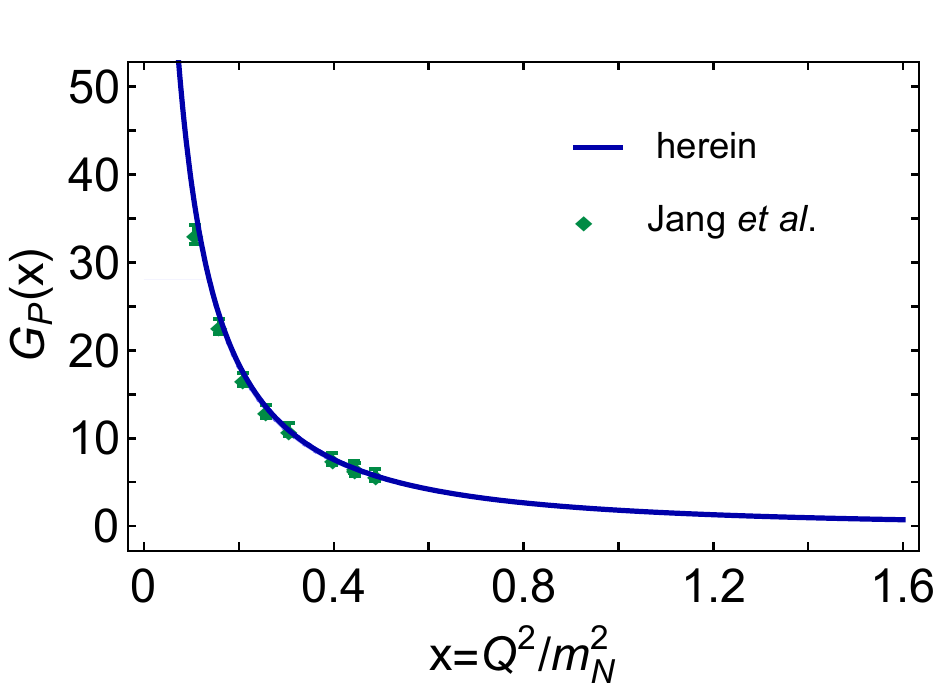}
\vspace*{1ex}

\leftline{\hspace*{0.5em}{\large{\textsf{B}}}}
\vspace*{-5ex}
\includegraphics[clip, width=0.42\textwidth]{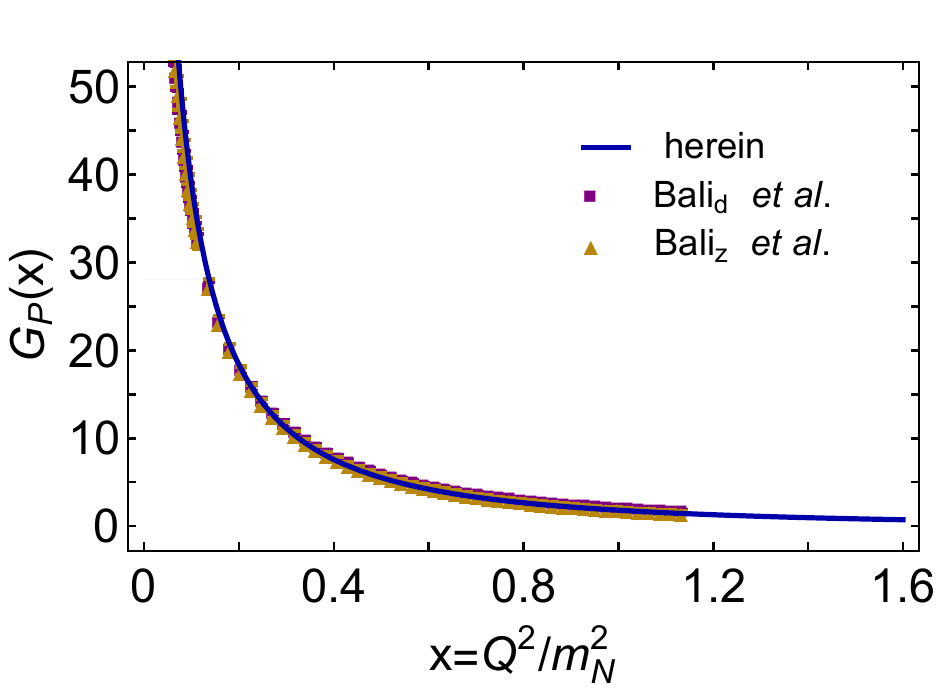}
\caption{\label{FigGPx}
\emph{Upper panel}\,--\,{\sf A}.  Result for $G_P(x)$ calculated herein -- blue curve within lighter blue model-uncertainty band, compared with lQCD results from Ref.\,\cite{Jang:2019vkm} -- green diamonds. With respect to our central results, this comparison may be quantified by reporting the mean-$\chi^2$ value, which is 1.04.
\emph{Lower panel}\,--\,{\sf B}.  Similar comparison with lQCD results from Ref.\,\cite{Bali:2019yiy} -- purple boxes [dipole] and golden triangles [$z$ expansion]. The mean-$\chi^2$ values are 15.14 [dipole] and 3.84 [$z$ expansion].
}
\end{figure}

In Fig.\,\ref{FigGPx} we depict our prediction for the induced pseudoscalar form factor $G_P(x)$, as well as corresponding lQCD results. The model-uncertainty band surrounding this curve cannot be distinguished when using the scale necessary to draw the figure. 

Muon capture experiments ($\mu\,+\,p\,\to\,\nu_\mu\,+\,n$) determine the induced pseudoscalar charge
\begin{equation}
g_p^\ast = \frac{m_\mu}{2m_N}
G_p(Q^2 = 0.88\,m_\mu^2)\,.
\end{equation}
We obtain $g_p^\ast=8.80(23)$, which is slightly larger
than the recent  MuCap Collaboration value, $g_p^\ast=8.06(55)$\,\cite{Andreev:2012fj, Andreev:2015evt}, but agrees with the world
average value $g_p^\ast=8.79(1.92)$\,\cite{Bernard:2001rs}.

In Table~\ref{tablegr}, referring to Fig.\,\ref{figcurrent}, we list again the relative strengths of each diagram's contribution to $G_P(0)$. Once again, diagram~(1), with the weak-boson striking the dressed-quark in association with a spectator scalar diquark, is overwhelmingly dominant. In this case, however, there is an active cancellation between the contributions from diagrams~(4),~(5) and~(6).

We find $g_A^d/g_A^u=-0.16(2)$ at the hadronic scale. This is a significant suppression of the magnitude of the $d$-quark component relative to that found in quark models, $g_A^d/g_A^u=-0.25$~\cite{He:1994gz}, and lQCD analyses, $g_A^d/g_A^u=-0.40(2)$~\cite{Bhattacharya:2016zcn} and $g_A^d/g_A^u=-0.58(3)$~\cite{Alexandrou:2019brg}.
Notably, whilst $g_A$ is a conserved charge, invariant under QCD evolution, the separation into component contributions from different quark flavours is not. This effect can potentially reconcile our flavour separation results with the lQCD predictions: our Faddeev wave function is defined at the hadronic scale, $\zeta_H \simeq 0.33\,$GeV \cite{Cui:2020dlm, Cui:2020tdf}, whereas the lQCD values are renormalised at $\zeta \simeq 2\,$GeV.
The size reduction shown in our framework owes to the presence of strong diquark correlations in our nucleon wave function, with the calculated value reflecting the relative strength of scalar and axial-vector diquarks: the isoscalar--scalar correlations are dominant, but the isovector--axial-vector diquarks have a measurable influence.

It is also interesting to find the pole position of $G_P$. Although we are not readily able to compute form factors in the time-like region, we can perform an extrapolation since the pion mass pole is very close to the origin. To do this, we use $1/G_P$, instead of $G_P$, and extrapolate to the region of small negative momentum.  The result is depicted in Fig.\,\ref{figgp2}: $1/G_P$ has a zero at $Q^2=-m_\pi^2$, verifying that $G_P$ has a pole at this position.

\begin{figure}[t]
\centerline{%
\includegraphics[clip, width=0.45\textwidth]{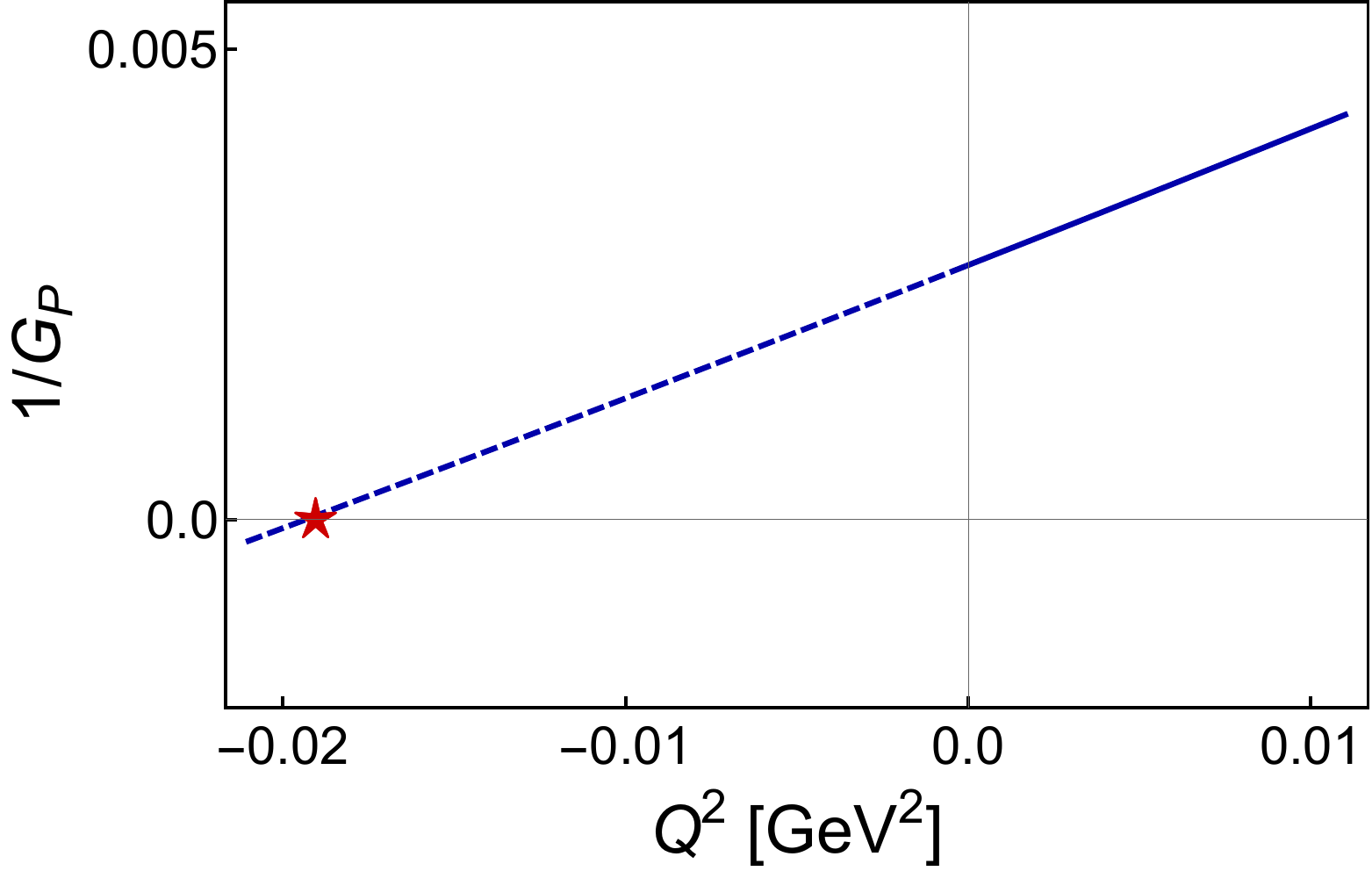}}
\caption{\label{figgp2}
Solid (blue) curve -- our computed $1/G_P$ in the spacelike region; dashed (blue) curve -- extrapolated $1/G_P$ into the timelike region; and (red) star -- pion mass $Q^2=-m_\pi^2$.
}
\end{figure}

The low momentum behaviour of the induced pseudoscalar form factor can be well approximated by using the pion pole dominance (PPD) assumption, which relates $G_P$, at low momenta, to the axial form factor $G_A$:
\begin{equation}
\label{ppd}
G_P \simeq \frac{4m_N^2G_A}{Q^2+m_\pi^2} \,.
\end{equation}
PPD is not an exact relation but there is every reason to expect that it is satisfied to a high level of precision -- see Eq.\,\eqref{PPDappendix} and the associated discussion.  Moreover, recent lQCD analyses \cite{Jang:2019vkm, Bali:2019yiy} find that PPD is satisfied with a discrepancy $\lesssim\,5\%$.

We have computed the PPD ratio
\begin{equation}
\label{ppdr}
R_{\rm PPD}	:=
\frac{4m_N^2G_A}{(Q^2+m_\pi^2)G_P}\,,
\end{equation}
and depict it in Fig.\,\ref{figppd}: $R_{\rm PPD}$ is very close to one; however, a deviation of about $1\%$ is found on $x\simeq 0$, which becomes smaller as $x$ gets larger. This situation is genuine and can be explained within the quark+diquark picture. First, we define a PCAC ratio according to Eq.\,\eqref{pcacp}
\begin{equation}
\label{pcacr}	
R_{\rm PCAC} :=
\frac{4m_N^2G_A}{Q^2G_P + 4m_qm_NG_5}\,,
\end{equation}
in the chiral limit $m_q=m_\pi=0\,{\rm GeV}$, $R_{\rm PCAC}$ and $R_{\rm PPD}$ are equivalent, and they should precisely be equal to 1:
\begin{equation}
\label{pcacr0}
R_{\rm PCAC}^{m_q=0}=R_{\rm PPD}^{m_q=0} =
\frac{4m_N^2G_A^{m_q=0}}{Q^2G_P^{m_q=0}} = 1 \,.
\end{equation}

Second, focusing on the axial current, the singular parts of the dressed-quark vertex, Eq.\,\eqref{axvx2}, seagull term, Eq.\,\eqref{axsg}, and diquark vertices, Eqs.\,\eqref{axdqvxAA}, \eqref{axdqvxSA}, do not contribute to $G_A$; and their contributing regular parts do not explicitly depend on the current quark mass $m_q$.
Thus, the chiral limit and the $m_q$-dependent expressions will generate the same axial form factor, \emph{i.e.} $G_A^{m_q=0}\equiv G_A$, provided that their parameters are the same.\footnote{In reality, the parameters should vary with the current quark mass, therefore, produce different $G_A$.}

\begin{figure}[!t]
\centerline{%
\includegraphics[clip, width=0.45\textwidth]{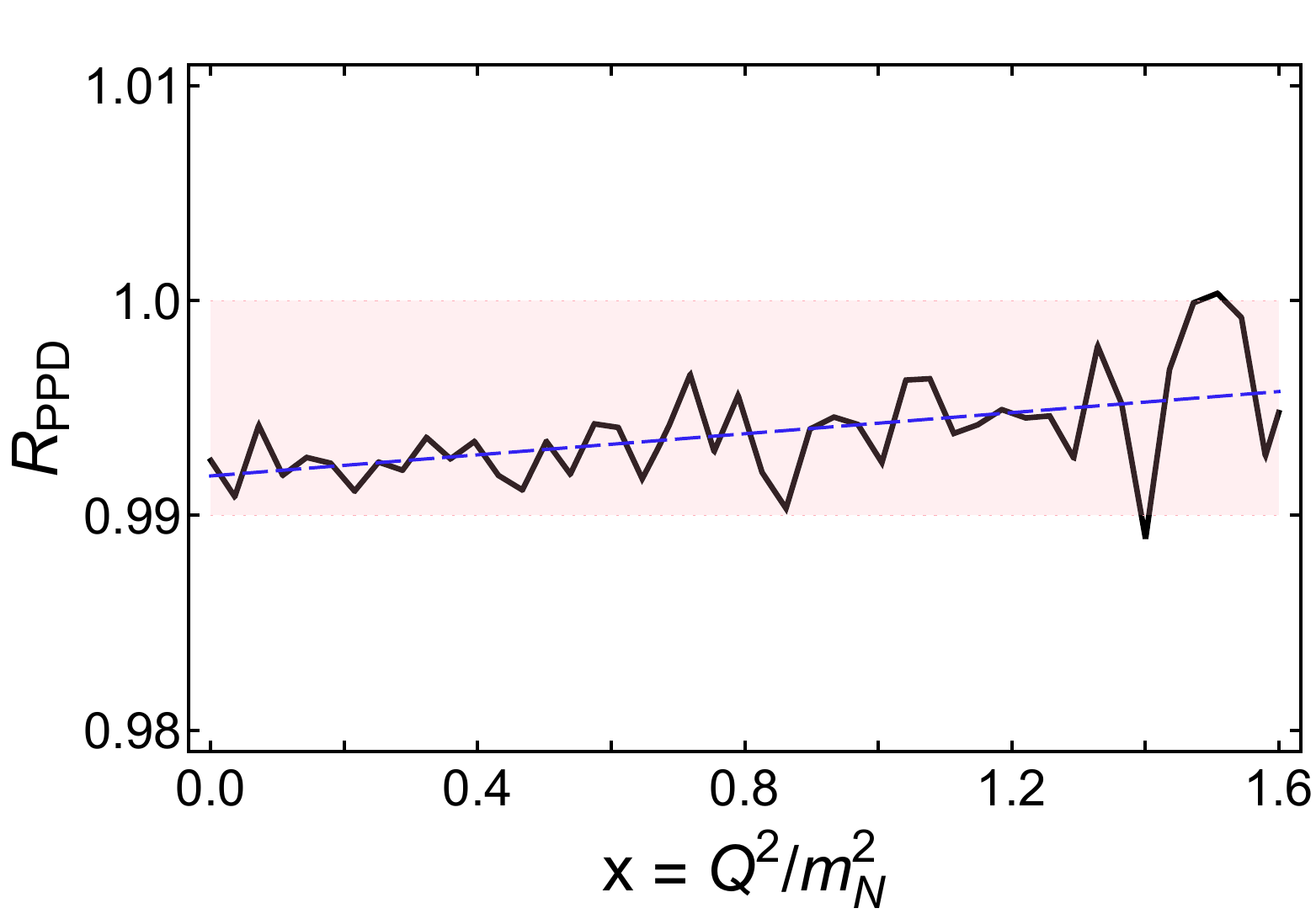}}
\caption{\label{figppd}
Solid (black) curve: $R_{\rm PPD}\in[0.98,1.01]$; and the dashed (blue) line: the linear fit of $R_{\rm PPD}$.
Small oscillations are due to the numerical precision.
}
\end{figure}

Third, regular and singular parts contribute to the induced pseudoscalar form factor, so one may write
\begin{equation}
G_P:= G_{P,{\rm regular}} + G_{P,{\rm singular}} \,.
\end{equation}
It is obvious that
{\allowdisplaybreaks\begin{subequations}
\begin{align}
G_{P,{\rm regular}}^{m_q=0} &= G_{P,{\rm regular}}\,\\
Q^2 G_{P,{\rm singular}}^{m_q=0} &= (Q^2+m_\pi^2)G_{P,{\rm singular}}\,;
\end{align}
\end{subequations}
hence, Eq.\,\eqref{ppdr} becomes
\begin{subequations}
\begin{align}
R_{\rm PPD} &=
\frac{4m_N^2G_A^{m_q=0}}{Q^2 G_{P}^{m_q=0}+m_\pi^2G_{P,{\rm regular}}}\\
&=
\frac{4m_N^2G_A^{m_q=0}}{Q^2 G_{P}^{m_q=0}+m_\pi^2G_{P,{\rm regular}}^{m_q=0}}\,.
\end{align}
\end{subequations}
Using Eq.\,\eqref{pcacr0}, one can find that the small deviation comes from the term $m_\pi^2G_{P,{\rm regular}}$, which approximately produces a $1\%$ contribution at $Q^2\simeq 0\,{\rm GeV}^2$; and it is also clear that $R_{\rm PPD}$ tends gradually to $1$ as $Q^2$ grows.}


\begin{figure}[t]
\leftline{\hspace*{0.5em}{\large{\textsf{A}}}}
\vspace*{-5ex}
\includegraphics[clip, width=0.42\textwidth]{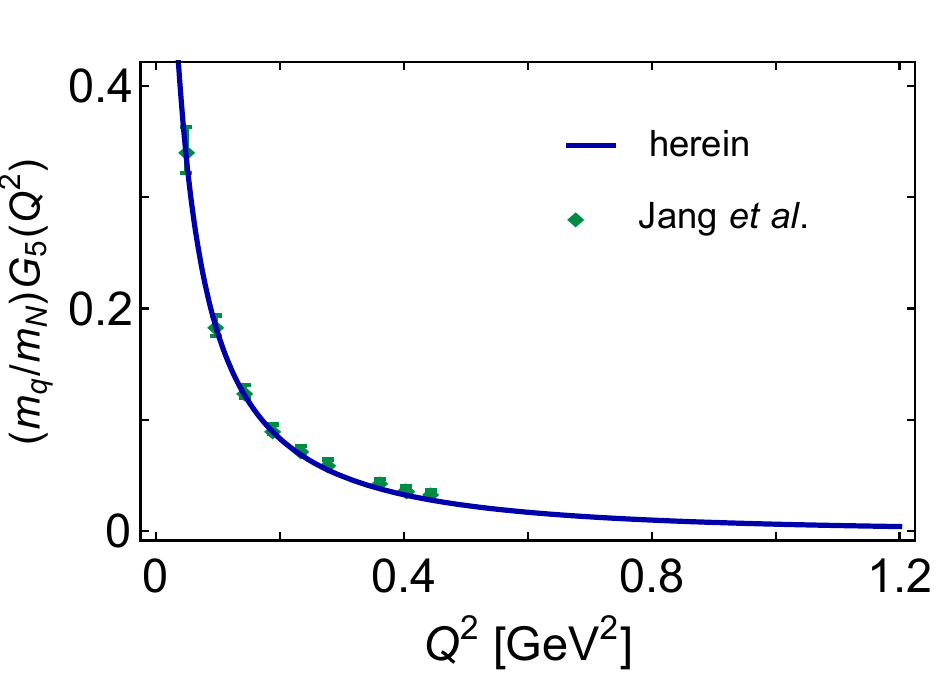}
\vspace*{1ex}

\leftline{\hspace*{0.5em}{\large{\textsf{B}}}}
\vspace*{-5ex}
\includegraphics[clip, width=0.42\textwidth]{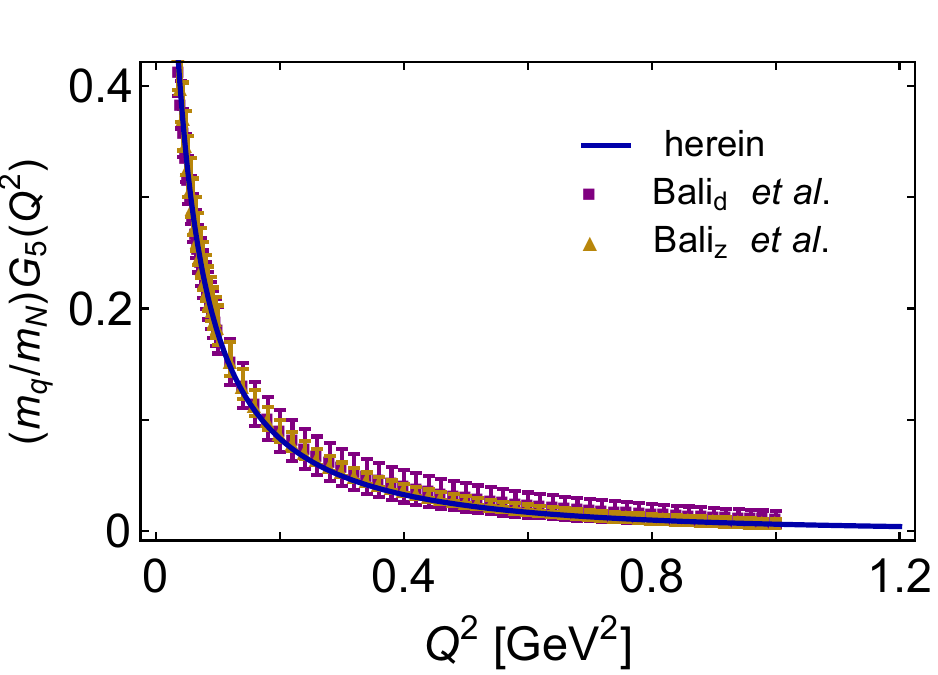}
\caption{\label{FigG5}
\emph{Upper panel}\,--\,{\sf A}.  Result for $(m_q/m_N)G_5(Q^2)$ calculated herein -- blue curve within lighter blue model-uncertainty band, compared with lQCD results from Ref.\,\cite{Jang:2019vkm} -- green diamonds.  With respect to our central results, this comparison may be quantified by reporting the mean-$\chi^2$ value, which is 5.37.
\emph{Lower panel}\,--\,{\sf B}.  Similar comparison with lQCD results in Ref.\,\cite{Bali:2019yiy} -- purple boxes [dipole] and golden triangles [$z$ expansion].  Here, the large lQCD uncertainties prevent a meaningful $\chi^2$ comparison with our prediction.
}
\end{figure}

\subsection{The pseudoscalar form factor, Goldberger-Treiman relation and PCAC}
\label{subsecgp}

The pseudoscalar form factor $G_5$, or equivalently the pion-nucleon form factor, $G_{\pi NN}$, cannot be directly measured except at the pion mass point $Q^2=-m_\pi^2$, where we can obtain the pion-nucleon coupling constant $g_{\pi NN}$. Our predicted value of $(m_q/m_N)G_5(Q^2)$ is depicted in Fig.~\ref{FigG5}, compared with recent lQCD results\,\cite{Jang:2019vkm, Bali:2019yiy} is also presented in the same figure.  We have also extrapolated the function $1/G_5$ from the space-like region onto a small domain of timelike momenta in order to determine its zero at $Q^2=-m_\pi^2$ and thus confirm that $G_5$ has a pole at this point, see Fig.\,\ref{figg52}.

\begin{figure}[!t]
\centerline{%
\includegraphics[clip, width=0.45\textwidth]{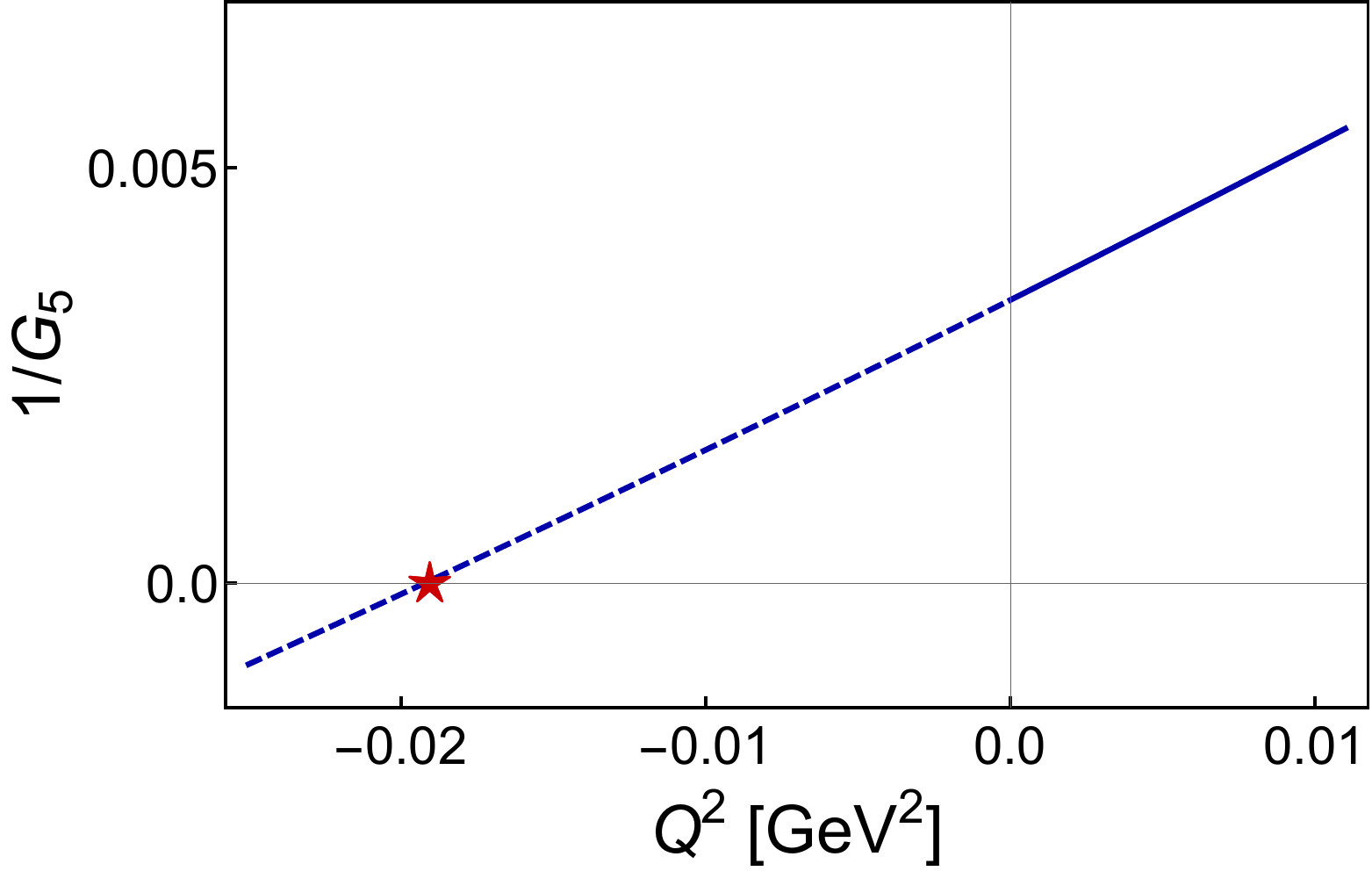}}
\caption{\label{figg52} Solid (blue) curve -- our calculated $1/G_5$ in the spacelike region; dashed (blue) curve -- extrapolated $1/G_5$ into the timelike region; and (red) star --  pion mass $Q^2=-m_\pi^2$.}
\end{figure}

One cannot claim a reliable calculation of the axial and pseudoscalar form factors if they do not satisfy the Goldberger-Treiman relation at the form factor level, Eq.\,\eqref{gtr}. We obtain
\begin{equation}
\frac{f_\pi}{m_N}G_{\pi NN}(0) = 1.25(3) = G_A(0) \,.
\end{equation}
(Recall, the theoretical uncertainty expresses the impact of $\pm5\%$ variations in the diquark masses. Clearly, our results satisfy the Goldberger-Treiman relation precisely.  

In Table~\ref{tablegr}, referring to Fig.\,\ref{figcurrent}, we list the relative strengths of each diagram's contribution to $G_5(0)$, \emph{i.e.} $G_{\pi NN}(0)$. The results for $G_5(0)$ are almost identical to those for $G_P(0)$. This can be understood in the following way: according to our argument on PPD, we know that the singular part of the axial current provides almost the total contribution ($\gtrsim 99\%$) to the induced pseudoscalar form factor $G_P$ at $Q^2\simeq 0\,{\rm GeV}^2$, and thus we can safely focus on the singular part of $J_{5\mu}$. Next, considering $G_P$'s projection, Eq.\,\eqref{gpproj}, we immediately find that $G_A$ does not contribute because of its transverse nature. Meanwhile, if one takes into account the detailed structure of the building blocks and currents, and compares Eq.\,\eqref{gpproj} with $G_5$'s projection, Eq.\,\eqref{g5proj}, one arrives at
{\allowdisplaybreaks
\begin{subequations}
\begin{align}
G_P &\simeq \frac{Q_\mu}{\tau^2}{\rm tr}_{\rm D}\big[J_{5\mu}\gamma_5\big]\times c_1,\\
G_5	&= \frac{1}{\tau}{\rm tr}_{\rm D}\big[J_{5}\gamma_5\big]\times c_2\\
&\simeq \frac{Q_\mu}{\tau^2}{\rm tr}_{\rm D}\big[J_{5\mu}\gamma_5\big]\times c_2'\\
&\simeq G_P\times(c_2'/c_1)\,,
\end{align}
\end{subequations}}
\hspace*{-0.4\parindent}when $Q^2\simeq 0\,{\rm GeV}^2$, where $c_1$, $c_2$ and $c_2'$ are constants. Since we want to compute the relative strengths, the values of $c_1$, $c_2$ and $c_2'$ do not contribute.

Regarding the pion-nucleon coupling, we predict $g_{\pi NN}/m_N= G_{\pi NN}(Q^2=-m_\pi^2)/m_N= 14.02(33)/{\rm GeV}$.
This value overlaps with that extracted from pion-nucleon scattering \cite{Baru:2011bw} ($g_{\pi NN}/m_N = 13.97(10)/{\rm GeV}$); and compares well with a determination based on the Granada 2013 $np$ and $pp$ scattering database \cite{NavarroPerez:2016eli} ($g_{\pi NN}/m_N = 14.11(3)/{\rm GeV}$) and a recent analysis of nucleon-nucleon scattering using effective field theory and related tools \cite{Reinert:2020mcu} ($g_{\pi NN}/m_N = 14.09(4)/{\rm GeV}$).
%
%
The lQCD study in Ref.\,\cite{Bali:2019yiy} produces $g_{\pi NN}/m_N = 13.77(85)/{\rm GeV}$ (dipole) and $g_{\pi NN}/m_N = 15.74(1.93)/{\rm GeV}$ (z-expansion).
All these comparisons are drawn in Fig.\,\ref{Figgpinn}\,A.

\begin{figure}[t]
\leftline{\hspace*{0.5em}{\large{\textsf{A}}}}
\vspace*{-5ex}
\includegraphics[clip, width=0.42\textwidth]{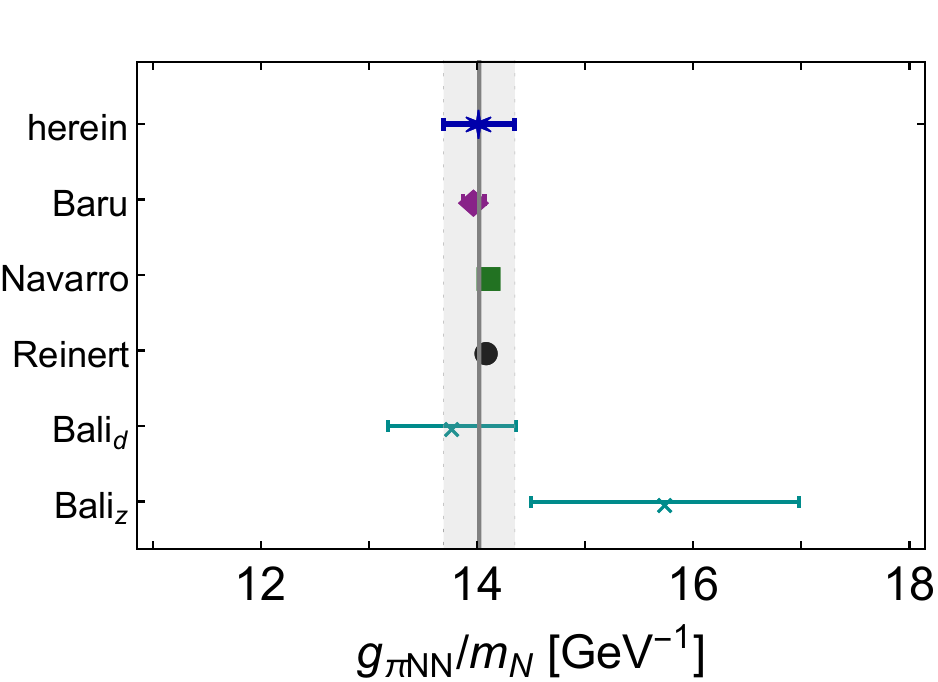}
\vspace*{2ex}

\leftline{\hspace*{0.5em}{\large{\textsf{B}}}}
\vspace*{-5ex}
\includegraphics[clip, width=0.42\textwidth]{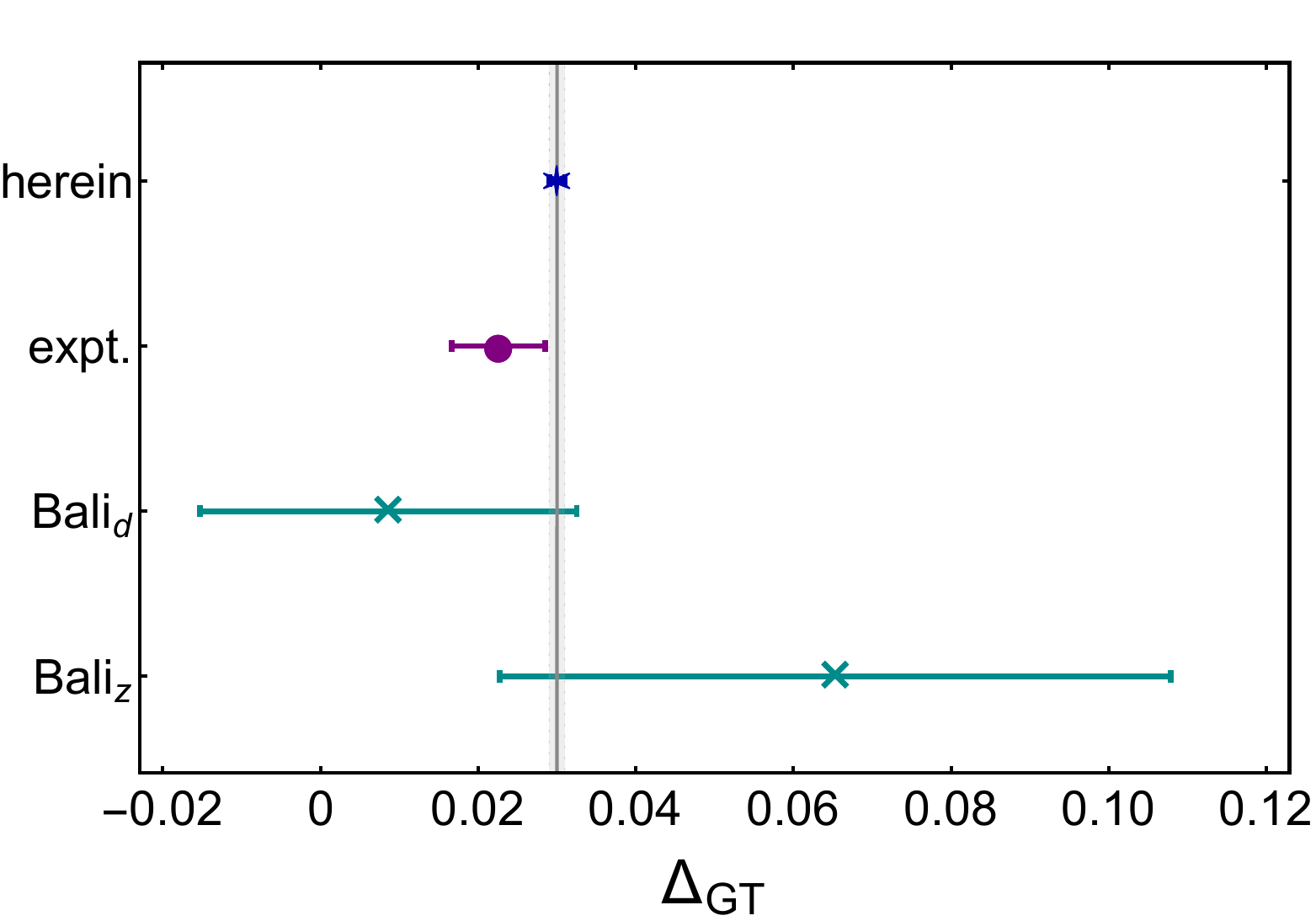}
\caption{\label{Figgpinn}
\emph{Upper panel}\,--\,{\sf A}.
Comparison of our prediction for $g_{\pi NN}/m_N$ (blue asterisk) with
values extracted from pion-nucleon scattering \cite{Baru:2011bw} (purple diamond),
Granada 2013 $np$ and $pp$ scattering database \cite{NavarroPerez:2016eli} (green square),
and nucleon-nucleon scattering \cite{Reinert:2020mcu} (black circle);
and lQCD results \cite{Bali:2019yiy} (cyan crosses).
\emph{Lower panel}\,--\,{\sf B}.  Comparison of our prediction for $\Delta_{\rm GT}$ (blue asterisk) with
the estimate in Ref.\,\cite{Nagy:2004tp} (purple circle)
and lQCD results \cite{Bali:2019yiy} (cyan crosses).
(In both panels, the vertical grey band marks the estimated uncertainty in our prediction.)
}
\end{figure}

It is also worth noting that on $-m_\pi^2 <Q^2 < 2\,m_N^2$, a good interpolation of our central result is provided by $(x=Q^2/m_N^2)$:
\begin{equation}
\label{GpiNNF}
G_{\pi NN}(x) = \frac{(13.52 - 2.291 x)m_N }{1+2.383 x + 0.5563 x^2 + 1.434 x^3}\,.
\end{equation}
For comparison with meson-exchange models of the nucleon-nucleon interaction, a fair approximation to Eq.\,\eqref{GpiNNF} is obtained with the following dipole form:
\begin{equation}
\label{DipolepiNN}
G_{\pi NN}^d(x) = \frac{13.47 m_N}{ ( 1 + x /0.845^2 )^2 }\,,
\end{equation}
which corresponds to a $\pi NN$ dipole scale $\Lambda_{\pi N N} = 0.845 m_N = 0.79\,$GeV, \emph{viz}.\ a soft form factor.
(A similar value was obtained previously in a rudimentary quark+scalar-diquark model \cite{Bloch:1999rm}.)
Our prediction is $\sim 20$\% larger than, hence qualitatively equivalent to, the $\pi N N$ dipole mass inferred from a dynamical coupled-channels analysis of $\pi N$, $\gamma N$ interactions \cite{Kamano:2013iva}.  Future such studies may profit by implementing couplings and range parameters determined in analyses like ours.

We now can compute the so-called Goldberger-Treiman discrepancy:
\begin{subequations}
\begin{align}
\Delta_{\rm GT}	&=
1-\frac{G_A(0)}{\frac{f_\pi}{m_N}G_{\pi NN}(-m_\pi^2)} \\
&=
1-\frac{G_{\pi NN}(0)}{G_{\pi NN}(-m_\pi^2)}\,,
\end{align}
\end{subequations}
which measures the difference of $G_{\pi NN}$ values at $Q^2=0$ and the pion's on-shell mass, for a given current-quark mass; and thereby the distance from the chiral limit. We predict $\Delta_{\rm GT}=0.030(1)$, which is loosely consistent with the estimate in Ref.\,\cite{Nagy:2004tp}, \emph{viz}.\ 0.023(5), and matches, within errors, the lQCD results in Ref.\,\cite{Bali:2019yiy}. These comparisons are drawn in Fig\,\ref{Figgpinn}\,B.

\begin{figure}[t]
\centerline{%
\includegraphics[clip, width=0.45\textwidth]{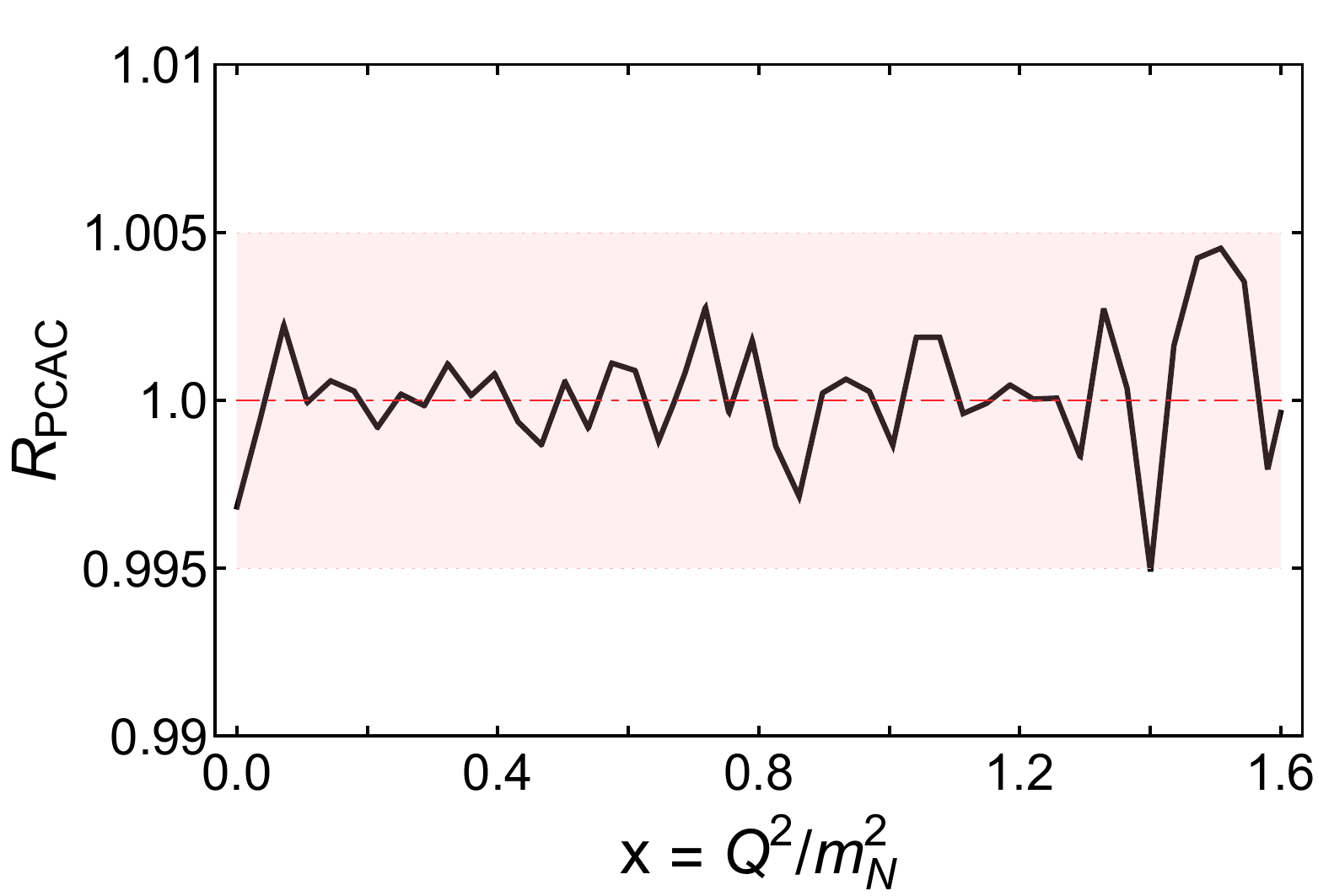}}
\caption{\label{figpcac}
Solid curve: computed result from Eq.\,\eqref{pcacr}: $R_{\rm PCAC}\in[0.995,1.005]$ on the entire domain.  (Fluctuations reflect the numerical precision of our calculation.)
}
\end{figure}

Finally, we check whether our computed $G_A(Q^2)$, $G_P(Q^2)$ and $G_5(Q^2)$ form factors satisfy the PCAC relation, Eq.\,\eqref{pcacp}. Figure~\ref{figpcac} displays the PCAC ratio, defined in Eq.~\eqref{pcacr}; one observes that the computed $R_{\rm PCAC}$ is practically indistinguishable from unity for the whole range of momentum-transfer depicted. Therefore, we have shown analytically but also numerically that our theoretical approach is consistent with fundamental symmetry requirements. It is worth emphasizing here that this result does not rely on any fine-tuned set of parameters; instead, it is automatically satisfied owing to our careful construction of the currents, discussed above.


\section{Summary}
\label{secsum}

Using a Poincar\'e-covariant quark+diquark Faddeev equation treatment of the nucleon and weak interaction currents that ensure consistency with relevant Ward-Green-Takahashi identities, we delivered a unified set of predictions for the nucleon's axial, induced pseudoscalar, and explicit pseudoscalar form factors.  The presentation included a detailed discussion of partial conservation of the nucleon axial current and associated Goldberger-Treiman relations, and all technical details relating to current constructions in the quark+diquark approach, including the seagull terms.

Concerning the axial form factor, $G_A$, we found that it can reliably be represented by a dipole form factor, normalised by the axial charge $g_A=1.25(3)$ and characterised by a mass-scale $M_A = 1.23(3) m_N$, where $m_N$ is the nucleon mass.  Moreover, the $Q^2$-behaviour of $G_A$ is in good agreement with recent lattice-QCD (lQCD) results \cite{Jang:2019vkm}.

Regarding the induced pseudoscalar form factor, we found that $G_P$ depends on the transferred momentum in such a way that it agrees favourably with data obtained from low-energy pion electroproduction \cite{Choi:1993vt} and also the lQCD results in Ref.\,\cite{Jang:2019vkm}.  Moreover, the pion pole dominance \emph{Ansatz} provides a sound estimate of the directly computed result. Additionally, the induced pseudoscalar charge $g_p^\ast$ is consistent with the value determined from a recent $\mu$-capture experiment \cite{Andreev:2012fj}.

With respect to $G_5$, recent lQCD computations \cite{Jang:2019vkm} agree with our theoretical result in the whole range of $Q^2$ studied. Furthermore, our prediction for the pion-nucleon coupling constant $g_{\pi NN}/m_N = G_{\pi NN}(Q^2=-m_\pi^2)/m_N=14.02(33)/{\rm GeV}$ agrees with a recent effective field theory analysis \cite{Reinert:2020mcu}.

We proved that our nucleon's axial, induced-pseudoscalar and pseudoscalar form factors analytically satisfy the PCAC relation and verified this numerically; establishing thereby that our theoretical approach is consistent with key symmetries in Nature.

It is here worth highlighting that we find $g_A^d/g_A^u=-0.16(2)$ at the hadronic scale. This is a significant suppression of the magnitude of the $d$-quark component relative to that found in nonrelativistic quark models. The size reduction owes to the presence of strong diquark correlations in our nucleon wave function, with the calculated value reflecting the relative strength of scalar and axial-vector diquarks: the isoscalar--scalar correlations are dominant, but the isovector--axial-vector diquarks have a measurable influence.

Finally, as noted elsewhere \cite{Chen:2020wuq}, a natural next step is to improve upon the quark+diquark approximation and use the more fundamental three-quark Faddeev equation approach to  the nucleon bound-state problem, extending the analysis in Ref.~\cite{Eichmann:2011pv}. In finding and implementing an approach to improving the expressions of emergent hadronic mass in both the Faddeev kernel and interaction current, one could deliver continuum predictions for all nucleon form factors that posses tighter links to QCD's Schwinger functions.


\acknowledgments
We are grateful to Y.-C.~Jang for providing us with the lattice results in Ref.\,\cite{Jang:2019vkm} and for constructive comments from Z.-F.~Cui, G.~Eichmann, M.~Engelhardt, A.~Lovato, U.~Mosel, M.~Oettel, E.~Ruiz-Arriola and N.~Santowsky.
Work supported by:
National Natural Science Foundation of China grant nos.\ 12135007 and 12047502;
%
%
DFG grant FI 970/11-1;
%
%
Chinese Ministry of Science and Technology \emph{International Expert Involvement Programme};
Ministerio Espa\~nol de Ciencia e Innovaci\'on (grant no.\,PID2019-107844GB-C22);
and Junta de Andaluc\'ia (contract nos.\ operativo FEDER Andaluc\'ia 2014-2020 UHU-1264517,  P18-FR-5057, PAIDI FQM-370).


\appendix
\setcounter{equation}{0}
\setcounter{figure}{0}
\setcounter{table}{0}
\renewcommand{\theequation}{\Alph{section}.\arabic{equation}}
\renewcommand{\thetable}{\Alph{section}.\arabic{table}}
\renewcommand{\thefigure}{\Alph{section}.\arabic{figure}}

\section{The QCD-kindred framework}
\label{appendixQCD}
The QCD-kindred model for mesons and baryons that we exploit herein was introduced in Refs.\,\cite{Ivanov:1998ms, Hecht:2000xa, Alkofer:2004yf} and refined in a series of subsequent analyses that may be traced from Refs.\,\cite{Cloet:2008re, Segovia:2014aza}.  Consistency between the various Schwinger function elements is guaranteed through their mutual interplay in the description and prediction of hadron observables.   Combined with the material above, the information in this appendix is sufficient for an interested practitioner to implement the framework for themself.

\subsection{Dressed quark propagator}
\label{subappendixqprop}
The dressed-quark propagator is:
{\allowdisplaybreaks
\begin{subequations}
\begin{align}
\label{Spsigma}
S(p) & =  -i \gamma\cdot p\, \sigma_V(p^2) + \sigma_S(p^2) \\
\label{SpAB}
& = 1/[i\gamma\cdot p\, A(p^2) + B(p^2)]\,.
\end{align}
\end{subequations}
Regarding light-quarks in QCD, the wave function renormalisation and dressed-quark mass:
\begin{equation}
\label{ZMdef}
Z(p^2)=1/A(p^2)\,,\;M(p^2)=B(p^2)/A(p^2)\,,
\end{equation}
respectively, receive significant momentum-dependent corrections at infrared momenta \cite{Lane:1974he, Politzer:1976tv, Bhagwat:2006tu, Binosi:2016wcx}: $Z(p^2)$ is suppressed and $M(p^2)$ enhanced.  These features are an expression of emergent hadronic mass \cite{Roberts:2020hiw}.

\begin{figure}[!t]
\centerline{%
\includegraphics[clip, width=0.45\textwidth]{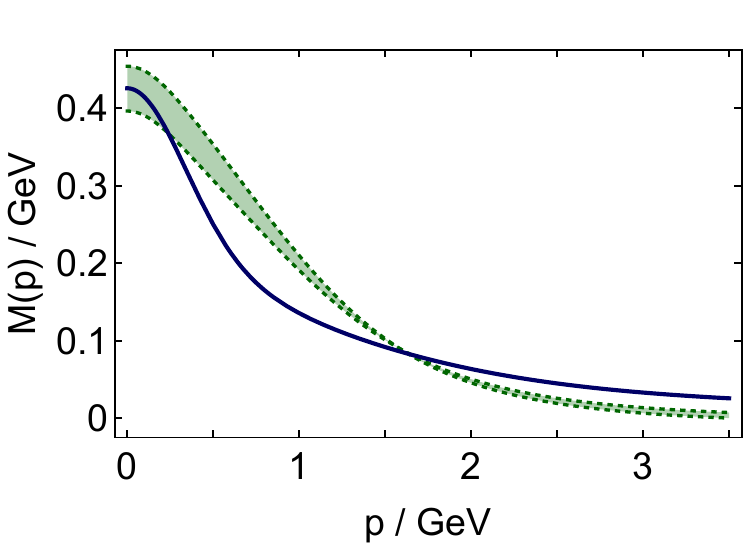}}
\caption{\label{MassPlot}
Solid curve (blue) -- quark mass function obtained from the dressed-quark propagator specified by Eqs.\,\eqref{EqSSSV}--\eqref{tableA}; and band (green) -- exemplary range of numerical results obtained by solving the gap equation with the modern kernels \cite{Chang:2013pq, Chang:2013epa}.}
\end{figure}

An efficacious parametrisation of $S(p)$, which exhibits the features described above, has been used extensively in hadron studies \cite{Chen:2017pse, Chen:2018nsg, Chen:2019fzn, Lu:2019bjs, Cui:2020rmu}.  It is defined via
{\allowdisplaybreaks
\begin{subequations}
\label{EqSSSV}
\begin{align}
\bar\sigma_S(x) & =  2\,\bar m \,{\cal F}(2 (x+\bar m^2)) \nonumber \\
& \quad + {\cal F}(b_1 x) \,{\cal F}(b_3 x) \,
\left[b_0 + b_2 {\cal F}(\epsilon x)\right]\,,\label{ssm} \\
\label{svm} \bar\sigma_V(x) & =  \frac{1}{x+\bar m^2}\, \left[ 1 - {\cal F}(2
(x+\bar m^2))\right]\,,
\end{align}
\end{subequations}}
\hspace*{-0.5\parindent}with $x=p^2/\lambda^2$, $\bar m$ = $m/\lambda$,
\begin{equation}
\label{defcalF}
{\cal F}(x)= \frac{1-\mbox{\rm e}^{-x}}{x}  \,,
\end{equation}
$\bar\sigma_S(x) = \lambda\,\sigma_S(p^2)$ and $\bar\sigma_V(x) =
\lambda^2\,\sigma_V(p^2)$.
The mass-scale, $\lambda=0.566\,$GeV, and
parameter values
\begin{equation}
\label{tableA}
\begin{array}{ccccc}
   \bar m& b_0 & b_1 & b_2 & b_3 \\\hline
   0.00897 & 0.131 & 2.90 & 0.603 & 0.185
\end{array}\;,
\end{equation}
associated with Eqs.\,\eqref{EqSSSV} were fixed in analyses of light-meson observables \cite{Burden:1995ve, Hecht:2000xa}.  ($\epsilon=10^{-4}$ in Eq.\ (\ref{ssm}) acts solely to decouple the large- and intermediate-$p^2$ domains.)

The dimensionless $u=d$ current-mass in Eq.\,(\ref{tableA}) corresponds to
$m_q=5.08\,{\rm MeV}$ 
and the propagator yields the following Euclidean constituent-quark mass, defined by solving $p^2=M^2(p^2)$:
$M_q^E = 0.33\,{\rm GeV}$.
The ratio $M_q^E/m_q = 65$ is one expression of DCSB, a corollary of emergent hadronic mass, in the parametrisation of $S(p)$.  It highlights the infrared enhancement of the dressed-quark mass function.

The dressed-quark mass function generated by Eqs.\,\eqref{EqSSSV}\,--\,\eqref{tableA} is drawn in Fig.\,\ref{MassPlot}.  Although simple and introduced long beforehand, the parametrisation is evidently a sound representation of contemporary numerical results.  (The numerical solutions drawn in Fig.\,\ref{MassPlot} were obtained in the chiral limit, which is why the (green) band falls below the parametrisation at larger $p$.)

As with the diquark propagators in Eq.\,\eqref{Eqqqprop}, the expressions in Eq.\,\eqref{EqSSSV} ensure dressed-quark confinement via the violation of reflection positivity (see, \emph{e.g}.\ Ref.\,\cite{Horn:2016rip}, Sec.\,3).
}


\subsection{Correlation amplitudes}
\label{subappendixdqbsa}

As mentioned in Sec.\,\ref{secfaddeev}, for the nucleon we only need to include the two positive-parity diquark correlations: the isoscalar-scalar ($0^+$) and isovector-pseudo\-vector ($1^+$) diquarks. Their dominant structures are:
{\allowdisplaybreaks
\begin{subequations}
\label{qqBSAs}
\begin{align}
\label{scBSAs}
\Gamma^{0^+}(k;K) & = g_{0^+} \, \gamma_5 C\, t^0_f \,\vec{H}_c \,{\mathpzc F}(k^2/\omega_{0^+}^2) \,, \\
\label{axBSAs}
{\Gamma}_\mu^{1^+}(k;K)
    & = i g_{1^+} \, \gamma_\mu C \, \vec{t}_f\, \vec{H}_c \,{\mathpzc F}(k^2/\omega_{1^+}^2)\,,
\end{align}
\end{subequations}
where:
$K$ is the total momentum of the correlation, $k$ is a two-body relative momentum, ${\mathpzc F}$ is the function in Eq.\,\eqref{defcalF}, $\omega_{J^P}$ is a size parameter, and $g_{J^P}$ is a coupling into the channel, which is fixed by normalisation;
$C=\gamma_2\gamma_4$ is the charge-conjugation matrix; $t^0$ and $\vec{t}=(t^1,t^2,t^3)$ are the flavour matrices:
\begin{subequations}
\label{dqf}
\begin{align}
t^0_f &= \tfrac{i}{\surd 2}\tau^2\,,	\\
t^1_f &= \tfrac{1}{2}(\tau^0+\tau^3)\,,\\
t^2_f &= \tfrac{1}{\surd 2}\tau^1\,,\\
t^3_f &= \tfrac{1}{2}(\tau^0-\tau^3)\,,
\end{align}
\end{subequations}
$\tau^0=\,$diag$[1,1]$, $\{\tau^j,j=1,2,3\}$ are the Pauli matrices;
and $\vec{H}_c = \{i\lambda_c^7, -i\lambda_c^5,i\lambda_c^2\}$, with $\{\lambda_c^k,k=1,\ldots,8\}$ being Gell-Mann matrices in colour space, expresses the diquarks' colour antitriplet character.

The amplitudes in Eqs.\,\eqref{qqBSAs} are canonically normalised:
\begin{subequations}
\label{CanNorm}
\begin{align}
2 K_\mu & = \left. \frac{\partial }{\partial Q_\mu} \,  \Pi(K;Q)\right|_{Q=K}^{K^2 = - m_{J^P}^2},\\
\nonumber
\Pi(K;Q) & = {\rm tr_{CDF}} \int \frac{d^4 k}{(2\pi)^4} \bar \Gamma(k;-K) S(k+Q/2) \\
& \quad \times \Gamma(k;K) S^{\rm T}(-k+Q/2)\,, \label{eqPi}
\end{align}
\end{subequations}
where $\bar\Gamma(k;K) = C^\dagger \Gamma(-k;K) C$.  When the amplitudes involved carry Lorentz indices $\mu$, $\nu$, the left-hand-side of Eq.\,\eqref{eqPi} also includes a factor $\delta_{\mu\nu}$.  Plainly, the coupling strength in each channel, $g_{J^P}$ in Eq.\,\eqref{qqBSAs}, is fixed by the associated value of $\omega_{J^P}$.
}


\subsection{Diquark propagators and couplings}
\label{subappendixdqprop}

A propagator is associated with each quark+quark correlation in Fig.\,\ref{figFaddeev}.  We use \cite{Segovia:2014aza}:
{\allowdisplaybreaks
\begin{subequations}
\label{Eqqqprop}
\begin{align}
\Delta^{0^+}(K) & = \frac{1}{m_{0^+}^2} \, {\mathpzc F}(k^2/\omega_{0^+}^2)\,,\\
\Delta^{1^+}_{\mu\nu}(K) & = \left[ \delta_{\mu\nu} + \frac{K_\mu K_\nu}{m_{1^+}^2} \right]
 \frac{1}{m_{1^+}^2} \, {\mathpzc F}(k^2/\omega_{1^+}^2)\,.
\end{align}
\end{subequations}
These forms ensure that the diquarks are confined within the baryons: whilst the propagators are free-particle-like at spacelike momenta, they are pole-free on the timelike axis.  This is enough to ensure confinement via the violation of reflection positivity (see, \emph{e.g.},\ Ref.\,\cite{Horn:2016rip}, Sec.\,3).

The diquark masses and widths are related via \cite{Segovia:2014aza}
\begin{equation}
m_{J^P}^2 = 2 \,\omega_{J^P}^2.
\end{equation}
This correspondence accentuates the diquarks' free-particle-like propagation characteristics within the baryon.

Using Eq.\,\eqref{dqmasses} and Eqs.\,\eqref{qqBSAs}, \eqref{CanNorm}, one finds
\begin{equation}
g_{0^+} = 14.8\,,\,\,\,\,\,
g_{1^+} = 12.7\,,
\end{equation}
Since it is the coupling-squared which appears in the Faddeev kernels, $0^+$ diquarks will dominate the Faddeev amplitudes of $J=1/2$ baryons; but $1^+$ diquarks must also play a measurable role because $g^2_{1^+}/g^2_{0^+} \simeq 0.7$.
}


\subsection{Faddeev amplitudes}
Solving the Faddeev equation, Fig.\,\ref{figFaddeev}, yields both the mass-squared and bound-state amplitude of all baryons with a given value of $J^P$.  The form of the Faddeev amplitude fixes the channel.  A baryon is described by
\begin{align}
\Psi^B & = \psi^B_1 + \psi^B_2 + \psi^B_3\,,
\end{align}
where the subscript identifies the bystander quark, \emph{viz}. the quark that is not participating in a diquark correlation.  $\psi^B_{1,2}$ are obtained from $\psi^B_3=:\psi^B$ by cyclic permutations of all quark labels.

For the nucleon ($B=N$),
{\allowdisplaybreaks
\begin{align}
\nonumber
& \psi^N(p_i,\alpha_i,\sigma_i) \\
\nonumber
 = & [\Gamma^{0^+}(k;K)]^{\alpha_1 \alpha_2}_{\sigma_1 \sigma_2} \, \Delta^{0^+}(K) \,[\Psi^{0^+}(\ell;P) u(P)]^{\alpha_3}_{\sigma_3} +\\
 &  [{\Gamma}^{1^+}_\mu(k;K)]^{\alpha_1 \alpha_2}_{\sigma_1 \sigma_2} \, \Delta^{1^+ }_{\mu\nu}(K) \, [{\Psi}_{\nu }^{1^+}(\ell;P) u(P)]^{\alpha_3}_{\sigma_3} \,,
\label{FaddeevAmp}
\end{align}
where
$(p_i,\sigma_i,\alpha_i)$ are the momentum, spin and isospin labels of the quarks comprising the bound state;
$P=p_1 + p_2 + p_3=p_d+p_q$ is the baryon's total momentum;
$k=(p_1-p_2)/2$, $K=p_1+p_2=p_d$, $\ell = (-K + 2 p_3)/3$;
and $u(P)$ is a Euclidean spinor (see Ref.\,\cite{Segovia:2014aza}, Appendix\,B for details).
The remaining terms in Eq.\,\eqref{FaddeevAmp} are the following matrix-valued functions:
\begin{subequations}
\label{8sa}
\begin{align}
\Psi^{0^+}(\ell;P) & = \sum_{k=1}^2 {\mathpzc s}_{k}(\ell^2,\ell\cdot P)\,  {\mathpzc S}^k(\ell;P)
\, \frac{\lambda^0_c}{\sqrt{3}}
\, s^0_f \,, \\
{\Psi}_{\nu}^{1^+}(\ell;P)  & = \sum_{k=1}^6 {\mathpzc a}_{k}(\ell^2,\ell\cdot P)\, \gamma_5 {\mathpzc A}^k_\nu(\ell;P)
\,  \frac{\lambda^0_c}{\sqrt{3}}
\, \vec{s}_f\,,
\end{align}
\end{subequations}
where
\begin{align}
\label{diracbasis}
\nonumber
{\mathpzc S}^1 & = {\mathbf I}_{\rm D} \,,\;
{\mathpzc S}^2  = i \gamma\cdot\hat\ell - \hat\ell \cdot\hat P {\mathbf I}_{\rm D}\,, \\
{\mathpzc A}^1_\nu & =  \gamma\cdot\ell^\perp \hat P_\nu\,,\;
{\mathpzc A}^2_\nu  = - i \hat P_\nu {\mathbf I}_{\rm D}\,,\;
{\mathpzc A}^3_\nu  = \gamma\cdot\hat\ell^\perp \hat\ell^\perp_\nu\,, \\
\nonumber
{\mathpzc A}^4_\nu & = i\hat \ell_\nu^\perp {\mathbf I}_{\rm D}\,,\;
{\mathpzc A}^5_\nu = \gamma_\nu^\perp - {\mathpzc A}^3_\nu\,,\;
{\mathpzc A}^6_\nu = i \gamma_\nu^\perp \gamma\cdot\hat\ell^\perp - {\mathpzc A}^4_\nu\,,
\end{align}
are the Dirac basis matrices, with $\hat\ell^2=1$, $\hat P^2 = -1$, $\ell^\perp = \hat\ell_\nu +\hat\ell\cdot\hat P \hat P_\nu$, $\gamma^\perp = \gamma_\nu +\gamma\cdot\hat P \hat P_\nu$; $\lambda^0_c/\sqrt{3}$ is the colour matrix with $\lambda^0_c={\rm diag}[1,1,1]$; $s^0_f$ and $\vec{s}_f$ are the flavour matrices of the quark+diquark amplitudes, which are obtained by removing the diquarks' flavour matrices \eqref{dqf} from the nucleon's full amplitude:
\begin{subequations}
\label{qfmat}
\begin{align}
s_f^0 &= \tau^0\,,\\
s_f^1 &= \tfrac{1}{\surd 6}(\tau^1-i\tau^2)\,,\\
s_f^2 &= -\tfrac{1}{\surd 3}\tau^3\,,\\
s_f^3 &= -\tfrac{1}{\surd 6}(\tau^1+i\tau^2)\,.
\end{align}
\end{subequations}
}


\subsection{Faddeev wave functions}
The (unamputated) Faddeev wave function $\Phi^{J^P}$ can be computed from Eqs.\,\eqref{FaddeevAmp}\,--\,\eqref{qfmat} by simply attaching the appropriate dressed-quark and diquark propagators.  It may also be written in the form of Eq.\,\eqref{8sa}, with different scalar functions $\tilde{{\mathpzc s}}_{k}$ and $\tilde{{\mathpzc a}}_{k}$:
{\allowdisplaybreaks
\begin{subequations}
\label{8sawf}
\begin{align}
\nonumber
\Phi^{0^+}(\ell;P)
& = S(\ell+\eta\,P)\Delta^{0^+}(\hat{\eta}\,P-\ell)\Psi^{0^+}(\ell;P) \\
= &\sum_{k=1}^2 \tilde{{\mathpzc s}}_{k}(\ell^2,\ell\cdot P)\,  {\mathpzc S}^k(\ell;P)
\, \frac{\lambda^0_c}{\sqrt{3}}
\, s^0_f \,, \\
{\Phi}_{\mu}^{1^+}(\ell;P) & = S(\ell+\eta\,P)\Delta^{1^+}_{\mu\nu}(\hat{\eta}\,P-\ell){\Psi}_{\nu}^{1^+}(\ell;P) \nonumber \\
= &\sum_{k=1}^6 \tilde{{\mathpzc a}}_{k}(\ell^2,\ell\cdot P)\, \gamma_5 {\mathpzc A}^k_\nu(\ell;P)
\,  \frac{\lambda^0_c}{\sqrt{3}}
\, \vec{s}_f\,,
\end{align}
\end{subequations}
where $\eta, \hat{\eta}\in[0,1]$ and $\eta+\hat{\eta}=1$

Both the Faddeev amplitude and wave function are Poincar\'e covariant: each of the scalar functions that appears is frame independent, but the chosen frame determines just how the elements should be combined. Hence, the manner by which the dressed quarks’ orbital angular momentum $L$ and spin $S$ add to form a particular $J^P$ combination is frame dependent: $L$ and $S$ are not independently Poincar\'e invariant.

The Dirac tensors of different $J^P$ are:
{\allowdisplaybreaks
\begin{subequations}
\label{Lidentifications8}
\begin{align}
^2\!S: & \quad {\mathpzc S}^1, {\mathpzc A}^2_\nu, ({\mathpzc A}^3_\nu+{\mathpzc A}^5_\nu) \,;\\
%
^2\!P: & \quad {\mathpzc S}^2, {\mathpzc A}^1_\nu, ({\mathpzc A}^4_\nu+{\mathpzc A}^6_\nu)\,;\\
%
^4\!P: & \quad (2{\mathpzc A}^4_\nu-{\mathpzc A}^6_\nu)/3\,;\\
%
^4\!D:  & \quad  (2{\mathpzc A}^3_\nu-{\mathpzc A}^5_\nu)/3  \,;
\end{align}
\end{subequations}}
\hspace*{-0.5\parindent}\emph{viz}.\ the scalar functions associated with these Dirac matrix combinations in a Faddeev wave function possess the indicated angular momentum correlation between the quark and diquark.
Those functions are:
\begin{subequations}
\label{LFunctionIdentifications}
\begin{align}
\label{LFunctionIdentificationsa}
^2\!S: & \quad \tilde{\mathpzc s}_1, \tilde{\mathpzc a}_2,
    (\tilde{\mathpzc a}_3+2\tilde{\mathpzc a}_5)/3\,;\\
^2\!P: & \quad \tilde{\mathpzc s}_2, \tilde{\mathpzc a}_1,
    (\tilde{\mathpzc a}_4+2\tilde{\mathpzc a}_6)/3\,;\\
^4\!P: & \quad (\tilde{\mathpzc a}_4-\tilde{\mathpzc a}_6)\,;\\
^4\!D: & \quad (\tilde{\mathpzc a}_3 - \tilde{\mathpzc a}_5) \,.
\end{align}
\end{subequations}
}


\section{Current diagrams}
\label{appcurrentdia}
To compute the form factors, one must specify how the probe couples to the constituents of the composite hadrons. Herein, this amounts to specifying the couplings of the axial $(5\mu)$ or pseudoscalar $(5)$ current to the dressed quarks and diquarks.

In Fig.\,\ref{figcurrent}, we have separated the different contributions to the currents into six terms. \emph{N.B}.\ Diagrams 1, 2 and 3 represent the impulse-approximation: they are one-loop integrals, which we evaluate by Gaussian quadrature.  The remainder, Diagram 4, 5 and 6 describe the probe's coupling to the Faddeev kernel: they are two-loop diagrams, whose evaluation requires the use of Monte-Carlo methods.  For explicit calculations, we work in a Breit frame: $P_i=K-Q/2$, $P_f=P+Q/2$ and $K=(0,0,0,i\sqrt{m_N^2+Q^2/4})$.

\subsection{Diagram 1}
{\allowdisplaybreaks
Probe coupling directly to the bystander quark:
\begin{align}
\label{curr1}
J_{5(\mu)}^{\rm q}(K,Q)=
J_{5(\mu)}^{{\rm q},0^+}(K,Q) + J_{5(\mu)}^{{\rm q},1^+}(K,Q)\,,
\end{align}
with
\begin{align}
\label{curr1s}
\nonumber
J_{5(\mu)}^{{\rm q},0^+}(K,Q)=
&\int_p\,
\bar{\Psi}^{0^+}(p'_f;-P_f)S(p_{q+})\Gamma_{5(\mu)}^{j}(p_{q+},p_{q-})\\
&\times S(p_{q-})\Delta^{0^+}(p_d)\Psi^{0^+}(p'_i;P_i)\,,
\end{align}
and
\begin{align}
\label{curr1a}
\nonumber
J_{5(\mu)}^{{\rm q},1^+}(K,Q)=
&\int_p\,
\bar{\Psi}^{1^+}_\alpha(p'_f;-P_f)S(p_{q+})
\Gamma_{5(\mu)}^{j}(p_{q+},p_{q-})\\
&\times S(p_{q-})\Delta^{1^+}_{\alpha\beta}(p_d)\Psi_\beta^{1^+}(p'_i;P_i)\,,
\end{align}
where $\Gamma^{j}_{5(\mu)}$ is the dressed-quark's pseudoscalar (axial-vector) vertex, Eq.\,\eqref{psvx2} or \eqref{axvx2}; and, with $p$ the loop momentum:
\begin{subequations}
\begin{align}
p'_i &= p-\hat{\eta}\,Q/2\,,\\
p'_f &= p+\hat{\eta}\,Q/2\,,\\
p_{q-} &= p'_i + \eta\,P_i\,,\\
p_{q+} &= p'_f + \eta\,P_f\,,\\
p_d &= \hat{\eta}\,P_i -p'_i\,\\
&= \hat{\eta}\,P_f -p'_f\,.
\end{align}
\end{subequations}

\subsection{Diagram 2}

Probe coupling to an axial-vector diquark correlation:\footnote{Owing to their flavour structures, neither the axial-vector nor the pseudoscalar probe couples to the scalar diquark, see Sec.\,\ref{subsecdqvx}.}
\begin{align}
\label{curr2}
\nonumber
J_{5(\mu)}^{{\rm dq},aa}&(K,Q)=
\int_p\,
\bar{\Psi}^{1^+}_{\alpha}(p''_f;-P_f)
\Delta^{1^+}_{\alpha\rho}(p_{d+})\times\\
&\Gamma^{aa}_{5(\mu),\rho\sigma}(p_{d+},p_{d-})\Delta^{1^+}_{\sigma\beta}(p_{d-})S(p_q)
\Psi^{1^+}_\beta(p''_i;P_i)\,,
\end{align}
where $\Gamma^{aa}_{5(\mu),\rho\sigma}$ is the axial-vector diquark correlation's pseudoscalar (axial-vector) vertex, Eq.\,\eqref{psdqvxAA} or \eqref{axdqvxAA}; and the momenta are
\begin{subequations}
\begin{align}
p''_i &= p+\eta Q/2\,,\\
p''_f &= p-\eta Q/2\,,\\
p_{d-} &= \hat{\eta}\,P_i - p''_i\,,\\
p_{d+} &= \hat{\eta}\,P_f - p''_f\,,\\
p_q &= p''_i + \eta\,P_i \\
&= p''_f + \eta\,P_f\,.
\end{align}
\end{subequations}

\subsection{Diagram 3}

Probe-induced transition between scalar and axial-vector diquarks.  It is the sum of the following transitions: charged currents, $\{dd\}_{1^+}\leftrightarrow [ud]_{0^+}$, $\{uu\}_{1^+}\leftrightarrow [ud]_{0^+}$; and neutral currents, $\{ud\}_{1^+}\leftrightarrow [ud]_{0^+}$, where
\begin{align}
\label{curr3sa}
\nonumber
& J_{5(\mu)}^{{\rm dq},sa}(K,Q)
= \int_p\,
\bar{\Psi}^{0^+}(p''_f;-P_f)\Delta^{0^+}(p_{d+}) \\
& \; \times\Gamma_{5(\mu),\sigma}^{sa}(p_{d+},p_{d-})
\Delta^{1^+}_{\sigma\beta}(p_{d-})S(p_q)
\Psi^{1^+}_\beta(p''_i;P_i)\,,
\end{align}
and
\begin{align}
\label{curr3as}
\nonumber
& J_{5(\mu)}^{{\rm dq},as}(K,Q)
= \int_p\,
\bar{\Psi}^{1^+}_\alpha(p''_f;-P_f)\Delta^{1^+}_{\alpha\rho}(p_{d+}) \\
&\;\times \Gamma_{5(\mu),\rho}^{as}(p_{d+},p_{d-})
\Delta^{0^+}(p_{d-})S(p_q)
\Psi^{0^+}(p''_i;P_i)\,,
\end{align}
where $\Gamma_{5(\mu),\sigma}^{sa}$ and $\Gamma_{5(\mu),\rho}^{as}$ are the probe-induced transition vertices in Eqs.\,\eqref{psdqvxSA}, \eqref{axdqvxSA}.

\subsection{Diagram 4}

Probe coupling to the quark that is exchanged as one diquark breaks up and another is formed:
\begin{align}
\label{curr4}
\nonumber
J_{5(\mu)}^{\rm ex}&(K,Q)=\sum_{J_1^{P_1},J_2^{P_2}=0^+,1^+}\int_{p}\int_{k}
\bar{\Phi}^{J_2^{P_2}}_f\Gamma^{J_1^{P_1}}(\tilde{k}_r)\\
&\times\big\{S(\tilde{q}')\Gamma_{5(\mu)}^{j}(\tilde{q}',\tilde{q})S(\tilde{q})\big\}^{\rm T}
\bar{\Gamma}^{J_2^{P_2}}(\tilde{p}'_r)\Phi^{J_1^{P_1}}_i\,.
\end{align}
The process of quark exchange provides the attraction required in the Faddeev equation to bind the nucleon. It also ensures the Faddeev amplitude has the correct antisymmetry under the exchange of any two dressed-quarks. These essential features are absent in models with elementary (noncomposite) diquarks. The complete contribution is obtained by summing over $J_1^{P_1}$ and $J_2^{P_2}$, which can each take the values $0^+$, $1^+$.

\subsection{Diagrams 5 and 6}

Two-loop seagull diagrams appearing as partners to Diagram 4:
\begin{align}
\label{curr5}
\nonumber
J^{\rm sg}_{5(\mu)}(K,Q)
=	
&\sum_{J_1^{P_1},J_2^{P_2}=0^+,1^+}\int_p\int_k
\bar{\Phi}^{J_2^{P_2}}_f \chi^{j,J_1^{P_1}}_{5(\mu),[{\rm sg}]}(k_1,Q)\\
&\times S^{\rm T}(\tilde{q}')\bar{\Gamma}^{J_2^{P_2}}(\tilde{p}'_r)
\Phi^{J_1^{P_1}}_i\,,
\end{align}
for Diagram 5;
and
\begin{align}
\label{curr6}
\nonumber
J^{\overline{\rm sg}}_{5(\mu)}(K,Q)
=&\sum_{J_1^{P_1},J_2^{P_2}=0^+,1^+}\int_p\int_k
\bar{\Phi}^{J_2^{P_2}}_f
\Gamma^{J_1^{P_1}}(\tilde{k}_r)S^{\rm T}(\tilde{q})\\
&\times \bar{\chi}^{j,J_2^{P_2}}_{5(\mu),[{\rm sg}]}(k_2,Q)
\Phi^{J_1^{P_1}}_i\,,
\end{align}
for Diagram 6; the momenta are:
\begin{subequations}
\begin{align}
\tilde{p}_{q-} &= k+\eta P_i\,,\\
\tilde{p}_{q+} &= p+\eta P_f\,,\\
k_1 &= \frac{\tilde{p}_{q+}-\tilde{q}'}{2}\,,\\
k_2 &= \frac{\tilde{p}_{q-}-\tilde{q}}{2}\,;	
\end{align}
\end{subequations}
and, again, $J_1^{P_1}$ and $J_2^{P_2}$ are summed.
}


\section{Colour and flavour coefficients}
\label{appcoecf}

The nucleon's Faddeev equation and the current diagrams of Appendix\,\ref{appcurrentdia} need to be augmented with appropriate colour and flavour coefficients.
Using the colour and flavour matrices of the diquark amplitudes, Eqs.\,\eqref{qqBSAs}, and the quark+diquark amplitudes, Eqs.\,\eqref{8sa}, projecting them onto the isospinors of the proton ${\mathrm p}=(1,0)^{\rm T}$ or the neutron ${\mathrm n}=(0,1)^{\rm T}$, we can write the nucleon's Faddeev equation, pictured in Fig.\,\ref{figFaddeev}:
{\allowdisplaybreaks
\begin{align}
\label{nFaddeev}
\nonumber
&\left(\begin{array}{c}
\Psi^{0^+}(p;P)\\
\Psi^{1^+}_\mu(p;P)
\end{array}\right) \\
=
&\int_k	
\left(\begin{array}{cc}
(-\frac{1}{2}){\mathpzc K}^{0^+0^+} & (\frac{\sqrt{3}}{2}){\mathpzc K}_{\,\,\nu}^{0^+1^+} \\
(\frac{\sqrt{3}}{2}){\mathpzc K}_{\,\,\mu}^{1^+0^+} & (\frac{1}{2}){\mathpzc K}_{\,\,\mu\nu}^{1^+1^+}
\end{array}\right)
\left(\begin{array}{c}
\Phi^{0^+}(k;P)\\
\Phi^{1^+}_\nu(k;P)
\end{array}\right)\,,
\end{align}
where $-\frac{1}{2}$, $\frac{\sqrt{3}}{2}$, $\frac{\sqrt{3}}{2}$, $\frac{1}{2}$ are the associated colour-flavour coefficients;
\begin{align}
\nonumber
{\mathpzc K}^{J_1^{P_1}J_2^{P_2}}&\equiv {\mathpzc K}^{J_1^{P_1}J_2^{P_2}}(p,k,P)\\
&:=\Gamma^{J_2^{P_2}}(k_r)S^{\rm T}(q)\bar{\Gamma}^{J_1^{P_1}}(p_r)\,,
\end{align}
with the momenta
\begin{subequations}
\begin{align}
\tilde{p}_q &= p+\eta\,P\,,\\
\tilde{k}_q &= k+\eta\,P\,,\\
\tilde{p}_d &= -p + \hat{\eta}\,P\,,\\
\tilde{k}_d &= -k + \hat{\eta}\,P\,,\\
q &= \tilde{p}_d - \tilde{k}_q\,,\\
p_r &= \frac{\tilde{k}_q-q}{2}\,,\\
k_r &= \frac{\tilde{p}_q-q}{2}\,.
\end{align}
\end{subequations}

For the form-factor diagrams of Fig.\,\ref{figcurrent}, their flavour factors are obtained by using Eqs.\,\eqref{dqf} and \eqref{qfmat}. We have
\begin{align}
\label{cfdia1}
\sum_{m,n=0}^3\bigg[\delta^{mn}(s^m_f)^\dagger\,
\big(\frac{\tau^j}{2}\big)\,
(s^n_f)\bigg]\,,
\end{align}
for the probe-quark diagram (Diagram 1);
\begin{align}
\label{cfdia23}
\sum_{m,n=0}^3\bigg[
(s^m_f)^\dagger\,
(s^n_f)\,
2{\rm tr}_{\rm F}\big[(t^m_f)^\dagger\,(t^n_f)\,\big(\frac{\tau^j}{2}\big)^\dagger\big]
\bigg]\,,
\end{align}
for the probe-diquark diagram (Diagram 2 and 3);
and
\begin{align}
\label{cfdia4}
\sum_{m,n=0}^3\bigg[
(s^m_f)^\dagger\,
(t^n_f)\,\big(\frac{\tau^j}{2}\big)^\dagger\,
(t^m_f)^\dagger\,
(s^n_f)\bigg]\,,
\end{align}
for the exchange diagram (Diagram 4).

The seagull case is a little more complicated: we need to treat the bystander and exchange quarks' legs separately. From the seagull diagram (Diagram 5) we have
\begin{equation}
\label{cfdia5ex}
\sum_{m,n=0}^3\bigg[
(s^m_f)^\dagger \,(t^n_f)\,
\big(\frac{\tau^j}{2}\big)^\dagger\,(t^m_f)^\dagger
\,(s^n_f)\bigg]\,,
\end{equation}
for the exchange leg; and
\begin{equation}
\label{cfdia5by}
\sum_{m,n=0}^3\bigg[
(s^m_f)^\dagger \,\big(\frac{\tau^j}{2}\big)\,
(t^n_f)\,(t^m_f)^\dagger
\,(s^n_f)\bigg]\,,
\end{equation}
for the bystander leg. From the diagram of the seagull's conjugation (Diagram 6), we have
\begin{equation}
\label{cfdia6ex}
\sum_{m,n=0}^3\bigg[
(s^m_f)^\dagger \,(t^n_f)\,
\big(\frac{\tau^j}{2}\big)^\dagger\,(t^m_f)^\dagger
\,(s^n_f)\bigg]\,,
\end{equation}
for the exchange leg; and
\begin{equation}
\label{cfdia6by}
\sum_{m,n=0}^3\bigg[
(s^m_f)^\dagger \,
(t^n_f)\,(t^m_f)^\dagger
\,\big(\frac{\tau^j}{2}\big)\,(s^n_f)\bigg]\,,
\end{equation}
for the bystander leg.
}

At last, again, we need to project these matrices, Eqs.\,\eqref{cfdia1}\,--\,\eqref{cfdia6by}, onto the isospinors of the proton ${\mathrm p}=(1,0)^{\rm T}$ or the neutron ${\mathrm n}=(0,1)^{\rm T}$ to obtain the flavour coefficients.

The colour factors of these diagrams are the same as the electromagnetic case: for the impulse-approximation diagrams (Diagrams 1\,--\,3), their colour coefficients are ``$1$''; and for the exchange or seagull diagrams (Diagrams 4\,--\,6), their colour coefficients are ``$-1$''.


\section{Proof of PCAC}
\label{subsecpcac1}
%
%
We verify here that the interaction current we have constructed is sufficient to ensure the identity in Eq.\,\eqref{pcacn} is preserved.
Observe first that $J^j_{5\mu}(K,Q)$ and $J^j_{5}(K,Q)$ are both a sum of six terms (drawn in Fig.\,\ref{figcurrent}):
\begin{align}
\label{currsum}
\nonumber
J^j_{5(\mu)}
= &J_{5(\mu)}^{\rm q} + J_{5(\mu)}^{{\rm dq},aa}
+ (J_{5(\mu)}^{{\rm dq},sa} +J_{5(\mu)}^{{\rm dq},as})\\
+& J_{5(\mu)}^{\rm ex} + J_{5(\mu)}^{\rm sg} 	+
J_{5(\mu)}^{\rm \overline{sg}}\,,
\end{align}
each one of which must be considered.  Note, too, that we will consider either the neutral ($\tau^3$) or the charged ($\tau^{1\pm i2}$) currents.  In the isospin-symmetry limit, their flavour coefficients are the same.


\subsection*{Diagram 1: current coupling to quark line}

For Diagram 1 in Fig.\,\ref{figcurrent}, contracting Eq.\,\eqref{curr1s} with $Q_\mu$ and using Eq.\,\eqref{axwti}, we obtain\footnote{For the sake of completeness, we will write out all the colour-flavour factors explicitly.}
\begin{align}
\nonumber
&Q_\mu J_{5\mu}^{{\rm q},0^+}(K,Q) + 2im_q J_{5}^{{\rm q},0^+}(K,Q) \\
\nonumber
=& \frac{1}{2}\int_p\,
\bar{\Psi}^{0^+}(p'_f;-P_f)\,S(p_{q+})\big[Q_\mu\Gamma_{5\mu}(p_{q+},p_{q-})+\\
\nonumber
& 2im_q\Gamma_{5}(p_{q+},p_{q-})\big] S(p_{q-})\Delta^{0^+}(p_d)\,
\Psi^{0^+}(p'_i;P_i)\\
\label{diapcac1a}
\nonumber
=& \frac{1}{2}\int_p\,
\bar{\Psi}^{0^+}(p'_f;-P_f)\,i\gamma_5\,\Phi^{0^+}(p'_i;P_i) +\\
& \frac{1}{2}\int_p\,
\bar{\Phi}^{0^+}(p'_f;-P_f)\,i\gamma_5\,\Psi^{0^+}(p'_i;P_i)\,,
\end{align}
where $\int_p:=\int d^4p/(2\pi)^4$ and $1/2$ is the colour-flavour coefficient computed via Eq.\,\eqref{cfdia1}. The Dirac parts of the dressed-quark's axial-vector and pseudoscalar vertices \eqref{axvx} and \eqref{psvx}
are denoted by $\Gamma_{5\mu}$ and $\Gamma_{5}$: $\Gamma_{5\mu}^j=:(\tau^j/2)\Gamma_{5\mu}$ and $\Gamma_{5}^j=:(\tau^j/2)\Gamma_{5}$.

Next, using the nucleon's Faddeev equation, Eq.\,\eqref{nFaddeev}, to substitute $\bar{\Psi}^{0^+}(p'_f;-P_f)$ and $\Psi^{0^+}(p'_i;P_i)$ in Eq.\,\eqref{diapcac1a},
some algebra leads to
\begin{align}
\label{pcacdia1s}
\nonumber
Q_\mu &J_{5\mu}^{{\rm q},0^+}(K,Q) + 2im_q J_{5}^{{\rm q},0^+}(K,Q) \\
\nonumber
=& (-\frac{1}{4})\int_p\int_k
\bar{\Phi}^{0^+}_f\,\Gamma^{0^+}(\tilde{k}_r)S^{\rm T}(\tilde{q})
\bar{\Gamma}^{0^+}(\tilde{p}_r)i\gamma_5\,\Phi^{0^+}_i\\
\nonumber
+& (-\frac{1}{4})\int_p\int_k
\bar{\Phi}^{0^+}_f\,i\gamma_5\Gamma^{0^+}(\tilde{k}'_r)S^{\rm T}(\tilde{q}')
\bar{\Gamma}^{0^+}(\tilde{p}'_r)\,\Phi^{0^+}_i\\
\nonumber
+& (\frac{\sqrt{3}}{4})\int_p\int_k
\bar{\Phi}^{1^+}_{\alpha,f}\,\Gamma^{0^+}(\tilde{k}_r)S^{\rm T}(\tilde{q})
\bar{\Gamma}^{1^+}_\alpha(\tilde{p}_r)i\gamma_5\,\Phi^{0^+}_i\\
+& (\frac{\sqrt{3}}{4})\int_p\int_k
\bar{\Phi}^{0^+}_f\,i\gamma_5\Gamma^{1^+}_\beta(\tilde{k}'_r)S^{\rm T}(\tilde{q}')
\bar{\Gamma}^{0^+}(\tilde{p}'_r)\,\Phi^{1^+}_{\beta,i}\,,
\end{align}
where we use the
abbreviations $\bar{\Phi}^{J^P}_f\equiv\bar{\Phi}^{J^P}(p;-P_f)$, $\Phi^{J^P}_i\equiv\Phi^{J^P}(k;P_i)$, $\bar{\Psi}^{J^P}_f\equiv\bar{\Psi}^{J^P}(p;-P_f)$ and
$\Psi^{J^P}_i\equiv\Psi^{J^P}(k;P_i)$; and the momenta are
\begin{subequations}
\begin{align}
\tilde{k}_r &= \frac{1}{2}((k+\hat{\eta}\,Q)+2p+(3\eta-1)P_f)\,,\\
\tilde{p}_r &= \frac{1}{2}(p+2(k+\hat{\eta}\,Q)+(3\eta-1)P_f)\,,\\
\tilde{q} &= -p-(k+\hat{\eta}\,Q)+(1-2\eta)P_f\,,\\
\tilde{k}'_r &= \tilde{k}_r -Q\,,\\
\tilde{p}'_r &= \tilde{p}_r -Q\,,\\
\tilde{q}' &= \tilde{q} +Q\,.
\end{align}
\end{subequations}

Similarly, for the axial-vector part, \emph{i.e.} $J_{5\mu}^{{\rm q},1^+}$ and $J_{5}^{{\rm q},1^+}$ in Eq.\,\eqref{curr1a}, we obtain
\begin{align}
\label{pcacdia1a}
\nonumber
 Q_\mu &J_{5\mu}^{{\rm q},1^+}(K,Q) + 2im_q J_{5}^{{\rm q},1^+}(K,Q) \\
\nonumber
=& (-\frac{\sqrt{3}}{12})\int_p\int_k
\bar{\Phi}^{0^+}_f\,\Gamma^{1^+}_\beta(\tilde{k}_r)S^{\rm T}(\tilde{q})
\bar{\Gamma}^{0^+}(\tilde{p}_r)i\gamma_5\,\Phi^{1^+}_{\beta,i}\\
\nonumber
+& (-\frac{\sqrt{3}}{12})\int_p\int_k
\bar{\Phi}^{1^+}_{\alpha,f}\,i\gamma_5\Gamma^{0^+}(\tilde{k}'_r)S^{\rm T}(\tilde{q}')
\bar{\Gamma}^{1^+}_\alpha(\tilde{p}'_r)\,\Phi^{0^+}_i\\
\nonumber
+& (-\frac{1}{12})\int_p\int_k
\bar{\Phi}^{1^+}_{\alpha,f}\,\Gamma^{1^+}_\beta(\tilde{k}_r)S^{\rm T}(\tilde{q})
\bar{\Gamma}^{1^+}_\alpha(\tilde{p}_r)i\gamma_5\,\Phi^{1^+}_{\beta,i}\\
+& (-\frac{1}{12})\int_p\int_k
\bar{\Phi}^{1^+}_{\alpha,f}\,i\gamma_5\Gamma^{1^+}_\beta(\tilde{k}'_r)S^{\rm T}(\tilde{q}')
\bar{\Gamma}^{1^+}_\alpha(\tilde{p}'_r)\,\Phi^{1^+}_{\beta,i}\,.
\end{align}

Adding together Eqs.~\eqref{pcacdia1s} and~\eqref{pcacdia1a}, and taking into account Eq.\,\eqref{curr1}, we find
\begin{align}
\label{pcacdia1}
\nonumber
 Q_\mu &J_{5\mu}^{{\rm q}}(K,Q) + 2im_q J_{5}^{{\rm q}}(K,Q)	\\
=& \sum_{J_1^{P_1}J_2^{P_2}=0^+,1^+}\big( Q_\mu J_{5\mu}^{{\rm q},J_1^{P_1}J_2^{P_2}} + 2im_q J_{5}^{{\rm q},J_1^{P_1},J_2^{P_2}} \big)\,,
\end{align}
where
\begin{align}
\nonumber
Q_\mu &J_{5\mu}^{{\rm q},0^+0^+} + 2im_q J_{5}^{{\rm q},0^+0^+}\\
\nonumber
=& \int_p\int_k	
\bar{\Phi}^{0^+}_{f}\,\bigg[(-\frac{1}{4})
\Gamma^{0^+}(\tilde{k}_r)S^{\rm T}(\tilde{q})
\bar{\Gamma}^{0^+}(\tilde{p}_r)i\gamma_5 +\\
& (-\frac{1}{4})i\gamma_5\Gamma^{0^+}(\tilde{k}'_r)S^{\rm T}(\tilde{q}')
\bar{\Gamma}^{0^+}(\tilde{p}'_r)\bigg]\,\Phi^{0^+}_{i}\,;
\end{align}
\begin{align}
\nonumber
 Q_\mu &J_{5\mu}^{{\rm q},0^+1^+} + 2im_q J_{5}^{{\rm q},0^+1^+}\\
\nonumber
=& \int_p\int_k	
\bar{\Phi}^{0^+}_{f}\,\bigg[(-\frac{\sqrt{3}}{12})
\Gamma^{1^+}_\beta(\tilde{k}_r)S^{\rm T}(\tilde{q})
\bar{\Gamma}^{0^+}(\tilde{p}_r)i\gamma_5 +\\
& (\frac{\sqrt{3}}{4})i\gamma_5\Gamma^{1^+}_\beta(\tilde{k}'_r)S^{\rm T}(\tilde{q}')
\bar{\Gamma}^{0^+}(\tilde{p}'_r)\bigg]\,\Phi^{1^+}_{\beta,i}\,;
\end{align}
\begin{align}
\nonumber
 Q_\mu &J_{5\mu}^{{\rm q},1^+0^+} + 2im_q J_{5}^{{\rm q},1^+0^+}\\
\nonumber
=& \int_p\int_k	
\bar{\Phi}^{1^+}_{\alpha,f}\,\bigg[(\frac{\sqrt{3}}{4})
\Gamma^{0^+}(\tilde{k}_r)S^{\rm T}(\tilde{q})
\bar{\Gamma}^{1^+}_\alpha(\tilde{p}_r)i\gamma_5 +\\
& (-\frac{\sqrt{3}}{12})i\gamma_5\Gamma^{0^+}(\tilde{k}'_r)S^{\rm T}(\tilde{q}')
\bar{\Gamma}^{1^+}_\alpha(\tilde{p}'_r)\bigg]\,\Phi^{0^+}_{i}\,;
\end{align}
and
\begin{align}
\nonumber
 Q_\mu &J_{5\mu}^{{\rm q},1^+1^+} + 2im_q J_{5}^{{\rm q},1^+1^+}\\
\nonumber
=& \int_p\int_k	
\bar{\Phi}^{1^+}_{\alpha,f}\,\bigg[(-\frac{1}{12})
\Gamma^{1^+}_\beta(\tilde{k}_r)S^{\rm T}(\tilde{q})
\bar{\Gamma}^{1^+}_\alpha(\tilde{p}_r)i\gamma_5 +\\
& (-\frac{1}{12})i\gamma_5\Gamma^{1^+}_\beta(\tilde{k}'_r)S^{\rm T}(\tilde{q}')
\bar{\Gamma}^{1^+}_\alpha(\tilde{p}'_r)\bigg]\,\Phi^{1^+}_{\beta,i}\,.
\end{align}


\subsection*{Diagrams 2 and 3: current coupling to diquark line}

For diagram 2, using Eq.\,\eqref{curr2} and the corresponding AXWTI~\eqref{dqwtiaa}, we have
\begin{align}
\label{pcacdia2}
\nonumber
 &Q_\mu J_{5\mu}^{{\rm dq},aa} +2im_qJ_{5}^{{\rm dq},aa}\\
\nonumber
=&(\frac{2}{3})\int_p\,
\nonumber
\bar{\Psi}^{1^+}_{\alpha}(p_f'';-P_f)
\Delta^{1^+}_{\alpha\rho}(p_{d+})
\big[Q_\mu\Gamma^{aa}_{5\mu,\rho\sigma}(p_{d+},p_{d-})\\
\nonumber
+&2im_q\Gamma^{aa}_{5,\rho\sigma}(p_{d+},p_{d-})\big]
\Delta^{1^+}_{\sigma\beta}(p_{d-})S(p_q)\,
\Psi^{1^+}_{\beta}(p''_i;P_i)\\
=& 0\,.
\end{align}
Similarly, using Eqs.\,\eqref{curr3sa}, \eqref{curr3as} and the corresponding AXWTI, Eq.\,\eqref{dqwtisa}, the $0^+-1^+$ transition of diagram 3 is expressed by
\begin{align}
\label{pcacdia3sa}
\nonumber
 & Q_\mu J_{5\mu}^{{\rm dq},sa} +2im_qJ_{5}^{{\rm dq},sa}\\
\nonumber
=&(\frac{\sqrt{3}}{3})\int_p\,
\nonumber
\bar{\Psi}^{0^+}_{f}(p''_f;-P_f)
\Delta^{0^+}(p_{d+})
\big[Q_\mu\Gamma^{sa}_{5\mu,\sigma}(p_{d+},p_{d-})\\
\nonumber
+&2im_q\Gamma^{sa}_{5,\sigma}(p_{d+},p_{d-})\big]
\Delta^{1^+}_{\sigma\beta}(p_{d-})S(p_q)\,
\Psi^{1^+}_{\beta}(p''_i;P_i)\\
=& 0\,,
\end{align}
and
\begin{align}
\label{pcacdia3as}
\nonumber
 & Q_\mu J_{5\mu}^{{\rm dq},as} +2im_qJ_{5}^{{\rm dq},as}\\
\nonumber
=&(\frac{\sqrt{3}}{3})\int_p\,
\nonumber
\bar{\Psi}^{1^+}_{\alpha}(p''_f;-P_f)\,
\Delta^{1^+}(p_{d+})_{\alpha\rho}
\big[Q_\mu\Gamma^{as}_{5\mu,\rho}(p_{d+},p_{d-})\\
\nonumber
+&2im_q\Gamma^{as}_{5,\rho}(p_{d+},p_{d-})\big]
\Delta^{0^+}(p_{d-})S(p_q)\,
\Psi^{0^+}(p''_i;P_i)\\
=& 0\,.
\end{align}
The factors $2/3$, $\sqrt{3}/3$ and $\sqrt{3}/3$ are combined colour/flavour coefficients computed via Eq.\,\eqref{cfdia23}.


\subsection*{Diagram 4: current coupling to exchange-quark}

For the quark exchange diagram 4, using Eq.\,\eqref{curr4} and AXWTI, Eq.\,\eqref{axwti}, we obtain
\begin{align}
\label{pcacdia4}
\nonumber
 Q_\mu &J_{5\mu}^{{\rm ex}}(K,Q) + 2im_q J_{5}^{{\rm ex}}(K,Q)	\\
=& \sum_{J_1^{P_1},J_2^{P_2}=0^+,1^+}\big( Q_\mu J_{5\mu}^{{\rm ex},J_1^{P_1}J_2^{P_2}} + 2im_q J_{5}^{{\rm ex},J_1^{P_1}J_2^{P_2}} \big)\,,
\end{align}
where
{\allowdisplaybreaks
\begin{align}
\nonumber
 Q_\mu &J_{5\mu}^{{\rm ex},0^+0^+} + 2im_q J_{5}^{{\rm ex},0^+0^+}\\
\nonumber
=& \int_p\int_k	
\bar{\Phi}^{0^+}_{f}\,\bigg[(\frac{1}{4})
\Gamma^{0^+}(\tilde{k}_r)i\gamma_5^{\rm T}S^{\rm T}(\tilde{q}')
\bar{\Gamma}^{0^+}(\tilde{p}'_r) +\\
& (\frac{1}{4})\Gamma^{0^+}(\tilde{k}_r)S^{\rm T}(\tilde{q})i\gamma_5^{\rm T}
\bar{\Gamma}^{0^+}(\tilde{p}'_r)\bigg]\,\Phi^{0^+}_{i}\,;
\end{align}
\begin{align}
\nonumber
 Q_\mu &J_{5\mu}^{{\rm ex},0^+1^+} + 2im_q J_{5}^{{\rm ex},0^+1^+}\\
\nonumber
=& \int_p\int_k	
\bar{\Phi}^{0^+}_{f}\,\bigg[(\frac{\sqrt{3}}{12})
\Gamma^{1^+}_\beta(\tilde{k}_r)i\gamma_5^{\rm T}S^{\rm T}(\tilde{q}')
\bar{\Gamma}^{0^+}(\tilde{p}'_r) +\\
& (\frac{\sqrt{3}}{12})\Gamma^{1^+}_\beta(\tilde{k}_r)S^{\rm T}(\tilde{q})i\gamma_5^{\rm T}
\bar{\Gamma}^{0^+}(\tilde{p}'_r)\bigg]\,\Phi^{1^+}_{\beta,i}\,;
\end{align}
\begin{align}
\nonumber
 Q_\mu &J_{5\mu}^{{\rm ex},1^+0^+} + 2im_q J_{5}^{{\rm ex},1^+0^+}\\
\nonumber
=& \int_p\int_k	
\bar{\Phi}^{1^+}_{\alpha,f}\,\bigg[(\frac{\sqrt{3}}{12})
\Gamma^{0^+}(\tilde{k}_r)i\gamma_5^{\rm T}S^{\rm T}(\tilde{q}')
\bar{\Gamma}^{1^+}_\alpha(\tilde{p}'_r) +\\
& (\frac{\sqrt{3}}{12})\Gamma^{0^+}(\tilde{k}_r)S^{\rm T}(\tilde{q})i\gamma_5^{\rm T}
\bar{\Gamma}^{1^+}_\alpha(\tilde{p}'_r)\bigg]\,\Phi^{0^+}_{i}\,;
\end{align}
and
\begin{align}
\nonumber
 Q_\mu &J_{5\mu}^{{\rm ex},1^+1^+} + 2im_q J_{5}^{{\rm ex},1^+1^+}\\
\nonumber
=& \int_p\int_k	
\bar{\Phi}^{1^+}_{\alpha,f}\,\bigg[(\frac{5}{12})
\Gamma^{1^+}_\beta(\tilde{k}_r)i\gamma_5^{\rm T}S^{\rm T}(\tilde{q}')
\bar{\Gamma}^{1^+}_\alpha(\tilde{p}'_r) +\\
& (\frac{5}{12})\Gamma^{1^+}_\beta(\tilde{k}_r)S^{\rm T}(\tilde{q})i\gamma_5^{\rm T}
\bar{\Gamma}^{1^+}_\alpha(\tilde{p}'_r)\bigg]\,\Phi^{1^+}_{\beta,i}\,.
\end{align}
The flavour coefficients are calculated via Eq.\,\eqref{cfdia4}.}


\subsection*{Diagram 5: first seagull contribution}

For the seagull diagram 5, using Eq.\,\eqref{curr5} and the seagull's AXWTI, Eq.\,\eqref{sgwti}, we get
{\allowdisplaybreaks\begin{align}
\label{pcacdia5}
\nonumber
 Q_\mu &J_{5\mu}^{{\rm sg}}(K,Q) + 2im_q J_{5}^{{\rm sg}}(K,Q)	\\
=& \sum_{J_1^{P_1},J_2^{P_2}=0^+,1^+}
\big( Q_\mu J_{5\mu}^{{\rm sg},J_1^{P_1}J_2^{P_2}} + 2im_q J_{5}^{{\rm sg},J_1^{P_1}J_2^{P_2}} \big)\,,
\end{align}
where
\begin{align}
\label{pcacdia5ss}
\nonumber
 Q_\mu &J_{5\mu}^{{\rm sg},0^+0^+} + 2im_q J_{5}^{{\rm sg},0^+0^+}\\
\nonumber
=& \int_p\int_k	
\bar{\Phi}^{0^+}_{f}\,\bigg[(\frac{1}{4})
i\gamma_5\Gamma^{0^+}(\tilde{k}'_r)S^{\rm T}(\tilde{q}')
\bar{\Gamma}^{0^+}(\tilde{p}'_r) +\\
& (-\frac{1}{4})\Gamma^{0^+}(\tilde{k}_r)i\gamma_5^{\rm T}S^{\rm T}(\tilde{q}')
\bar{\Gamma}^{0^+}(\tilde{p}'_r)\bigg]\,\Phi^{0^+}_{i}\,;
\end{align}
\begin{align}
\label{pcacdia5sa}
\nonumber
 Q_\mu &J_{5\mu}^{{\rm sg},0^+1^+} + 2im_q J_{5}^{{\rm sg},0^+1^+}\\
\nonumber
=& \int_p\int_k	
\bar{\Phi}^{0^+}_{f}\,\bigg[(-\frac{\sqrt{3}}{4})
i\gamma_5\Gamma^{1^+}_\beta(\tilde{k}'_r)S^{\rm T}(\tilde{q}')
\bar{\Gamma}^{0^+}(\tilde{p}'_r) +\\
& (-\frac{\sqrt{3}}{12})\Gamma^{1^+}_\beta(\tilde{k}_r)i\gamma_5^{\rm T}S^{\rm T}(\tilde{q}')
\bar{\Gamma}^{0^+}(\tilde{p}'_r)\bigg]\,\Phi^{1^+}_{\beta,i}\,;
\end{align}
\begin{align}
\label{pcacdia5as}
\nonumber
 Q_\mu &J_{5\mu}^{{\rm sg},1^+0^+} + 2im_q J_{5}^{{\rm sg},1^+0^+}\\
\nonumber
=& \int_p\int_k	
\bar{\Phi}^{0^+}_{\alpha,f}\,\bigg[(\frac{\sqrt{3}}{12})
i\gamma_5\Gamma^{0^+}(\tilde{k}'_r)S^{\rm T}(\tilde{q}')
\bar{\Gamma}^{1^+}_\alpha(\tilde{p}'_r) +\\
& (-\frac{\sqrt{3}}{12})\Gamma^{0^+}(\tilde{k}_r)i\gamma_5^{\rm T}S^{\rm T}(\tilde{q}')
\bar{\Gamma}^{1^+}_\alpha(\tilde{p}'_r)\bigg]\,\Phi^{0^+}_{i}\,;
\end{align}
and
\begin{align}
\label{pcacdia5aa}
\nonumber
 Q_\mu &J_{5\mu}^{{\rm sg},1^+1^+} + 2im_q J_{5}^{{\rm sg},1^+1^+}\\
\nonumber
=& \int_p\int_k	
\bar{\Phi}^{1^+}_{\alpha,f}\,\bigg[(\frac{1}{12})
i\gamma_5\Gamma^{1^+}_\beta(\tilde{k}'_r)S^{\rm T}(\tilde{q}')
\bar{\Gamma}^{1^+}_\alpha(\tilde{p}'_r) +\\
& (-\frac{5}{12})\Gamma^{1^+}_\beta(\tilde{k}_r)i\gamma_5^{\rm T}S^{\rm T}(\tilde{q}')
\bar{\Gamma}^{1^+}_\alpha(\tilde{p}'_r)\bigg]\,\Phi^{1^+}_{\beta,i}\,.
\end{align}}

The colour/flavour coefficients in the first lines of Eqs.\,\eqref{pcacdia5ss}\,--\,\eqref{pcacdia5aa} are calculated via Eq.\,\eqref{cfdia5by}, \emph{i.e.} the bystander legs of the seagulls; and the coefficients in the second lines are calculated via Eq.\,\eqref{cfdia5ex}, the exchange legs of the seagulls.


\subsection*{Diagram 6: second seagull contribution}

For the conjugated seagull contribution, diagram 6, using Eq.\,\eqref{curr6} and the AXWTI of the seagulls' conjugations, one finds
{\allowdisplaybreaks
\begin{align}
\label{pcacdia6}
\nonumber
 Q_\mu &J_{5\mu}^{{\rm \overline{sg}}}(K,Q) + 2im_q J_{5}^{{\rm \overline{sg}}}(K,Q)	\\
=& \sum_{J_1^{P_1},J_2^{P_2}=0^+,1^+}
\big( Q_\mu J_{5\mu}^{{\rm \overline{sg}},J_1^{P_1}J_2^{P_2}} + 2im_q J_{5}^{{\rm \overline{sg}},J_1^{P_1}J_2^{P_2}} \big)\,,
\end{align}
where
\begin{align}
\label{pcacdia6ss}
\nonumber
 Q_\mu &J_{5\mu}^{{\rm \overline{sg}},0^+0^+} + 2im_q J_{5}^{{\rm \overline{sg}},0^+0^+}\\
\nonumber
=& \int_p\int_k	
\bar{\Phi}^{0^+}_{f}\,\bigg[(\frac{1}{4})
\Gamma^{0^+}(\tilde{k}_r)S^{\rm T}(\tilde{q})
\bar{\Gamma}^{0^+}(\tilde{p}_r)i\gamma_5 +\\
& (-\frac{1}{4})\Gamma^{0^+}(\tilde{k}_r)S^{\rm T}(\tilde{q})i\gamma_5^{\rm T}
\bar{\Gamma}^{0^+}(\tilde{p}'_r)\bigg]\,\Phi^{0^+}_{i}\,;
\end{align}
\begin{align}
\label{pcacdia6sa}
\nonumber
 Q_\mu &J_{5\mu}^{{\rm \overline{sg}},0^+1^+} + 2im_q J_{5}^{{\rm \overline{sg}},0^+1^+}\\
\nonumber
=& \int_p\int_k	
\bar{\Phi}^{0^+}_{f}\,\bigg[(\frac{\sqrt{3}}{12})
\Gamma^{1^+}_\beta(\tilde{k}_r)S^{\rm T}(\tilde{q})
\bar{\Gamma}^{0^+}(\tilde{p}_r)i\gamma_5 +\\
& (-\frac{\sqrt{3}}{12})\Gamma^{1^+}_\beta(\tilde{k}_r)S^{\rm T}(\tilde{q})i\gamma_5^{\rm T}
\bar{\Gamma}^{0^+}(\tilde{p}'_r)\bigg]\,\Phi^{1^+}_{\beta,i}\,;
\end{align}
\begin{align}
\label{pcacdia6as}
\nonumber
 Q_\mu &J_{5\mu}^{{\rm \overline{sg}},1^+0^+} + 2im_q J_{5}^{{\rm \overline{sg}},1^+0^+}\\
\nonumber
=& \int_p\int_k	
\bar{\Phi}^{1^+}_{\alpha,f}\,\bigg[(-\frac{\sqrt{3}}{4})
\Gamma^{0^+}(\tilde{k}_r)S^{\rm T}(\tilde{q})
\bar{\Gamma}^{1^+}_\alpha(\tilde{p}_r)i\gamma_5 +\\
& (-\frac{\sqrt{3}}{12})\Gamma^{0^+}(\tilde{k}_r)S^{\rm T}(\tilde{q})i\gamma_5^{\rm T}
\bar{\Gamma}^{1^+}_\alpha(\tilde{p}'_r)\bigg]\,\Phi^{0^+}_{i}\,;
\end{align}
and
\begin{align}
\label{pcacdia6aa}
\nonumber
 Q_\mu &J_{5\mu}^{{\rm \overline{sg}},1^+1^+} + 2im_q J_{5}^{{\rm \overline{sg}},1^+1^+}\\
\nonumber
=& \int_p\int_k	
\bar{\Phi}^{1^+}_{\alpha,f}\,\bigg[(\frac{1}{12})
\Gamma^{1^+}_\beta(\tilde{k}_r)S^{\rm T}(\tilde{q})
\bar{\Gamma}^{1^+}_\alpha(\tilde{p}_r)i\gamma_5 +\\
& (-\frac{5}{12})\Gamma^{1^+}_\beta(\tilde{k}_r)S^{\rm T}(\tilde{q})i\gamma_5^{\rm T}
\bar{\Gamma}^{1^+}_\alpha(\tilde{p}'_r)\bigg]\,\Phi^{1^+}_{\beta,i}\,.
\end{align}}

The colour/flavour coefficients in the first lines of Eqs.\,\eqref{pcacdia6ss}\,--\,\eqref{pcacdia6aa} are calculated via Eq.\,\eqref{cfdia6by}, \emph{i.e.} the bystander legs of the seagulls' conjugations; and the coefficients in the second lines are calculated via Eq.\,\eqref{cfdia6ex}, the exchange legs.

\subsection*{Sum of all contributions}

Using Eqs.\,\eqref{currsum}, \eqref{pcacdia1}, \eqref{pcacdia2}\,--\,\eqref{pcacdia4}, \eqref{pcacdia5} and \eqref{pcacdia6}, it is straightforward to obtain their sum:
\begin{align}
\nonumber
Q_\mu &J^j_{5\mu}(K,Q) + 2im_q J^j_{5}(K,Q) = \sum_{J_1^{P_1},J_2^{P_2}=0^+,1^+}\\
\nonumber
\bigg[&\big(Q_\mu J_{5\mu}^{{\rm q},J_1^{P_1}J_2^{P_2}}(K,Q) + 2im_q J_{5}^{{\rm q},J_1^{P_1}J_2^{P_2}}(K,Q)	\big)\\
\nonumber
+&\big(Q_\mu J_{5\mu}^{{\rm ex},J_1^{P_1}J_2^{P_2}}(K,Q) + 2im_q J_{5}^{{\rm ex},J_1^{P_1}J_2^{P_2}}(K,Q)	\big)\\
\nonumber
+&\big(Q_\mu J_{5\mu}^{{\rm sg},J_1^{P_1}J_2^{P_2}}(K,Q) + 2im_q J_{5}^{{\rm sg},J_1^{P_1}J_2^{P_2}}(K,Q)	\big)\\
\nonumber
+&\big(Q_\mu J_{5\mu}^{{\rm \overline{sg}},J_1^{P_1}J_2^{P_2}}(K,Q) + 2im_q J_{5}^{{\rm \overline{sg}},J_1^{P_1}J_2^{P_2}}(K,Q)	\big)\bigg]\\
=& \,0\,,
\end{align}
where $j=3$ for the neutral current, or $j=1\pm i2$ for the charged currents.


There are interesting features behind the details of this proof.
First, Eqs.\,\eqref{pcacdia2}\,--\,\eqref{pcacdia3as} explicitly show that the three processes of Diagrams 2 and 3 satisfy PCAC separately. This is natural, and expected, since the axial-vector and pseudoscalar seagull terms do not involve a diquark part, in contrast to their electromagnetic counterparts.
Second, the contributions from the coupling to the quark leg are cancelled by the contributions from the bystander legs of the seagull terms, and their conjugations. Furthermore, the contributions of the quark-exchange diagrams are cancelled by the those from the exchange legs of the seagulls, and their conjugations. This is analogous to the electromagnetic case \cite{Oettel:1999gc}.
%
Third, in an analysis of Nambu--Jona-Lasinio like models, it was found that PCAC can be ensured by including an isoscalar-vector diquark correlation in the nucleon in addition to scalar and axial-vector diquarks \cite{Ishii:2000zy}.  Our analysis has shown that whilst this may be useful in some cases, it is not necessary.

\section{Impacts of pion pole terms}
\label{Apppionpole}
Regarding $G_A(Q^2)$, consider Eq.\,\eqref{gaproj}.  Evidently, $G_A(Q^2)$ only receives contributions from axial-vector $Q$-transverse pieces of the nucleon current.  Reviewing our construction, the axial-vector terms are
Eqs.\,\eqref{axvx2},
\eqref{axsg}, \eqref{axsgcg},
\eqref{axdqvxAA},
\eqref{axdqvxSA}.
Each of the associated pion-pole parts is $Q$-longitudinal.  Thus, $G_A(Q^2)$ is unaffected by the pion-pole terms in our current construction.  It is completely determined by the regular parts alone.  This is apparent in Fig.\,\ref{FigGAx}.

Turn now to Eq.\,\eqref{pcacp}, an identity valid $\forall Q^2$.  In the chiral limit, it entails:
\begin{equation}
\label{GPGAidentity}
G_P(Q^2) = \frac{4 m_N^2}{Q^2} G_A(Q^2) \,.
\end{equation}
This is an exact statement of chiral symmetry and the pattern by which it is broken at the level of the nucleon.  It is true when using any and all symmetry-consistent current constructions; consequently, true in our formulation.

Eq.\,\eqref{GPGAidentity} states that $G_{A}$ and $G_P$ are not independent functions in the chiral limit.  Instead, the behaviour of $G_A$ is completely determined by that of $Q^2 G_P(Q^2)$ and vice versa.   This can be understood at a deeper level by recognising that Eq.\,\eqref{GPGAidentity} is equivalent to the symmetry constraint
\begin{equation}
{\rm tr}_{\rm D} Q_\mu \gamma_5 J_{5\mu}(Q) = 0\,.
\end{equation}
In words, this means that the current, $J_{5\mu}(Q)$, may contain both regular and $Q_\mu/Q^2$  pole terms; but when contracted with $Q_\mu$, the resulting regular and pole pieces combine to cancel exactly.  An analogous process takes place at the dressed-quark level in QCD, as shown for the dressed-quark+axial-vector vertex in Ref.\,\cite[Eqs.\,(10)-(13)]{Maris:1997hd}.

Consequently, in the chiral limit, $J_{5\mu}(Q)$ is purely $Q$-transverse.  Further, the $1/Q^2$  in Eq.\,\eqref{GPGAidentity} is completely absorbed into the transverse projection operator in Eq.\,\eqref{gaproj}.  One may interpret this as follows: in the chiral limit, the pion pole contributions to the nucleon axial-vector current serve to cancel any and all longitudinal axial-vector couplings of the nucleon, ensuring that only $Q$-transverse interactions survive.
In this sense, the essentially dynamical chiral-limit $Q^2$-dependence of $G_P (Q^2)$ is also completely determined by the regular parts of our current construction, with the  $1/Q^2$  in Eq.\,\eqref{GPGAidentity} expressing what is effectively a kinematic factor.

At physical values of the light-quark current masses, Eq.\,\eqref{GPGAidentity} returns to the identity in  Eq.\,\eqref{pcacp}.  In consequence, the nucleon axial-vector current is no longer purely $Q$-transverse.  It possesses a $Q$-divergence that is measured by $2m_q G_5 (Q^2 )$.  This means that, at physical current-quark masses, the pion pole contribution to the nucleon axial-vector current is insufficient to completely cancel all longitudinal axial-vector couplings of the nucleon; and the surviving longitudinal components are measured by the pseudoscalar form factor, $G_5 (Q^2)$.

Focusing now on $m_q G_5(Q^2)$, which is obtained using Eq.\,\eqref{g5proj}, one notes that every term which can contribute to the $J_5(Q)$ current includes an overall multiplicative factor of $m_\pi^2/[Q^2+m_\pi^2]$.  The associated (pole) pieces are represented in
Eqs.\,\eqref{psvx2},
\eqref{pssg}, \eqref{pssgcg},
\eqref{psdqvxAA},
\eqref{psdqvxSA}.
%
The form of Eq.\,\eqref{psvx2} is fixed by DCSB at the dressed-quark level.
The forms of Eqs.\,\eqref{pssg}, \eqref{pssgcg}, \eqref{psdqvxAA}, \eqref{psdqvxSA} are determined by the axial-vector Ward-Green-Takahashi identity at the nucleon level, compliance with which is itself driven by the dressed-quark-level expression of DCSB.
Of these five terms, Eq.\,\eqref{psvx2}, which generates the diagram~1 part in Fig.\,\ref{figcurrent}, produces the dominant contribution at all values of $Q^2$.  All other contributions are soft, providing symmetry-constrained contributions on $Q^2\simeq 0$, but dropping away more rapidly than the Eq.\,\eqref{psvx2} contributions as $Q^2$ increases.  Thus, ratios of the type $|{\rm column}~n/{\rm column}~1|$, $n=2,\ldots,6$, formed from the entries in row~3 of Table~\ref{tablegr}, provide upper bounds on the magnitudes of every such $G_5 (Q^2)$-related diagram contribution ratio $\forall Q^2 \geq 0$.

Eq.\,\eqref{pcacp} also explains the success of the pion pole dominance (PPD) \emph{Ansatz} -- Eq.\,\eqref{ppd} -- for $G_P(Q^2)$ on $Q^2\geq 0$.  In fact, using Eq.\,\eqref{gpinn} and the Goldberger-Treiman relation -- Eq.\eqref{gtr}, whose accuracy is displayed in Fig.\,\ref{Figgpinn}B,
Eq.\,\eqref{pcacp} entails
\begin{align}
G_P & (Q^2 ) =
\frac{4m_N^2}{Q^2} G_A (Q^2 ) \nonumber \\
& \quad \times \left[ 1 - \frac{m_\pi^2}{Q^2+m_\pi^2}\frac{G_{\pi NN}(Q^2)/G_{\pi NN}(0)}{G_{A}(Q^2)/G_{A}(0)} \right] .\label{PPDappendix}
\end{align}
These identities are to be compared with the PPD assumption, Eq.\,\eqref{ppd}:
\begin{subequations}
\begin{align}
G_P^\pi (Q^2 ) & =\frac{4m_N^2}{Q^2+m_\pi^2} G_A (Q^2 ) \\
&= \frac{4m_N^2}{Q^2}  G_A (Q^2 ) \left[ 1 - \frac{m_\pi^2}{Q^2+m_\pi^2} \right].
\label{GP2}
\end{align}
\end{subequations}

Forming the ratio of these two expressions, one finds
\begin{subequations}
\begin{align}
\delta_{G_P}(Q^2) & := \frac{G_P(Q^2)}{G_P^\pi (Q^2 )} - 1  \\
& = \frac{m_\pi^2}{Q^2} \left[
1 - \frac{G_{\pi NN}(Q^2)/G_{\pi NN}(0)}{G_{A}(Q^2)/G_{A}(0)}
\right].
\end{align}
\end{subequations}
Now,
\begin{equation}
\delta_{G_P}(Q^2\simeq 0) =  \tfrac{1}{6} m_\pi^2 \left[ r_{\pi NN}^2 - r_A^2  \right],
\end{equation}
where $r_{\pi NN}$, $r_A$, are the associated form factor radii.  Inserting typical values for these quantities, then
\begin{equation}
|\delta_{G_P}(Q^2\simeq 0) | < 0.01\,.
\end{equation}
Furthermore, using the fact that both $G_A (Q^2)$ and $G_{\pi NN} (Q^2)$ possess the same power-law behaviour at large $Q^2$, and approximating each by a dipole function, Eqs.\,\eqref{DipoleGA}, \eqref{DipolepiNN}, then
\begin{equation}
\delta_{G_P}(Q^2) \stackrel{m_\pi^2/Q^2 \simeq 0}{=}
\frac{m_\pi^2}{Q^2} \left[1- \frac{\Lambda_{\pi NN}^4}{M_A^4} \right].
\end{equation}
Hence, the ratio very rapidly approaches unity as $Q^2$ increases.  Additional details are presented in connection with Eq.\,\eqref{ppdr}.

Collecting these remarks, one arrives at the following observations.
(${\mathpzc a}$)
$G_A (Q^2 )$ is regular on $Q^2 \geq -m_\pi^2$ and receives no contributions from any term of the form $Q_\mu/[Q^2+m_\pi^2]$ in general and using our current construction.
($\mathpzc b$) The product
$Q^2 G_P(Q^2)$ is directly proportional to $G_A (Q^2 )$ in the chiral limit; hence, likewise, receives no pion pole contributions.  At the physical current-quark mass, $Q^2 G_P (Q^2 )$ retains its regular part but receives, in addition, the same pion pole contributions as $G_5 (Q^2 )$: these corrections vanish as $m_\pi^2/[Q^2+m_\pi^2]$.
($\mathpzc c$) 	The pion pole pieces of our current construction are largely lodged in $G_5 (Q^2 )$.  The feed-in to $G_P (Q^2 )$, required by Eq.\,\eqref{pcacp}, is indicated by Table~\ref{tablegr}.  Given that diagram~1 in Fig.\,\ref{figcurrent} is always dominant, then the Table~\ref{tablegr} ratios $|{\rm column}~n/{\rm column}~1|$, $n=2,\ldots,6$, are upper bounds on the magnitude of the given ratio $\forall Q^2>0$.



\begin{thebibliography}{130}%
\makeatletter
\providecommand \@ifxundefined [1]{%
 \@ifx{#1\undefined}
}%
\providecommand \@ifnum [1]{%
 \ifnum #1\expandafter \@firstoftwo
 \else \expandafter \@secondoftwo
 \fi
}%
\providecommand \@ifx [1]{%
 \ifx #1\expandafter \@firstoftwo
 \else \expandafter \@secondoftwo
 \fi
}%
\providecommand \natexlab [1]{#1}%
\providecommand \enquote  [1]{``#1''}%
\providecommand \bibnamefont  [1]{#1}%
\providecommand \bibfnamefont [1]{#1}%
\providecommand \citenamefont [1]{#1}%
\providecommand \href@noop [0]{\@secondoftwo}%
\providecommand \href [0]{\begingroup \@sanitize@url \@href}%
\providecommand \@href[1]{\@@startlink{#1}\@@href}%
\providecommand \@@href[1]{\endgroup#1\@@endlink}%
\providecommand \@sanitize@url [0]{\catcode `\\12\catcode `\$12\catcode
  `\&12\catcode `\#12\catcode `\^12\catcode `\_12\catcode `\%12\relax}%
\providecommand \@@startlink[1]{}%
\providecommand \@@endlink[0]{}%
\providecommand \url  [0]{\begingroup\@sanitize@url \@url }%
\providecommand \@url [1]{\endgroup\@href {#1}{\urlprefix }}%
\providecommand \urlprefix  [0]{URL }%
\providecommand \Eprint [0]{\href }%
\providecommand \doibase [0]{https://doi.org/}%
\providecommand \selectlanguage [0]{\@gobble}%
\providecommand \bibinfo  [0]{\@secondoftwo}%
\providecommand \bibfield  [0]{\@secondoftwo}%
\providecommand \translation [1]{[#1]}%
\providecommand \BibitemOpen [0]{}%
\providecommand \bibitemStop [0]{}%
\providecommand \bibitemNoStop [0]{.\EOS\space}%
\providecommand \EOS [0]{\spacefactor3000\relax}%
\providecommand \BibitemShut  [1]{\csname bibitem#1\endcsname}%
\let\auto@bib@innerbib\@empty
\bibitem [{\citenamefont {Hofstadter}\ and\ \citenamefont
  {McAllister}(1955)}]{Hofstadter:1955ae}%
  \BibitemOpen
  \bibfield  {author} {\bibinfo {author} {\bibfnamefont {R.}~\bibnamefont
  {Hofstadter}}\ and\ \bibinfo {author} {\bibfnamefont {R.~W.}\ \bibnamefont
  {McAllister}},\ }\href@noop {} {\bibfield  {journal} {\bibinfo  {journal}
  {Phys. Rev.}\ }\textbf {\bibinfo {volume} {98}},\ \bibinfo {pages} {217}
  (\bibinfo {year} {1955})}\BibitemShut {NoStop}%
\bibitem [{\citenamefont {Jones}\ \emph {et~al.}(2000)\citenamefont {Jones}
  \emph {et~al.}}]{Jones:1999rz}%
  \BibitemOpen
  \bibfield  {author} {\bibinfo {author} {\bibfnamefont {M.~K.}\ \bibnamefont
  {Jones}} \emph {et~al.},\ }\href@noop {} {\bibfield  {journal} {\bibinfo
  {journal} {Phys. Rev. Lett.}\ }\textbf {\bibinfo {volume} {84}},\ \bibinfo
  {pages} {1398} (\bibinfo {year} {2000})}\BibitemShut {NoStop}%
\bibitem [{\citenamefont {Gayou}\ \emph {et~al.}(2002)\citenamefont {Gayou}
  \emph {et~al.}}]{Gayou:2001qd}%
  \BibitemOpen
  \bibfield  {author} {\bibinfo {author} {\bibfnamefont {O.}~\bibnamefont
  {Gayou}} \emph {et~al.},\ }\href@noop {} {\bibfield  {journal} {\bibinfo
  {journal} {Phys. Rev. Lett.}\ }\textbf {\bibinfo {volume} {88}},\ \bibinfo
  {pages} {092301} (\bibinfo {year} {2002})}\BibitemShut {NoStop}%
\bibitem [{\citenamefont {Punjabi}\ \emph {et~al.}(2005)\citenamefont {Punjabi}
  \emph {et~al.}}]{Punjabi:2005wq}%
  \BibitemOpen
  \bibfield  {author} {\bibinfo {author} {\bibfnamefont {V.}~\bibnamefont
  {Punjabi}} \emph {et~al.},\ }\href@noop {} {\bibfield  {journal} {\bibinfo
  {journal} {Phys. Rev. C}\ }\textbf {\bibinfo {volume} {71}},\ \bibinfo
  {pages} {055202} (\bibinfo {year} {2005})},\ \bibinfo {note} {[Erratum-ibid.
  C\,\textbf{71}, 069902 (2005)]}\BibitemShut {NoStop}%
\bibitem [{\citenamefont {Puckett}\ \emph {et~al.}(2010)\citenamefont {Puckett}
  \emph {et~al.}}]{Puckett:2010ac}%
  \BibitemOpen
  \bibfield  {author} {\bibinfo {author} {\bibfnamefont {A.~J.~R.}\
  \bibnamefont {Puckett}} \emph {et~al.},\ }\href@noop {} {\bibfield  {journal}
  {\bibinfo  {journal} {Phys. Rev. Lett.}\ }\textbf {\bibinfo {volume} {104}},\
  \bibinfo {pages} {242301} (\bibinfo {year} {2010})}\BibitemShut {NoStop}%
\bibitem [{\citenamefont {Puckett}\ \emph {et~al.}(2012)\citenamefont
  {Puckett}, \citenamefont {Brash}, \citenamefont {Gayou}, \citenamefont
  {Jones}, \citenamefont {Pentchev} \emph {et~al.}}]{Puckett:2011xg}%
  \BibitemOpen
  \bibfield  {author} {\bibinfo {author} {\bibfnamefont {A.~J.~R.}\
  \bibnamefont {Puckett}}, \bibinfo {author} {\bibfnamefont {E.~J.}\
  \bibnamefont {Brash}}, \bibinfo {author} {\bibfnamefont {O.}~\bibnamefont
  {Gayou}}, \bibinfo {author} {\bibfnamefont {M.~K.}\ \bibnamefont {Jones}},
  \bibinfo {author} {\bibfnamefont {L.}~\bibnamefont {Pentchev}}, \emph
  {et~al.},\ }\href@noop {} {\bibfield  {journal} {\bibinfo  {journal} {Phys.
  Rev. C}\ }\textbf {\bibinfo {volume} {85}},\ \bibinfo {pages} {045203}
  (\bibinfo {year} {2012})}\BibitemShut {NoStop}%
\bibitem [{\citenamefont {Cates}\ \emph {et~al.}(2011)\citenamefont {Cates},
  \citenamefont {de~Jager}, \citenamefont {Riordan},\ and\ \citenamefont
  {Wojtsekhowski}}]{Cates:2011pz}%
  \BibitemOpen
  \bibfield  {author} {\bibinfo {author} {\bibfnamefont {G.}~\bibnamefont
  {Cates}}, \bibinfo {author} {\bibfnamefont {C.}~\bibnamefont {de~Jager}},
  \bibinfo {author} {\bibfnamefont {S.}~\bibnamefont {Riordan}},\ and\ \bibinfo
  {author} {\bibfnamefont {B.}~\bibnamefont {Wojtsekhowski}},\ }\href@noop {}
  {\bibfield  {journal} {\bibinfo  {journal} {Phys. Rev. Lett.}\ }\textbf
  {\bibinfo {volume} {106}},\ \bibinfo {pages} {252003} (\bibinfo {year}
  {2011})}\BibitemShut {NoStop}%
\bibitem [{\citenamefont {Puckett}\ \emph {et~al.}(2017)\citenamefont {Puckett}
  \emph {et~al.}}]{Puckett:2017flj}%
  \BibitemOpen
  \bibfield  {author} {\bibinfo {author} {\bibfnamefont {A.~J.~R.}\
  \bibnamefont {Puckett}} \emph {et~al.},\ }\href@noop {} {\bibfield  {journal}
  {\bibinfo  {journal} {Phys. Rev. C}\ }\textbf {\bibinfo {volume} {96}},\
  \bibinfo {pages} {055203} (\bibinfo {year} {2017})},\ \bibinfo {note}
  {[erratum: Phys. Rev. C \textbf{98}, 019907 (2018)]}\BibitemShut {NoStop}%
\bibitem [{\citenamefont {Ahrens}\ \emph {et~al.}(1988)\citenamefont {Ahrens}
  \emph {et~al.}}]{Ahrens:1988rr}%
  \BibitemOpen
  \bibfield  {author} {\bibinfo {author} {\bibfnamefont {L.~A.}\ \bibnamefont
  {Ahrens}} \emph {et~al.},\ }\href@noop {} {\bibfield  {journal} {\bibinfo
  {journal} {Phys. Lett. B}\ }\textbf {\bibinfo {volume} {202}},\ \bibinfo
  {pages} {284} (\bibinfo {year} {1988})}\BibitemShut {NoStop}%
\bibitem [{\citenamefont {Bodek}\ \emph {et~al.}(2008)\citenamefont {Bodek},
  \citenamefont {Avvakumov}, \citenamefont {Bradford},\ and\ \citenamefont
  {Budd}}]{Bodek:2007vi}%
  \BibitemOpen
  \bibfield  {author} {\bibinfo {author} {\bibfnamefont {A.}~\bibnamefont
  {Bodek}}, \bibinfo {author} {\bibfnamefont {S.}~\bibnamefont {Avvakumov}},
  \bibinfo {author} {\bibfnamefont {R.}~\bibnamefont {Bradford}},\ and\
  \bibinfo {author} {\bibfnamefont {H.~S.}\ \bibnamefont {Budd}},\ }\href@noop
  {} {\bibfield  {journal} {\bibinfo  {journal} {J. Phys. Conf. Ser.}\ }\textbf
  {\bibinfo {volume} {110}},\ \bibinfo {pages} {082004} (\bibinfo {year}
  {2008})}\BibitemShut {NoStop}%
\bibitem [{\citenamefont {Mosel}(2016)}]{Mosel:2016cwa}%
  \BibitemOpen
  \bibfield  {author} {\bibinfo {author} {\bibfnamefont {U.}~\bibnamefont
  {Mosel}},\ }\href@noop {} {\bibfield  {journal} {\bibinfo  {journal} {Ann.
  Rev. Nucl. Part. Sci.}\ }\textbf {\bibinfo {volume} {66}},\ \bibinfo {pages}
  {171} (\bibinfo {year} {2016})}\BibitemShut {NoStop}%
\bibitem [{\citenamefont {Meyer}\ \emph {et~al.}(2016)\citenamefont {Meyer},
  \citenamefont {Betancourt}, \citenamefont {Gran},\ and\ \citenamefont
  {Hill}}]{Meyer:2016oeg}%
  \BibitemOpen
  \bibfield  {author} {\bibinfo {author} {\bibfnamefont {A.~S.}\ \bibnamefont
  {Meyer}}, \bibinfo {author} {\bibfnamefont {M.}~\bibnamefont {Betancourt}},
  \bibinfo {author} {\bibfnamefont {R.}~\bibnamefont {Gran}},\ and\ \bibinfo
  {author} {\bibfnamefont {R.~J.}\ \bibnamefont {Hill}},\ }\href@noop {}
  {\bibfield  {journal} {\bibinfo  {journal} {Phys. Rev. D}\ }\textbf {\bibinfo
  {volume} {93}},\ \bibinfo {pages} {113015} (\bibinfo {year}
  {2016})}\BibitemShut {NoStop}%
\bibitem [{\citenamefont {Choi}\ \emph {et~al.}(1993)\citenamefont {Choi} \emph
  {et~al.}}]{Choi:1993vt}%
  \BibitemOpen
  \bibfield  {author} {\bibinfo {author} {\bibfnamefont {S.}~\bibnamefont
  {Choi}} \emph {et~al.},\ }\href@noop {} {\bibfield  {journal} {\bibinfo
  {journal} {Phys. Rev. Lett.}\ }\textbf {\bibinfo {volume} {71}},\ \bibinfo
  {pages} {3927} (\bibinfo {year} {1993})}\BibitemShut {NoStop}%
\bibitem [{\citenamefont {Bernard}\ \emph {et~al.}(1994)\citenamefont
  {Bernard}, \citenamefont {Meissner},\ and\ \citenamefont
  {Kaiser}}]{Bernard:1994pk}%
  \BibitemOpen
  \bibfield  {author} {\bibinfo {author} {\bibfnamefont {V.}~\bibnamefont
  {Bernard}}, \bibinfo {author} {\bibfnamefont {U.~G.}\ \bibnamefont
  {Meissner}},\ and\ \bibinfo {author} {\bibfnamefont {N.}~\bibnamefont
  {Kaiser}},\ }\href@noop {} {\bibfield  {journal} {\bibinfo  {journal} {Phys.
  Rev. Lett.}\ }\textbf {\bibinfo {volume} {72}},\ \bibinfo {pages} {2810}
  (\bibinfo {year} {1994})}\BibitemShut {NoStop}%
\bibitem [{\citenamefont {Fuchs}\ and\ \citenamefont
  {Scherer}(2003)}]{Fuchs:2003vw}%
  \BibitemOpen
  \bibfield  {author} {\bibinfo {author} {\bibfnamefont {T.}~\bibnamefont
  {Fuchs}}\ and\ \bibinfo {author} {\bibfnamefont {S.}~\bibnamefont
  {Scherer}},\ }\href@noop {} {\bibfield  {journal} {\bibinfo  {journal} {Phys.
  Rev. C}\ }\textbf {\bibinfo {volume} {68}},\ \bibinfo {pages} {055501}
  (\bibinfo {year} {2003})}\BibitemShut {NoStop}%
\bibitem [{\citenamefont {Tomalak}(2021)}]{Tomalak:2020zlv}%
  \BibitemOpen
  \bibfield  {author} {\bibinfo {author} {\bibfnamefont {O.}~\bibnamefont
  {Tomalak}},\ }\href@noop {} {\bibfield  {journal} {\bibinfo  {journal} {Phys.
  Rev. D}\ }\textbf {\bibinfo {volume} {103}},\ \bibinfo {pages} {013006}
  (\bibinfo {year} {2021})}\BibitemShut {NoStop}%
\bibitem [{\citenamefont {Mendenhall}\ \emph {et~al.}(2013)\citenamefont
  {Mendenhall} \emph {et~al.}}]{Mendenhall:2012tz}%
  \BibitemOpen
  \bibfield  {author} {\bibinfo {author} {\bibfnamefont {M.~P.}\ \bibnamefont
  {Mendenhall}} \emph {et~al.},\ }\href@noop {} {\bibfield  {journal} {\bibinfo
   {journal} {Phys. Rev. C}\ }\textbf {\bibinfo {volume} {87}},\ \bibinfo
  {pages} {032501} (\bibinfo {year} {2013})}\BibitemShut {NoStop}%
\bibitem [{\citenamefont {Mund}\ \emph {et~al.}(2013)\citenamefont {Mund},
  \citenamefont {Maerkisch}, \citenamefont {Deissenroth}, \citenamefont
  {Krempel}, \citenamefont {Schumann}, \citenamefont {Abele}, \citenamefont
  {Petoukhov},\ and\ \citenamefont {Soldner}}]{Mund:2012fq}%
  \BibitemOpen
  \bibfield  {author} {\bibinfo {author} {\bibfnamefont {D.}~\bibnamefont
  {Mund}}, \bibinfo {author} {\bibfnamefont {B.}~\bibnamefont {Maerkisch}},
  \bibinfo {author} {\bibfnamefont {M.}~\bibnamefont {Deissenroth}}, \bibinfo
  {author} {\bibfnamefont {J.}~\bibnamefont {Krempel}}, \bibinfo {author}
  {\bibfnamefont {M.}~\bibnamefont {Schumann}}, \bibinfo {author}
  {\bibfnamefont {H.}~\bibnamefont {Abele}}, \bibinfo {author} {\bibfnamefont
  {A.}~\bibnamefont {Petoukhov}},\ and\ \bibinfo {author} {\bibfnamefont
  {T.}~\bibnamefont {Soldner}},\ }\href@noop {} {\bibfield  {journal} {\bibinfo
   {journal} {Phys. Rev. Lett.}\ }\textbf {\bibinfo {volume} {110}},\ \bibinfo
  {pages} {172502} (\bibinfo {year} {2013})}\BibitemShut {NoStop}%
\bibitem [{\citenamefont {Brown}\ \emph {et~al.}(2018)\citenamefont {Brown}
  \emph {et~al.}}]{Brown:2017mhw}%
  \BibitemOpen
  \bibfield  {author} {\bibinfo {author} {\bibfnamefont {M.~A.~P.}\
  \bibnamefont {Brown}} \emph {et~al.},\ }\href@noop {} {\bibfield  {journal}
  {\bibinfo  {journal} {Phys. Rev. C}\ }\textbf {\bibinfo {volume} {97}},\
  \bibinfo {pages} {035505} (\bibinfo {year} {2018})}\BibitemShut {NoStop}%
\bibitem [{\citenamefont {Darius}\ \emph {et~al.}(2017)\citenamefont {Darius}
  \emph {et~al.}}]{Darius:2017arh}%
  \BibitemOpen
  \bibfield  {author} {\bibinfo {author} {\bibfnamefont {G.}~\bibnamefont
  {Darius}} \emph {et~al.},\ }\href@noop {} {\bibfield  {journal} {\bibinfo
  {journal} {Phys. Rev. Lett.}\ }\textbf {\bibinfo {volume} {119}},\ \bibinfo
  {pages} {042502} (\bibinfo {year} {2017})}\BibitemShut {NoStop}%
\bibitem [{\citenamefont {Castro}\ and\ \citenamefont
  {Dominguez}(1977)}]{Castro:1977ep}%
  \BibitemOpen
  \bibfield  {author} {\bibinfo {author} {\bibfnamefont {J.~J.}\ \bibnamefont
  {Castro}}\ and\ \bibinfo {author} {\bibfnamefont {C.~A.}\ \bibnamefont
  {Dominguez}},\ }\href@noop {} {\bibfield  {journal} {\bibinfo  {journal}
  {Phys. Rev. Lett.}\ }\textbf {\bibinfo {volume} {39}},\ \bibinfo {pages}
  {440} (\bibinfo {year} {1977})}\BibitemShut {NoStop}%
\bibitem [{\citenamefont {Bernard}\ \emph {et~al.}(2001)\citenamefont
  {Bernard}, \citenamefont {Hemmert},\ and\ \citenamefont
  {Meissner}}]{Bernard:2000et}%
  \BibitemOpen
  \bibfield  {author} {\bibinfo {author} {\bibfnamefont {V.}~\bibnamefont
  {Bernard}}, \bibinfo {author} {\bibfnamefont {T.~R.}\ \bibnamefont
  {Hemmert}},\ and\ \bibinfo {author} {\bibfnamefont {U.-G.}\ \bibnamefont
  {Meissner}},\ }\href@noop {} {\bibfield  {journal} {\bibinfo  {journal}
  {Nucl. Phys. A}\ }\textbf {\bibinfo {volume} {686}},\ \bibinfo {pages} {290}
  (\bibinfo {year} {2001})}\BibitemShut {NoStop}%
\bibitem [{\citenamefont {Andreev}\ \emph {et~al.}(2007)\citenamefont {Andreev}
  \emph {et~al.}}]{Andreev:2007wg}%
  \BibitemOpen
  \bibfield  {author} {\bibinfo {author} {\bibfnamefont {V.~A.}\ \bibnamefont
  {Andreev}} \emph {et~al.},\ }\href@noop {} {\bibfield  {journal} {\bibinfo
  {journal} {Phys. Rev. Lett.}\ }\textbf {\bibinfo {volume} {99}},\ \bibinfo
  {pages} {032002} (\bibinfo {year} {2007})}\BibitemShut {NoStop}%
\bibitem [{\citenamefont {Andreev}\ \emph {et~al.}(2013)\citenamefont {Andreev}
  \emph {et~al.}}]{Andreev:2012fj}%
  \BibitemOpen
  \bibfield  {author} {\bibinfo {author} {\bibfnamefont {V.}~\bibnamefont
  {Andreev}} \emph {et~al.},\ }\href@noop {} {\bibfield  {journal} {\bibinfo
  {journal} {Phys. Rev. Lett.}\ }\textbf {\bibinfo {volume} {110}},\ \bibinfo
  {pages} {012504} (\bibinfo {year} {2013})}\BibitemShut {NoStop}%
\bibitem [{\citenamefont {Andreev}\ \emph {et~al.}(2015)\citenamefont {Andreev}
  \emph {et~al.}}]{Andreev:2015evt}%
  \BibitemOpen
  \bibfield  {author} {\bibinfo {author} {\bibfnamefont {V.~A.}\ \bibnamefont
  {Andreev}} \emph {et~al.},\ }\href@noop {} {\bibfield  {journal} {\bibinfo
  {journal} {Phys. Rev. C}\ }\textbf {\bibinfo {volume} {91}},\ \bibinfo
  {pages} {055502} (\bibinfo {year} {2015})}\BibitemShut {NoStop}%
\bibitem [{\citenamefont {Capitani}\ \emph {et~al.}(2019)\citenamefont
  {Capitani}, \citenamefont {Della~Morte}, \citenamefont {Djukanovic},
  \citenamefont {von Hippel}, \citenamefont {Hua}, \citenamefont {J\"ager},
  \citenamefont {Junnarkar}, \citenamefont {Meyer}, \citenamefont {Rae},\ and\
  \citenamefont {Wittig}}]{Capitani:2017qpc}%
  \BibitemOpen
  \bibfield  {author} {\bibinfo {author} {\bibfnamefont {S.}~\bibnamefont
  {Capitani}}, \bibinfo {author} {\bibfnamefont {M.}~\bibnamefont
  {Della~Morte}}, \bibinfo {author} {\bibfnamefont {D.}~\bibnamefont
  {Djukanovic}}, \bibinfo {author} {\bibfnamefont {G.~M.}\ \bibnamefont {von
  Hippel}}, \bibinfo {author} {\bibfnamefont {J.}~\bibnamefont {Hua}}, \bibinfo
  {author} {\bibfnamefont {B.}~\bibnamefont {J\"ager}}, \bibinfo {author}
  {\bibfnamefont {P.~M.}\ \bibnamefont {Junnarkar}}, \bibinfo {author}
  {\bibfnamefont {H.~B.}\ \bibnamefont {Meyer}}, \bibinfo {author}
  {\bibfnamefont {T.~D.}\ \bibnamefont {Rae}},\ and\ \bibinfo {author}
  {\bibfnamefont {H.}~\bibnamefont {Wittig}},\ }\href@noop {} {\bibfield
  {journal} {\bibinfo  {journal} {Int. J. Mod. Phys. A}\ }\textbf {\bibinfo
  {volume} {34}},\ \bibinfo {pages} {1950009} (\bibinfo {year}
  {2019})}\BibitemShut {NoStop}%
\bibitem [{\citenamefont {Gupta}\ \emph {et~al.}(2017)\citenamefont {Gupta},
  \citenamefont {Jang}, \citenamefont {Lin}, \citenamefont {Yoon},\ and\
  \citenamefont {Bhattacharya}}]{Rajan:2017lxk}%
  \BibitemOpen
  \bibfield  {author} {\bibinfo {author} {\bibfnamefont {R.}~\bibnamefont
  {Gupta}}, \bibinfo {author} {\bibfnamefont {Y.-C.}\ \bibnamefont {Jang}},
  \bibinfo {author} {\bibfnamefont {H.-W.}\ \bibnamefont {Lin}}, \bibinfo
  {author} {\bibfnamefont {B.}~\bibnamefont {Yoon}},\ and\ \bibinfo {author}
  {\bibfnamefont {T.}~\bibnamefont {Bhattacharya}},\ }\href@noop {} {\bibfield
  {journal} {\bibinfo  {journal} {Phys. Rev. D}\ }\textbf {\bibinfo {volume}
  {96}},\ \bibinfo {pages} {114503} (\bibinfo {year} {2017})}\BibitemShut
  {NoStop}%
\bibitem [{\citenamefont {Alexandrou}\ \emph {et~al.}(2017)\citenamefont
  {Alexandrou}, \citenamefont {Constantinou}, \citenamefont {Hadjiyiannakou},
  \citenamefont {Jansen}, \citenamefont {Kallidonis}, \citenamefont {Koutsou},\
  and\ \citenamefont {Vaquero Aviles-Casco}}]{Alexandrou:2017hac}%
  \BibitemOpen
  \bibfield  {author} {\bibinfo {author} {\bibfnamefont {C.}~\bibnamefont
  {Alexandrou}}, \bibinfo {author} {\bibfnamefont {M.}~\bibnamefont
  {Constantinou}}, \bibinfo {author} {\bibfnamefont {K.}~\bibnamefont
  {Hadjiyiannakou}}, \bibinfo {author} {\bibfnamefont {K.}~\bibnamefont
  {Jansen}}, \bibinfo {author} {\bibfnamefont {C.}~\bibnamefont {Kallidonis}},
  \bibinfo {author} {\bibfnamefont {G.}~\bibnamefont {Koutsou}},\ and\ \bibinfo
  {author} {\bibfnamefont {A.}~\bibnamefont {Vaquero Aviles-Casco}},\
  }\href@noop {} {\bibfield  {journal} {\bibinfo  {journal} {Phys. Rev. D}\
  }\textbf {\bibinfo {volume} {96}},\ \bibinfo {pages} {054507} (\bibinfo
  {year} {2017})}\BibitemShut {NoStop}%
\bibitem [{\citenamefont {Shintani}\ \emph {et~al.}(2019)\citenamefont
  {Shintani}, \citenamefont {Ishikawa}, \citenamefont {Kuramashi},
  \citenamefont {Sasaki},\ and\ \citenamefont {Yamazaki}}]{Shintani:2018ozy}%
  \BibitemOpen
  \bibfield  {author} {\bibinfo {author} {\bibfnamefont {E.}~\bibnamefont
  {Shintani}}, \bibinfo {author} {\bibfnamefont {K.-I.}\ \bibnamefont
  {Ishikawa}}, \bibinfo {author} {\bibfnamefont {Y.}~\bibnamefont {Kuramashi}},
  \bibinfo {author} {\bibfnamefont {S.}~\bibnamefont {Sasaki}},\ and\ \bibinfo
  {author} {\bibfnamefont {T.}~\bibnamefont {Yamazaki}},\ }\href@noop {}
  {\bibfield  {journal} {\bibinfo  {journal} {Phys. Rev. D}\ }\textbf {\bibinfo
  {volume} {99}},\ \bibinfo {pages} {014510} (\bibinfo {year} {2019})},\
  \bibinfo {note} {[Erratum: Phys. Rev. D \textbf{102}, 019902
  (2020)]}\BibitemShut {NoStop}%
\bibitem [{\citenamefont {Bali}\ \emph {et~al.}(2019)\citenamefont {Bali},
  \citenamefont {Collins}, \citenamefont {Gruber}, \citenamefont {Sch\"afer},
  \citenamefont {Wein},\ and\ \citenamefont {Wurm}}]{Bali:2018qus}%
  \BibitemOpen
  \bibfield  {author} {\bibinfo {author} {\bibfnamefont {G.~S.}\ \bibnamefont
  {Bali}}, \bibinfo {author} {\bibfnamefont {S.}~\bibnamefont {Collins}},
  \bibinfo {author} {\bibfnamefont {M.}~\bibnamefont {Gruber}}, \bibinfo
  {author} {\bibfnamefont {A.}~\bibnamefont {Sch\"afer}}, \bibinfo {author}
  {\bibfnamefont {P.}~\bibnamefont {Wein}},\ and\ \bibinfo {author}
  {\bibfnamefont {T.}~\bibnamefont {Wurm}},\ }\href@noop {} {\bibfield
  {journal} {\bibinfo  {journal} {Phys. Lett. B}\ }\textbf {\bibinfo {volume}
  {789}},\ \bibinfo {pages} {666} (\bibinfo {year} {2019})}\BibitemShut
  {NoStop}%
\bibitem [{\citenamefont {Ishikawa}\ \emph {et~al.}(2018)\citenamefont
  {Ishikawa}, \citenamefont {Kuramashi}, \citenamefont {Sasaki}, \citenamefont
  {Tsukamoto}, \citenamefont {Ukawa},\ and\ \citenamefont
  {Yamazaki}}]{Ishikawa:2018rew}%
  \BibitemOpen
  \bibfield  {author} {\bibinfo {author} {\bibfnamefont {K.-I.}\ \bibnamefont
  {Ishikawa}}, \bibinfo {author} {\bibfnamefont {Y.}~\bibnamefont {Kuramashi}},
  \bibinfo {author} {\bibfnamefont {S.}~\bibnamefont {Sasaki}}, \bibinfo
  {author} {\bibfnamefont {N.}~\bibnamefont {Tsukamoto}}, \bibinfo {author}
  {\bibfnamefont {A.}~\bibnamefont {Ukawa}},\ and\ \bibinfo {author}
  {\bibfnamefont {T.}~\bibnamefont {Yamazaki}},\ }\href@noop {} {\bibfield
  {journal} {\bibinfo  {journal} {Phys. Rev. D}\ }\textbf {\bibinfo {volume}
  {98}},\ \bibinfo {pages} {074510} (\bibinfo {year} {2018})}\BibitemShut
  {NoStop}%
\bibitem [{\citenamefont {Jang}\ \emph
  {et~al.}(2020{\natexlab{a}})\citenamefont {Jang}, \citenamefont {Gupta},
  \citenamefont {Yoon},\ and\ \citenamefont {Bhattacharya}}]{Jang:2019vkm}%
  \BibitemOpen
  \bibfield  {author} {\bibinfo {author} {\bibfnamefont {Y.-C.}\ \bibnamefont
  {Jang}}, \bibinfo {author} {\bibfnamefont {R.}~\bibnamefont {Gupta}},
  \bibinfo {author} {\bibfnamefont {B.}~\bibnamefont {Yoon}},\ and\ \bibinfo
  {author} {\bibfnamefont {T.}~\bibnamefont {Bhattacharya}},\ }\href@noop {}
  {\bibfield  {journal} {\bibinfo  {journal} {Phys. Rev. Lett.}\ }\textbf
  {\bibinfo {volume} {124}},\ \bibinfo {pages} {072002} (\bibinfo {year}
  {2020}{\natexlab{a}})}\BibitemShut {NoStop}%
\bibitem [{\citenamefont {Bali}\ \emph {et~al.}(2020)\citenamefont {Bali},
  \citenamefont {Barca}, \citenamefont {Collins}, \citenamefont {Gruber},
  \citenamefont {L{\"o}ffler}, \citenamefont {Sch{\"a}fer}, \citenamefont
  {S{\"o}ldner}, \citenamefont {Wein}, \citenamefont {Weish{\"a}upl},\ and\
  \citenamefont {Wurm}}]{Bali:2019yiy}%
  \BibitemOpen
  \bibfield  {author} {\bibinfo {author} {\bibfnamefont {G.~S.}\ \bibnamefont
  {Bali}}, \bibinfo {author} {\bibfnamefont {L.}~\bibnamefont {Barca}},
  \bibinfo {author} {\bibfnamefont {S.}~\bibnamefont {Collins}}, \bibinfo
  {author} {\bibfnamefont {M.}~\bibnamefont {Gruber}}, \bibinfo {author}
  {\bibfnamefont {M.}~\bibnamefont {L{\"o}ffler}}, \bibinfo {author}
  {\bibfnamefont {A.}~\bibnamefont {Sch{\"a}fer}}, \bibinfo {author}
  {\bibfnamefont {W.}~\bibnamefont {S{\"o}ldner}}, \bibinfo {author}
  {\bibfnamefont {P.}~\bibnamefont {Wein}}, \bibinfo {author} {\bibfnamefont
  {S.}~\bibnamefont {Weish{\"a}upl}},\ and\ \bibinfo {author} {\bibfnamefont
  {T.}~\bibnamefont {Wurm}},\ }\href@noop {} {\bibfield  {journal} {\bibinfo
  {journal} {JHEP}\ }\textbf {\bibinfo {volume} {05}},\ \bibinfo {pages} {126
  (2020)}}\BibitemShut {NoStop}%
\bibitem [{\citenamefont {Alexandrou}\ \emph
  {et~al.}(2020{\natexlab{a}})\citenamefont {Alexandrou} \emph
  {et~al.}}]{Alexandrou:2020okk}%
  \BibitemOpen
  \bibfield  {author} {\bibinfo {author} {\bibfnamefont {C.}~\bibnamefont
  {Alexandrou}} \emph {et~al.},\ }\href@noop {} {\  (\bibinfo {year}
  {2020}{\natexlab{a}})},\ \bibinfo {note} {{\emph{Nucleon axial and
  pseudoscalar form factors from lattice QCD at the physical point} --
  arXiv:2011.13342 [hep-lat]}}\BibitemShut {NoStop}%
\bibitem [{\citenamefont {Capitani}\ \emph {et~al.}(2010)\citenamefont
  {Capitani}, \citenamefont {Knippschild}, \citenamefont {Della~Morte},\ and\
  \citenamefont {Wittig}}]{Capitani:2010sg}%
  \BibitemOpen
  \bibfield  {author} {\bibinfo {author} {\bibfnamefont {S.}~\bibnamefont
  {Capitani}}, \bibinfo {author} {\bibfnamefont {B.}~\bibnamefont
  {Knippschild}}, \bibinfo {author} {\bibfnamefont {M.}~\bibnamefont
  {Della~Morte}},\ and\ \bibinfo {author} {\bibfnamefont {H.}~\bibnamefont
  {Wittig}},\ }\href@noop {} {\bibfield  {journal} {\bibinfo  {journal} {PoS}\
  }\textbf {\bibinfo {volume} {LATTICE2010}},\ \bibinfo {pages} {147} (\bibinfo
  {year} {2010})}\BibitemShut {NoStop}%
\bibitem [{\citenamefont {Bernard}\ \emph {et~al.}(2002)\citenamefont
  {Bernard}, \citenamefont {Elouadrhiri},\ and\ \citenamefont
  {Meissner}}]{Bernard:2001rs}%
  \BibitemOpen
  \bibfield  {author} {\bibinfo {author} {\bibfnamefont {V.}~\bibnamefont
  {Bernard}}, \bibinfo {author} {\bibfnamefont {L.}~\bibnamefont
  {Elouadrhiri}},\ and\ \bibinfo {author} {\bibfnamefont {U.-G.}\ \bibnamefont
  {Meissner}},\ }\href@noop {} {\bibfield  {journal} {\bibinfo  {journal} {J.
  Phys. G}\ }\textbf {\bibinfo {volume} {28}},\ \bibinfo {pages} {R1} (\bibinfo
  {year} {2002})}\BibitemShut {NoStop}%
\bibitem [{\citenamefont {Bar}(2019)}]{Bar:2018xyi}%
  \BibitemOpen
  \bibfield  {author} {\bibinfo {author} {\bibfnamefont {O.}~\bibnamefont
  {Bar}},\ }\href@noop {} {\bibfield  {journal} {\bibinfo  {journal} {Phys.
  Rev. D}\ }\textbf {\bibinfo {volume} {99}},\ \bibinfo {pages} {054506}
  (\bibinfo {year} {2019})}\BibitemShut {NoStop}%
\bibitem [{\citenamefont {Bar}(2020)}]{Bar:2019igf}%
  \BibitemOpen
  \bibfield  {author} {\bibinfo {author} {\bibfnamefont {O.}~\bibnamefont
  {Bar}},\ }\href@noop {} {\bibfield  {journal} {\bibinfo  {journal} {Phys.
  Rev. D}\ }\textbf {\bibinfo {volume} {101}},\ \bibinfo {pages} {034515}
  (\bibinfo {year} {2020})}\BibitemShut {NoStop}%
\bibitem [{\citenamefont {Capitani}\ \emph {et~al.}(2012)\citenamefont
  {Capitani}, \citenamefont {Della~Morte}, \citenamefont {von Hippel},
  \citenamefont {Jager}, \citenamefont {Juttner}, \citenamefont {Knippschild},
  \citenamefont {Meyer},\ and\ \citenamefont {Wittig}}]{Capitani:2012gj}%
  \BibitemOpen
  \bibfield  {author} {\bibinfo {author} {\bibfnamefont {S.}~\bibnamefont
  {Capitani}}, \bibinfo {author} {\bibfnamefont {M.}~\bibnamefont
  {Della~Morte}}, \bibinfo {author} {\bibfnamefont {G.}~\bibnamefont {von
  Hippel}}, \bibinfo {author} {\bibfnamefont {B.}~\bibnamefont {Jager}},
  \bibinfo {author} {\bibfnamefont {A.}~\bibnamefont {Juttner}}, \bibinfo
  {author} {\bibfnamefont {B.}~\bibnamefont {Knippschild}}, \bibinfo {author}
  {\bibfnamefont {H.~B.}\ \bibnamefont {Meyer}},\ and\ \bibinfo {author}
  {\bibfnamefont {H.}~\bibnamefont {Wittig}},\ }\href@noop {} {\bibfield
  {journal} {\bibinfo  {journal} {Phys. Rev. D}\ }\textbf {\bibinfo {volume}
  {86}},\ \bibinfo {pages} {074502} (\bibinfo {year} {2012})}\BibitemShut
  {NoStop}%
\bibitem [{\citenamefont {Jang}\ \emph
  {et~al.}(2020{\natexlab{b}})\citenamefont {Jang}, \citenamefont {Gupta},
  \citenamefont {Bhattacharya}, \citenamefont {Park}, \citenamefont {Yoon},\
  and\ \citenamefont {Lin}}]{Jang:2020ygs}%
  \BibitemOpen
  \bibfield  {author} {\bibinfo {author} {\bibfnamefont {Y.-C.}\ \bibnamefont
  {Jang}}, \bibinfo {author} {\bibfnamefont {R.}~\bibnamefont {Gupta}},
  \bibinfo {author} {\bibfnamefont {T.}~\bibnamefont {Bhattacharya}}, \bibinfo
  {author} {\bibfnamefont {S.}~\bibnamefont {Park}}, \bibinfo {author}
  {\bibfnamefont {B.}~\bibnamefont {Yoon}},\ and\ \bibinfo {author}
  {\bibfnamefont {H.-W.}\ \bibnamefont {Lin}},\ }\href@noop {} {\bibfield
  {journal} {\bibinfo  {journal} {PoS}\ }\textbf {\bibinfo {volume}
  {LATTICE2019}},\ \bibinfo {pages} {131} (\bibinfo {year}
  {2020}{\natexlab{b}})}\BibitemShut {NoStop}%
\bibitem [{\citenamefont {Ottnad}(2021)}]{Ottnad:2020qbw}%
  \BibitemOpen
  \bibfield  {author} {\bibinfo {author} {\bibfnamefont {K.}~\bibnamefont
  {Ottnad}},\ }\href@noop {} {\bibfield  {journal} {\bibinfo  {journal} {Eur.
  Phys. J. A}\ }\textbf {\bibinfo {volume} {57}},\ \bibinfo {pages} {50}
  (\bibinfo {year} {2021})}\BibitemShut {NoStop}%
\bibitem [{\citenamefont {Roberts}\ and\ \citenamefont
  {Schmidt}(2000)}]{Roberts:2000aa}%
  \BibitemOpen
  \bibfield  {author} {\bibinfo {author} {\bibfnamefont {C.~D.}\ \bibnamefont
  {Roberts}}\ and\ \bibinfo {author} {\bibfnamefont {S.~M.}\ \bibnamefont
  {Schmidt}},\ }\href@noop {} {\bibfield  {journal} {\bibinfo  {journal} {Prog.
  Part. Nucl. Phys.}\ }\textbf {\bibinfo {volume} {45}},\ \bibinfo {pages} {S1}
  (\bibinfo {year} {2000})}\BibitemShut {NoStop}%
\bibitem [{\citenamefont {Alkofer}\ and\ \citenamefont {von
  Smekal}(2001)}]{Alkofer:2000wg}%
  \BibitemOpen
  \bibfield  {author} {\bibinfo {author} {\bibfnamefont {R.}~\bibnamefont
  {Alkofer}}\ and\ \bibinfo {author} {\bibfnamefont {L.}~\bibnamefont {von
  Smekal}},\ }\href@noop {} {\bibfield  {journal} {\bibinfo  {journal} {Phys.
  Rept.}\ }\textbf {\bibinfo {volume} {353}},\ \bibinfo {pages} {281} (\bibinfo
  {year} {2001})}\BibitemShut {NoStop}%
\bibitem [{\citenamefont {Maris}\ and\ \citenamefont
  {Roberts}(2003)}]{Maris:2003vk}%
  \BibitemOpen
  \bibfield  {author} {\bibinfo {author} {\bibfnamefont {P.}~\bibnamefont
  {Maris}}\ and\ \bibinfo {author} {\bibfnamefont {C.~D.}\ \bibnamefont
  {Roberts}},\ }\href@noop {} {\bibfield  {journal} {\bibinfo  {journal} {Int.
  J. Mod. Phys. E}\ }\textbf {\bibinfo {volume} {12}},\ \bibinfo {pages} {297}
  (\bibinfo {year} {2003})}\BibitemShut {NoStop}%
\bibitem [{\citenamefont {Pawlowski}(2007)}]{Pawlowski:2005xe}%
  \BibitemOpen
  \bibfield  {author} {\bibinfo {author} {\bibfnamefont {J.~M.}\ \bibnamefont
  {Pawlowski}},\ }\href@noop {} {\bibfield  {journal} {\bibinfo  {journal}
  {Annals Phys.}\ }\textbf {\bibinfo {volume} {322}},\ \bibinfo {pages} {2831}
  (\bibinfo {year} {2007})}\BibitemShut {NoStop}%
\bibitem [{\citenamefont {Bashir}\ \emph {et~al.}(2012)\citenamefont {Bashir}
  \emph {et~al.}}]{Bashir:2012fs}%
  \BibitemOpen
  \bibfield  {author} {\bibinfo {author} {\bibfnamefont {A.}~\bibnamefont
  {Bashir}} \emph {et~al.},\ }\href@noop {} {\bibfield  {journal} {\bibinfo
  {journal} {Commun. Theor. Phys.}\ }\textbf {\bibinfo {volume} {58}},\
  \bibinfo {pages} {79} (\bibinfo {year} {2012})}\BibitemShut {NoStop}%
\bibitem [{\citenamefont {Eichmann}\ \emph
  {et~al.}(2016{\natexlab{a}})\citenamefont {Eichmann}, \citenamefont
  {Sanchis-Alepuz}, \citenamefont {Williams}, \citenamefont {Alkofer},\ and\
  \citenamefont {Fischer}}]{Eichmann:2016yit}%
  \BibitemOpen
  \bibfield  {author} {\bibinfo {author} {\bibfnamefont {G.}~\bibnamefont
  {Eichmann}}, \bibinfo {author} {\bibfnamefont {H.}~\bibnamefont
  {Sanchis-Alepuz}}, \bibinfo {author} {\bibfnamefont {R.}~\bibnamefont
  {Williams}}, \bibinfo {author} {\bibfnamefont {R.}~\bibnamefont {Alkofer}},\
  and\ \bibinfo {author} {\bibfnamefont {C.~S.}\ \bibnamefont {Fischer}},\
  }\href@noop {} {\bibfield  {journal} {\bibinfo  {journal} {Prog. Part. Nucl.
  Phys.}\ }\textbf {\bibinfo {volume} {91}},\ \bibinfo {pages} {1} (\bibinfo
  {year} {2016}{\natexlab{a}})}\BibitemShut {NoStop}%
\bibitem [{\citenamefont {Fischer}(2019)}]{Fischer:2018sdj}%
  \BibitemOpen
  \bibfield  {author} {\bibinfo {author} {\bibfnamefont {C.~S.}\ \bibnamefont
  {Fischer}},\ }\href@noop {} {\bibfield  {journal} {\bibinfo  {journal} {Prog.
  Part. Nucl. Phys.}\ }\textbf {\bibinfo {volume} {105}},\ \bibinfo {pages} {1}
  (\bibinfo {year} {2019})}\BibitemShut {NoStop}%
\bibitem [{\citenamefont {Qin}\ and\ \citenamefont
  {Roberts}(2020)}]{Qin:2020rad}%
  \BibitemOpen
  \bibfield  {author} {\bibinfo {author} {\bibfnamefont {S.-X.}\ \bibnamefont
  {Qin}}\ and\ \bibinfo {author} {\bibfnamefont {C.~D.}\ \bibnamefont
  {Roberts}},\ }\href@noop {} {\bibfield  {journal} {\bibinfo  {journal} {Chin.
  Phys. Lett.}\ }\textbf {\bibinfo {volume} {37}},\ \bibinfo {pages} {121201}
  (\bibinfo {year} {2020})}\BibitemShut {NoStop}%
\bibitem [{\citenamefont {Chen}\ \emph {et~al.}(2018)\citenamefont {Chen},
  \citenamefont {El-Bennich}, \citenamefont {Roberts}, \citenamefont {Schmidt},
  \citenamefont {Segovia},\ and\ \citenamefont {Wan}}]{Chen:2017pse}%
  \BibitemOpen
  \bibfield  {author} {\bibinfo {author} {\bibfnamefont {C.}~\bibnamefont
  {Chen}}, \bibinfo {author} {\bibfnamefont {B.}~\bibnamefont {El-Bennich}},
  \bibinfo {author} {\bibfnamefont {C.~D.}\ \bibnamefont {Roberts}}, \bibinfo
  {author} {\bibfnamefont {S.~M.}\ \bibnamefont {Schmidt}}, \bibinfo {author}
  {\bibfnamefont {J.}~\bibnamefont {Segovia}},\ and\ \bibinfo {author}
  {\bibfnamefont {S.}~\bibnamefont {Wan}},\ }\href@noop {} {\bibfield
  {journal} {\bibinfo  {journal} {Phys. Rev. D}\ }\textbf {\bibinfo {volume}
  {97}},\ \bibinfo {pages} {034016} (\bibinfo {year} {2018})}\BibitemShut
  {NoStop}%
\bibitem [{\citenamefont {Chen}\ \emph
  {et~al.}(2019{\natexlab{a}})\citenamefont {Chen}, \citenamefont {Lu},
  \citenamefont {Binosi}, \citenamefont {Roberts}, \citenamefont
  {Rodr\'\i{}guez-Quintero},\ and\ \citenamefont {Segovia}}]{Chen:2018nsg}%
  \BibitemOpen
  \bibfield  {author} {\bibinfo {author} {\bibfnamefont {C.}~\bibnamefont
  {Chen}}, \bibinfo {author} {\bibfnamefont {Y.}~\bibnamefont {Lu}}, \bibinfo
  {author} {\bibfnamefont {D.}~\bibnamefont {Binosi}}, \bibinfo {author}
  {\bibfnamefont {C.~D.}\ \bibnamefont {Roberts}}, \bibinfo {author}
  {\bibfnamefont {J.}~\bibnamefont {Rodr\'\i{}guez-Quintero}},\ and\ \bibinfo
  {author} {\bibfnamefont {J.}~\bibnamefont {Segovia}},\ }\href@noop {}
  {\bibfield  {journal} {\bibinfo  {journal} {Phys. Rev. D}\ }\textbf {\bibinfo
  {volume} {99}},\ \bibinfo {pages} {034013} (\bibinfo {year}
  {2019}{\natexlab{a}})}\BibitemShut {NoStop}%
\bibitem [{\citenamefont {Chen}\ \emph
  {et~al.}(2019{\natexlab{b}})\citenamefont {Chen}, \citenamefont {Krein},
  \citenamefont {Roberts}, \citenamefont {Schmidt},\ and\ \citenamefont
  {Segovia}}]{Chen:2019fzn}%
  \BibitemOpen
  \bibfield  {author} {\bibinfo {author} {\bibfnamefont {C.}~\bibnamefont
  {Chen}}, \bibinfo {author} {\bibfnamefont {G.~I.}\ \bibnamefont {Krein}},
  \bibinfo {author} {\bibfnamefont {C.~D.}\ \bibnamefont {Roberts}}, \bibinfo
  {author} {\bibfnamefont {S.~M.}\ \bibnamefont {Schmidt}},\ and\ \bibinfo
  {author} {\bibfnamefont {J.}~\bibnamefont {Segovia}},\ }\href@noop {}
  {\bibfield  {journal} {\bibinfo  {journal} {Phys. Rev. D}\ }\textbf {\bibinfo
  {volume} {100}},\ \bibinfo {pages} {054009} (\bibinfo {year}
  {2019}{\natexlab{b}})}\BibitemShut {NoStop}%
\bibitem [{\citenamefont {Lu}\ \emph {et~al.}(2019)\citenamefont {Lu},
  \citenamefont {Chen}, \citenamefont {Cui}, \citenamefont {Roberts},
  \citenamefont {Schmidt}, \citenamefont {Segovia},\ and\ \citenamefont
  {Zong}}]{Lu:2019bjs}%
  \BibitemOpen
  \bibfield  {author} {\bibinfo {author} {\bibfnamefont {Y.}~\bibnamefont
  {Lu}}, \bibinfo {author} {\bibfnamefont {C.}~\bibnamefont {Chen}}, \bibinfo
  {author} {\bibfnamefont {Z.-F.}\ \bibnamefont {Cui}}, \bibinfo {author}
  {\bibfnamefont {C.~D.}\ \bibnamefont {Roberts}}, \bibinfo {author}
  {\bibfnamefont {S.~M.}\ \bibnamefont {Schmidt}}, \bibinfo {author}
  {\bibfnamefont {J.}~\bibnamefont {Segovia}},\ and\ \bibinfo {author}
  {\bibfnamefont {H.~S.}\ \bibnamefont {Zong}},\ }\href@noop {} {\bibfield
  {journal} {\bibinfo  {journal} {Phys. Rev. D}\ }\textbf {\bibinfo {volume}
  {100}},\ \bibinfo {pages} {034001} (\bibinfo {year} {2019})}\BibitemShut
  {NoStop}%
\bibitem [{\citenamefont {Cui}\ \emph {et~al.}(2020{\natexlab{a}})\citenamefont
  {Cui}, \citenamefont {Chen}, \citenamefont {Binosi}, \citenamefont {de~Soto},
  \citenamefont {Roberts}, \citenamefont {Rodr{\'{\i}}guez-Quintero},
  \citenamefont {Schmidt},\ and\ \citenamefont {Segovia}}]{Cui:2020rmu}%
  \BibitemOpen
  \bibfield  {author} {\bibinfo {author} {\bibfnamefont {Z.-F.}\ \bibnamefont
  {Cui}}, \bibinfo {author} {\bibfnamefont {C.}~\bibnamefont {Chen}}, \bibinfo
  {author} {\bibfnamefont {D.}~\bibnamefont {Binosi}}, \bibinfo {author}
  {\bibfnamefont {F.}~\bibnamefont {de~Soto}}, \bibinfo {author} {\bibfnamefont
  {C.~D.}\ \bibnamefont {Roberts}}, \bibinfo {author} {\bibfnamefont
  {J.}~\bibnamefont {Rodr{\'{\i}}guez-Quintero}}, \bibinfo {author}
  {\bibfnamefont {S.~M.}\ \bibnamefont {Schmidt}},\ and\ \bibinfo {author}
  {\bibfnamefont {J.}~\bibnamefont {Segovia}},\ }\href@noop {} {\bibfield
  {journal} {\bibinfo  {journal} {Phys. Rev. D}\ }\textbf {\bibinfo {volume}
  {102}},\ \bibinfo {pages} {014043} (\bibinfo {year}
  {2020}{\natexlab{a}})}\BibitemShut {NoStop}%
\bibitem [{\citenamefont {Fischer}\ and\ \citenamefont
  {Alkofer}(2003)}]{Fischer:2003rp}%
  \BibitemOpen
  \bibfield  {author} {\bibinfo {author} {\bibfnamefont {C.~S.}\ \bibnamefont
  {Fischer}}\ and\ \bibinfo {author} {\bibfnamefont {R.}~\bibnamefont
  {Alkofer}},\ }\href@noop {} {\bibfield  {journal} {\bibinfo  {journal} {Phys.
  Rev. D}\ }\textbf {\bibinfo {volume} {67}},\ \bibinfo {pages} {094020}
  (\bibinfo {year} {2003})}\BibitemShut {NoStop}%
\bibitem [{\citenamefont {Fischer}\ and\ \citenamefont
  {Williams}(2008)}]{Fischer:2008wy}%
  \BibitemOpen
  \bibfield  {author} {\bibinfo {author} {\bibfnamefont {C.~S.}\ \bibnamefont
  {Fischer}}\ and\ \bibinfo {author} {\bibfnamefont {R.}~\bibnamefont
  {Williams}},\ }\href@noop {} {\bibfield  {journal} {\bibinfo  {journal}
  {Phys. Rev. D}\ }\textbf {\bibinfo {volume} {78}},\ \bibinfo {pages} {074006}
  (\bibinfo {year} {2008})}\BibitemShut {NoStop}%
\bibitem [{\citenamefont {Eichmann}\ \emph {et~al.}(2008)\citenamefont
  {Eichmann}, \citenamefont {Alkofer}, \citenamefont {Cloet}, \citenamefont
  {Krassnigg},\ and\ \citenamefont {Roberts}}]{Eichmann:2008ae}%
  \BibitemOpen
  \bibfield  {author} {\bibinfo {author} {\bibfnamefont {G.}~\bibnamefont
  {Eichmann}}, \bibinfo {author} {\bibfnamefont {R.}~\bibnamefont {Alkofer}},
  \bibinfo {author} {\bibfnamefont {I.~C.}\ \bibnamefont {Cloet}}, \bibinfo
  {author} {\bibfnamefont {A.}~\bibnamefont {Krassnigg}},\ and\ \bibinfo
  {author} {\bibfnamefont {C.~D.}\ \bibnamefont {Roberts}},\ }\href@noop {}
  {\bibfield  {journal} {\bibinfo  {journal} {Phys. Rev. C}\ }\textbf {\bibinfo
  {volume} {77}},\ \bibinfo {pages} {042202(R)} (\bibinfo {year}
  {2008})}\BibitemShut {NoStop}%
\bibitem [{\citenamefont {Binosi}\ and\ \citenamefont
  {Papavassiliou}(2009)}]{Binosi:2009qm}%
  \BibitemOpen
  \bibfield  {author} {\bibinfo {author} {\bibfnamefont {D.}~\bibnamefont
  {Binosi}}\ and\ \bibinfo {author} {\bibfnamefont {J.}~\bibnamefont
  {Papavassiliou}},\ }\href@noop {} {\bibfield  {journal} {\bibinfo  {journal}
  {Phys. Rept.}\ }\textbf {\bibinfo {volume} {479}},\ \bibinfo {pages} {1}
  (\bibinfo {year} {2009})}\BibitemShut {NoStop}%
\bibitem [{\citenamefont {Williams}\ \emph {et~al.}(2016)\citenamefont
  {Williams}, \citenamefont {Fischer},\ and\ \citenamefont
  {Heupel}}]{Williams:2015cvx}%
  \BibitemOpen
  \bibfield  {author} {\bibinfo {author} {\bibfnamefont {R.}~\bibnamefont
  {Williams}}, \bibinfo {author} {\bibfnamefont {C.~S.}\ \bibnamefont
  {Fischer}},\ and\ \bibinfo {author} {\bibfnamefont {W.}~\bibnamefont
  {Heupel}},\ }\href@noop {} {\bibfield  {journal} {\bibinfo  {journal} {Phys.
  Rev. D}\ }\textbf {\bibinfo {volume} {93}},\ \bibinfo {pages} {034026}
  (\bibinfo {year} {2016})}\BibitemShut {NoStop}%
\bibitem [{\citenamefont {Huber}(2020{\natexlab{a}})}]{Huber:2018ned}%
  \BibitemOpen
  \bibfield  {author} {\bibinfo {author} {\bibfnamefont {M.~Q.}\ \bibnamefont
  {Huber}},\ }\href@noop {} {\bibfield  {journal} {\bibinfo  {journal} {Phys.
  Rept.}\ }\textbf {\bibinfo {volume} {879}},\ \bibinfo {pages} {1 } (\bibinfo
  {year} {2020}{\natexlab{a}})}\BibitemShut {NoStop}%
\bibitem [{\citenamefont {Eichmann}\ \emph {et~al.}(2010)\citenamefont
  {Eichmann}, \citenamefont {Alkofer}, \citenamefont {Krassnigg},\ and\
  \citenamefont {Nicmorus}}]{Eichmann:2009qa}%
  \BibitemOpen
  \bibfield  {author} {\bibinfo {author} {\bibfnamefont {G.}~\bibnamefont
  {Eichmann}}, \bibinfo {author} {\bibfnamefont {R.}~\bibnamefont {Alkofer}},
  \bibinfo {author} {\bibfnamefont {A.}~\bibnamefont {Krassnigg}},\ and\
  \bibinfo {author} {\bibfnamefont {D.}~\bibnamefont {Nicmorus}},\ }\href@noop
  {} {\bibfield  {journal} {\bibinfo  {journal} {Phys. Rev. Lett.}\ }\textbf
  {\bibinfo {volume} {104}},\ \bibinfo {pages} {201601} (\bibinfo {year}
  {2010})}\BibitemShut {NoStop}%
\bibitem [{\citenamefont {Sanchis-Alepuz}\ and\ \citenamefont
  {Fischer}(2014)}]{Sanchis-Alepuz:2014sca}%
  \BibitemOpen
  \bibfield  {author} {\bibinfo {author} {\bibfnamefont {H.}~\bibnamefont
  {Sanchis-Alepuz}}\ and\ \bibinfo {author} {\bibfnamefont {C.~S.}\
  \bibnamefont {Fischer}},\ }\href@noop {} {\bibfield  {journal} {\bibinfo
  {journal} {Phys. Rev. D}\ }\textbf {\bibinfo {volume} {90}},\ \bibinfo
  {pages} {096001} (\bibinfo {year} {2014})}\BibitemShut {NoStop}%
\bibitem [{\citenamefont {Sanchis-Alepuz}\ \emph {et~al.}(2014)\citenamefont
  {Sanchis-Alepuz}, \citenamefont {Fischer},\ and\ \citenamefont
  {Kubrak}}]{Sanchis-Alepuz:2014wea}%
  \BibitemOpen
  \bibfield  {author} {\bibinfo {author} {\bibfnamefont {H.}~\bibnamefont
  {Sanchis-Alepuz}}, \bibinfo {author} {\bibfnamefont {C.~S.}\ \bibnamefont
  {Fischer}},\ and\ \bibinfo {author} {\bibfnamefont {S.}~\bibnamefont
  {Kubrak}},\ }\href@noop {} {\bibfield  {journal} {\bibinfo  {journal} {Phys.
  Lett. B}\ }\textbf {\bibinfo {volume} {733}},\ \bibinfo {pages} {151}
  (\bibinfo {year} {2014})}\BibitemShut {NoStop}%
\bibitem [{\citenamefont {Qin}\ \emph {et~al.}(2018)\citenamefont {Qin},
  \citenamefont {Roberts},\ and\ \citenamefont {Schmidt}}]{Qin:2018dqp}%
  \BibitemOpen
  \bibfield  {author} {\bibinfo {author} {\bibfnamefont {S.-X.}\ \bibnamefont
  {Qin}}, \bibinfo {author} {\bibfnamefont {C.~D.}\ \bibnamefont {Roberts}},\
  and\ \bibinfo {author} {\bibfnamefont {S.~M.}\ \bibnamefont {Schmidt}},\
  }\href@noop {} {\bibfield  {journal} {\bibinfo  {journal} {Phys. Rev. D}\
  }\textbf {\bibinfo {volume} {97}},\ \bibinfo {pages} {114017} (\bibinfo
  {year} {2018})}\BibitemShut {NoStop}%
\bibitem [{\citenamefont {Qin}\ \emph {et~al.}(2019)\citenamefont {Qin},
  \citenamefont {Roberts},\ and\ \citenamefont {Schmidt}}]{Qin:2019hgk}%
  \BibitemOpen
  \bibfield  {author} {\bibinfo {author} {\bibfnamefont {S.-X.}\ \bibnamefont
  {Qin}}, \bibinfo {author} {\bibfnamefont {C.~D.}\ \bibnamefont {Roberts}},\
  and\ \bibinfo {author} {\bibfnamefont {S.~M.}\ \bibnamefont {Schmidt}},\
  }\href@noop {} {\bibfield  {journal} {\bibinfo  {journal} {Few Body Syst.}\
  }\textbf {\bibinfo {volume} {60}},\ \bibinfo {pages} {26} (\bibinfo {year}
  {2019})}\BibitemShut {NoStop}%
\bibitem [{\citenamefont {Oettel}\ \emph {et~al.}(1998)\citenamefont {Oettel},
  \citenamefont {Hellstern}, \citenamefont {Alkofer},\ and\ \citenamefont
  {Reinhardt}}]{Oettel:1998bk}%
  \BibitemOpen
  \bibfield  {author} {\bibinfo {author} {\bibfnamefont {M.}~\bibnamefont
  {Oettel}}, \bibinfo {author} {\bibfnamefont {G.}~\bibnamefont {Hellstern}},
  \bibinfo {author} {\bibfnamefont {R.}~\bibnamefont {Alkofer}},\ and\ \bibinfo
  {author} {\bibfnamefont {H.}~\bibnamefont {Reinhardt}},\ }\href@noop {}
  {\bibfield  {journal} {\bibinfo  {journal} {Phys. Rev. C}\ }\textbf {\bibinfo
  {volume} {58}},\ \bibinfo {pages} {2459} (\bibinfo {year}
  {1998})}\BibitemShut {NoStop}%
\bibitem [{\citenamefont {Segovia}\ \emph
  {et~al.}(2015{\natexlab{a}})\citenamefont {Segovia}, \citenamefont
  {Roberts},\ and\ \citenamefont {Schmidt}}]{Segovia:2015ufa}%
  \BibitemOpen
  \bibfield  {author} {\bibinfo {author} {\bibfnamefont {J.}~\bibnamefont
  {Segovia}}, \bibinfo {author} {\bibfnamefont {C.~D.}\ \bibnamefont
  {Roberts}},\ and\ \bibinfo {author} {\bibfnamefont {S.~M.}\ \bibnamefont
  {Schmidt}},\ }\href@noop {} {\bibfield  {journal} {\bibinfo  {journal} {Phys.
  Lett. B}\ }\textbf {\bibinfo {volume} {750}},\ \bibinfo {pages} {100}
  (\bibinfo {year} {2015}{\natexlab{a}})}\BibitemShut {NoStop}%
\bibitem [{\citenamefont {Eichmann}\ \emph
  {et~al.}(2016{\natexlab{b}})\citenamefont {Eichmann}, \citenamefont
  {Fischer},\ and\ \citenamefont {Sanchis-Alepuz}}]{Eichmann:2016hgl}%
  \BibitemOpen
  \bibfield  {author} {\bibinfo {author} {\bibfnamefont {G.}~\bibnamefont
  {Eichmann}}, \bibinfo {author} {\bibfnamefont {C.~S.}\ \bibnamefont
  {Fischer}},\ and\ \bibinfo {author} {\bibfnamefont {H.}~\bibnamefont
  {Sanchis-Alepuz}},\ }\href@noop {} {\bibfield  {journal} {\bibinfo  {journal}
  {Phys. Rev. D}\ }\textbf {\bibinfo {volume} {94}},\ \bibinfo {pages} {094033}
  (\bibinfo {year} {2016}{\natexlab{b}})}\BibitemShut {NoStop}%
\bibitem [{\citenamefont {Barabanov}\ \emph {et~al.}(2021)\citenamefont
  {Barabanov} \emph {et~al.}}]{Barabanov:2020jvn}%
  \BibitemOpen
  \bibfield  {author} {\bibinfo {author} {\bibfnamefont {M.~Y.}\ \bibnamefont
  {Barabanov}} \emph {et~al.},\ }\href@noop {} {\bibfield  {journal} {\bibinfo
  {journal} {Prog. Part. Nucl. Phys.}\ }\textbf {\bibinfo {volume} {116}},\
  \bibinfo {pages} {103835} (\bibinfo {year} {2021})}\BibitemShut {NoStop}%
\bibitem [{\citenamefont {Cahill}\ \emph {et~al.}(1987)\citenamefont {Cahill},
  \citenamefont {Roberts},\ and\ \citenamefont {Praschifka}}]{Cahill:1987qr}%
  \BibitemOpen
  \bibfield  {author} {\bibinfo {author} {\bibfnamefont {R.~T.}\ \bibnamefont
  {Cahill}}, \bibinfo {author} {\bibfnamefont {C.~D.}\ \bibnamefont
  {Roberts}},\ and\ \bibinfo {author} {\bibfnamefont {J.}~\bibnamefont
  {Praschifka}},\ }\href@noop {} {\bibfield  {journal} {\bibinfo  {journal}
  {Phys. Rev. D}\ }\textbf {\bibinfo {volume} {36}},\ \bibinfo {pages} {2804}
  (\bibinfo {year} {1987})}\BibitemShut {NoStop}%
\bibitem [{\citenamefont {Cahill}\ \emph {et~al.}(1989)\citenamefont {Cahill},
  \citenamefont {Roberts},\ and\ \citenamefont {Praschifka}}]{Cahill:1988dx}%
  \BibitemOpen
  \bibfield  {author} {\bibinfo {author} {\bibfnamefont {R.~T.}\ \bibnamefont
  {Cahill}}, \bibinfo {author} {\bibfnamefont {C.~D.}\ \bibnamefont
  {Roberts}},\ and\ \bibinfo {author} {\bibfnamefont {J.}~\bibnamefont
  {Praschifka}},\ }\href@noop {} {\bibfield  {journal} {\bibinfo  {journal}
  {Austral. J. Phys.}\ }\textbf {\bibinfo {volume} {42}},\ \bibinfo {pages}
  {129} (\bibinfo {year} {1989})}\BibitemShut {NoStop}%
\bibitem [{\citenamefont {Bloch}\ \emph {et~al.}(1999)\citenamefont {Bloch},
  \citenamefont {Roberts},\ and\ \citenamefont {Schmidt}}]{Bloch:1999vk}%
  \BibitemOpen
  \bibfield  {author} {\bibinfo {author} {\bibfnamefont {J.~C.~R.}\
  \bibnamefont {Bloch}}, \bibinfo {author} {\bibfnamefont {C.~D.}\ \bibnamefont
  {Roberts}},\ and\ \bibinfo {author} {\bibfnamefont {S.~M.}\ \bibnamefont
  {Schmidt}},\ }\href@noop {} {\bibfield  {journal} {\bibinfo  {journal} {Phys.
  Rev. C}\ }\textbf {\bibinfo {volume} {60}},\ \bibinfo {pages} {065208}
  (\bibinfo {year} {1999})}\BibitemShut {NoStop}%
\bibitem [{\citenamefont {Bender}\ \emph {et~al.}(2002)\citenamefont {Bender},
  \citenamefont {Detmold}, \citenamefont {Roberts},\ and\ \citenamefont
  {Thomas}}]{Bender:2002as}%
  \BibitemOpen
  \bibfield  {author} {\bibinfo {author} {\bibfnamefont {A.}~\bibnamefont
  {Bender}}, \bibinfo {author} {\bibfnamefont {W.}~\bibnamefont {Detmold}},
  \bibinfo {author} {\bibfnamefont {C.~D.}\ \bibnamefont {Roberts}},\ and\
  \bibinfo {author} {\bibfnamefont {A.~W.}\ \bibnamefont {Thomas}},\
  }\href@noop {} {\bibfield  {journal} {\bibinfo  {journal} {Phys. Rev. C}\
  }\textbf {\bibinfo {volume} {65}},\ \bibinfo {pages} {065203} (\bibinfo
  {year} {2002})}\BibitemShut {NoStop}%
\bibitem [{\citenamefont {Eichmann}\ \emph {et~al.}(2009)\citenamefont
  {Eichmann}, \citenamefont {Cloet}, \citenamefont {Alkofer}, \citenamefont
  {Krassnigg},\ and\ \citenamefont {Roberts}}]{Eichmann:2008ef}%
  \BibitemOpen
  \bibfield  {author} {\bibinfo {author} {\bibfnamefont {G.}~\bibnamefont
  {Eichmann}}, \bibinfo {author} {\bibfnamefont {I.~C.}\ \bibnamefont {Cloet}},
  \bibinfo {author} {\bibfnamefont {R.}~\bibnamefont {Alkofer}}, \bibinfo
  {author} {\bibfnamefont {A.}~\bibnamefont {Krassnigg}},\ and\ \bibinfo
  {author} {\bibfnamefont {C.~D.}\ \bibnamefont {Roberts}},\ }\href@noop {}
  {\bibfield  {journal} {\bibinfo  {journal} {Phys. Rev. C}\ }\textbf {\bibinfo
  {volume} {79}},\ \bibinfo {pages} {012202(R)} (\bibinfo {year}
  {2009})}\BibitemShut {NoStop}%
\bibitem [{\citenamefont {Roberts}\ \emph {et~al.}(2011)\citenamefont
  {Roberts}, \citenamefont {Chang}, \citenamefont {Cloet},\ and\ \citenamefont
  {Roberts}}]{Roberts:2011cf}%
  \BibitemOpen
  \bibfield  {author} {\bibinfo {author} {\bibfnamefont {H.~L.~L.}\
  \bibnamefont {Roberts}}, \bibinfo {author} {\bibfnamefont {L.}~\bibnamefont
  {Chang}}, \bibinfo {author} {\bibfnamefont {I.~C.}\ \bibnamefont {Cloet}},\
  and\ \bibinfo {author} {\bibfnamefont {C.~D.}\ \bibnamefont {Roberts}},\
  }\href@noop {} {\bibfield  {journal} {\bibinfo  {journal} {Few Body Syst.}\
  }\textbf {\bibinfo {volume} {51}},\ \bibinfo {pages} {1} (\bibinfo {year}
  {2011})}\BibitemShut {NoStop}%
\bibitem [{\citenamefont {Eichmann}\ and\ \citenamefont
  {Fischer}(2012)}]{Eichmann:2011pv}%
  \BibitemOpen
  \bibfield  {author} {\bibinfo {author} {\bibfnamefont {G.}~\bibnamefont
  {Eichmann}}\ and\ \bibinfo {author} {\bibfnamefont {C.~S.}\ \bibnamefont
  {Fischer}},\ }\href@noop {} {\bibfield  {journal} {\bibinfo  {journal} {Eur.
  Phys. J. A}\ }\textbf {\bibinfo {volume} {48}},\ \bibinfo {pages} {9}
  (\bibinfo {year} {2012})}\BibitemShut {NoStop}%
\bibitem [{\citenamefont {Munczek}(1995)}]{Munczek:1994zz}%
  \BibitemOpen
  \bibfield  {author} {\bibinfo {author} {\bibfnamefont {H.~J.}\ \bibnamefont
  {Munczek}},\ }\href@noop {} {\bibfield  {journal} {\bibinfo  {journal} {Phys.
  Rev. D}\ }\textbf {\bibinfo {volume} {52}},\ \bibinfo {pages} {4736}
  (\bibinfo {year} {1995})}\BibitemShut {NoStop}%
\bibitem [{\citenamefont {Bender}\ \emph {et~al.}(1996)\citenamefont {Bender},
  \citenamefont {Roberts},\ and\ \citenamefont {von Smekal}}]{Bender:1996bb}%
  \BibitemOpen
  \bibfield  {author} {\bibinfo {author} {\bibfnamefont {A.}~\bibnamefont
  {Bender}}, \bibinfo {author} {\bibfnamefont {C.~D.}\ \bibnamefont
  {Roberts}},\ and\ \bibinfo {author} {\bibfnamefont {L.}~\bibnamefont {von
  Smekal}},\ }\href@noop {} {\bibfield  {journal} {\bibinfo  {journal} {Phys.
  Lett. B}\ }\textbf {\bibinfo {volume} {380}},\ \bibinfo {pages} {7} (\bibinfo
  {year} {1996})}\BibitemShut {NoStop}%
\bibitem [{\citenamefont {Hellstern}\ \emph {et~al.}(1997)\citenamefont
  {Hellstern}, \citenamefont {Alkofer}, \citenamefont {Oettel},\ and\
  \citenamefont {Reinhardt}}]{Hellstern:1997pg}%
  \BibitemOpen
  \bibfield  {author} {\bibinfo {author} {\bibfnamefont {G.}~\bibnamefont
  {Hellstern}}, \bibinfo {author} {\bibfnamefont {R.}~\bibnamefont {Alkofer}},
  \bibinfo {author} {\bibfnamefont {M.}~\bibnamefont {Oettel}},\ and\ \bibinfo
  {author} {\bibfnamefont {H.}~\bibnamefont {Reinhardt}},\ }\href@noop {}
  {\bibfield  {journal} {\bibinfo  {journal} {Nucl. Phys. A}\ }\textbf
  {\bibinfo {volume} {627}},\ \bibinfo {pages} {679} (\bibinfo {year}
  {1997})}\BibitemShut {NoStop}%
\bibitem [{\citenamefont {Bloch}\ \emph {et~al.}(2000)\citenamefont {Bloch},
  \citenamefont {Roberts},\ and\ \citenamefont {Schmidt}}]{Bloch:1999rm}%
  \BibitemOpen
  \bibfield  {author} {\bibinfo {author} {\bibfnamefont {J.~C.~R.}\
  \bibnamefont {Bloch}}, \bibinfo {author} {\bibfnamefont {C.~D.}\ \bibnamefont
  {Roberts}},\ and\ \bibinfo {author} {\bibfnamefont {S.~M.}\ \bibnamefont
  {Schmidt}},\ }\href@noop {} {\bibfield  {journal} {\bibinfo  {journal} {Phys.
  Rev. C}\ }\textbf {\bibinfo {volume} {61}},\ \bibinfo {pages} {065207}
  (\bibinfo {year} {2000})}\BibitemShut {NoStop}%
\bibitem [{\citenamefont {Oettel}\ \emph
  {et~al.}(2000{\natexlab{a}})\citenamefont {Oettel}, \citenamefont {Alkofer},\
  and\ \citenamefont {von Smekal}}]{Oettel:2000jj}%
  \BibitemOpen
  \bibfield  {author} {\bibinfo {author} {\bibfnamefont {M.}~\bibnamefont
  {Oettel}}, \bibinfo {author} {\bibfnamefont {R.}~\bibnamefont {Alkofer}},\
  and\ \bibinfo {author} {\bibfnamefont {L.}~\bibnamefont {von Smekal}},\
  }\href@noop {} {\bibfield  {journal} {\bibinfo  {journal} {Eur. Phys. J. A}\
  }\textbf {\bibinfo {volume} {8}},\ \bibinfo {pages} {553} (\bibinfo {year}
  {2000}{\natexlab{a}})}\BibitemShut {NoStop}%
\bibitem [{\citenamefont {Roberts}\ \emph {et~al.}(2007)\citenamefont
  {Roberts}, \citenamefont {Bhagwat}, \citenamefont {H{\"o}ll},\ and\
  \citenamefont {Wright}}]{Roberts:2007jh}%
  \BibitemOpen
  \bibfield  {author} {\bibinfo {author} {\bibfnamefont {C.~D.}\ \bibnamefont
  {Roberts}}, \bibinfo {author} {\bibfnamefont {M.~S.}\ \bibnamefont
  {Bhagwat}}, \bibinfo {author} {\bibfnamefont {A.}~\bibnamefont {H{\"o}ll}},\
  and\ \bibinfo {author} {\bibfnamefont {S.~V.}\ \bibnamefont {Wright}},\
  }\href@noop {} {\bibfield  {journal} {\bibinfo  {journal} {Eur. Phys. J. ST}\
  }\textbf {\bibinfo {volume} {140}},\ \bibinfo {pages} {53} (\bibinfo {year}
  {2007})}\BibitemShut {NoStop}%
\bibitem [{\citenamefont {Segovia}\ \emph {et~al.}(2014)\citenamefont
  {Segovia}, \citenamefont {Cloet}, \citenamefont {Roberts},\ and\
  \citenamefont {Schmidt}}]{Segovia:2014aza}%
  \BibitemOpen
  \bibfield  {author} {\bibinfo {author} {\bibfnamefont {J.}~\bibnamefont
  {Segovia}}, \bibinfo {author} {\bibfnamefont {I.~C.}\ \bibnamefont {Cloet}},
  \bibinfo {author} {\bibfnamefont {C.~D.}\ \bibnamefont {Roberts}},\ and\
  \bibinfo {author} {\bibfnamefont {S.~M.}\ \bibnamefont {Schmidt}},\
  }\href@noop {} {\bibfield  {journal} {\bibinfo  {journal} {Few Body Syst.}\
  }\textbf {\bibinfo {volume} {55}},\ \bibinfo {pages} {1185} (\bibinfo {year}
  {2014})}\BibitemShut {NoStop}%
\bibitem [{\citenamefont {Burkert}\ and\ \citenamefont
  {Roberts}(2019)}]{Burkert:2017djo}%
  \BibitemOpen
  \bibfield  {author} {\bibinfo {author} {\bibfnamefont {V.~D.}\ \bibnamefont
  {Burkert}}\ and\ \bibinfo {author} {\bibfnamefont {C.~D.}\ \bibnamefont
  {Roberts}},\ }\href@noop {} {\bibfield  {journal} {\bibinfo  {journal} {Rev.
  Mod. Phys.}\ }\textbf {\bibinfo {volume} {91}},\ \bibinfo {pages} {011003}
  (\bibinfo {year} {2019})}\BibitemShut {NoStop}%
\bibitem [{\citenamefont {Liu}\ \emph {et~al.}(2022)\citenamefont {Liu},
  \citenamefont {Chen}, \citenamefont {Lu}, \citenamefont {Roberts},\ and\
  \citenamefont {Segovia}}]{Liu:2022ndb}%
  \BibitemOpen
  \bibfield  {author} {\bibinfo {author} {\bibfnamefont {L.}~\bibnamefont
  {Liu}}, \bibinfo {author} {\bibfnamefont {C.}~\bibnamefont {Chen}}, \bibinfo
  {author} {\bibfnamefont {Y.}~\bibnamefont {Lu}}, \bibinfo {author}
  {\bibfnamefont {C.~D.}\ \bibnamefont {Roberts}},\ and\ \bibinfo {author}
  {\bibfnamefont {J.}~\bibnamefont {Segovia}},\ }\href@noop {} {\  (\bibinfo
  {year} {2022})},\ \bibinfo {note} {{\emph{Composition of low-lying}
  $J=\tfrac{3}{2}^\pm \,\Delta$-\emph{baryons} -- arXiv:2203.12083
  [hep-ph]}}\BibitemShut {NoStop}%
\bibitem [{\citenamefont {Chen}\ \emph {et~al.}(2021)\citenamefont {Chen},
  \citenamefont {Fischer}, \citenamefont {Roberts},\ and\ \citenamefont
  {Segovia}}]{Chen:2020wuq}%
  \BibitemOpen
  \bibfield  {author} {\bibinfo {author} {\bibfnamefont {C.}~\bibnamefont
  {Chen}}, \bibinfo {author} {\bibfnamefont {C.~S.}\ \bibnamefont {Fischer}},
  \bibinfo {author} {\bibfnamefont {C.~D.}\ \bibnamefont {Roberts}},\ and\
  \bibinfo {author} {\bibfnamefont {J.}~\bibnamefont {Segovia}},\ }\href@noop
  {} {\bibfield  {journal} {\bibinfo  {journal} {Phys. Lett. B}\ }\textbf
  {\bibinfo {volume} {815}},\ \bibinfo {pages} {136150} (\bibinfo {year}
  {2021})}\BibitemShut {NoStop}%
\bibitem [{\citenamefont {Aguilar}\ \emph {et~al.}(2008)\citenamefont
  {Aguilar}, \citenamefont {Binosi},\ and\ \citenamefont
  {Papavassiliou}}]{Aguilar:2008xm}%
  \BibitemOpen
  \bibfield  {author} {\bibinfo {author} {\bibfnamefont {A.~C.}\ \bibnamefont
  {Aguilar}}, \bibinfo {author} {\bibfnamefont {D.}~\bibnamefont {Binosi}},\
  and\ \bibinfo {author} {\bibfnamefont {J.}~\bibnamefont {Papavassiliou}},\
  }\href@noop {} {\bibfield  {journal} {\bibinfo  {journal} {Phys. Rev. D}\
  }\textbf {\bibinfo {volume} {78}},\ \bibinfo {pages} {025010} (\bibinfo
  {year} {2008})}\BibitemShut {NoStop}%
\bibitem [{\citenamefont {Aguilar}\ and\ \citenamefont
  {Papavassiliou}(2010)}]{Aguilar:2009ke}%
  \BibitemOpen
  \bibfield  {author} {\bibinfo {author} {\bibfnamefont {A.~C.}\ \bibnamefont
  {Aguilar}}\ and\ \bibinfo {author} {\bibfnamefont {J.}~\bibnamefont
  {Papavassiliou}},\ }\href@noop {} {\bibfield  {journal} {\bibinfo  {journal}
  {Phys. Rev. D}\ }\textbf {\bibinfo {volume} {81}},\ \bibinfo {pages} {034003}
  (\bibinfo {year} {2010})}\BibitemShut {NoStop}%
\bibitem [{\citenamefont {Aguilar}\ \emph {et~al.}(2009)\citenamefont
  {Aguilar}, \citenamefont {Binosi}, \citenamefont {Papavassiliou},\ and\
  \citenamefont {Rodr{\'i}guez-Quintero}}]{Aguilar:2009nf}%
  \BibitemOpen
  \bibfield  {author} {\bibinfo {author} {\bibfnamefont {A.}~\bibnamefont
  {Aguilar}}, \bibinfo {author} {\bibfnamefont {D.}~\bibnamefont {Binosi}},
  \bibinfo {author} {\bibfnamefont {J.}~\bibnamefont {Papavassiliou}},\ and\
  \bibinfo {author} {\bibfnamefont {J.}~\bibnamefont
  {Rodr{\'i}guez-Quintero}},\ }\href@noop {} {\bibfield  {journal} {\bibinfo
  {journal} {Phys. Rev. D}\ }\textbf {\bibinfo {volume} {80}},\ \bibinfo
  {pages} {085018} (\bibinfo {year} {2009})}\BibitemShut {NoStop}%
\bibitem [{\citenamefont {Maas}(2013)}]{Maas:2011se}%
  \BibitemOpen
  \bibfield  {author} {\bibinfo {author} {\bibfnamefont {A.}~\bibnamefont
  {Maas}},\ }\href@noop {} {\bibfield  {journal} {\bibinfo  {journal} {Phys.
  Rept.}\ }\textbf {\bibinfo {volume} {524}},\ \bibinfo {pages} {203} (\bibinfo
  {year} {2013})}\BibitemShut {NoStop}%
\bibitem [{\citenamefont {Boucaud}\ \emph {et~al.}(2012)\citenamefont
  {Boucaud}, \citenamefont {Leroy}, \citenamefont {Le-Yaouanc}, \citenamefont
  {Micheli}, \citenamefont {Pene},\ and\ \citenamefont
  {Rodr{\'i}guez-Quintero}}]{Boucaud:2011ug}%
  \BibitemOpen
  \bibfield  {author} {\bibinfo {author} {\bibfnamefont {P.}~\bibnamefont
  {Boucaud}}, \bibinfo {author} {\bibfnamefont {J.~P.}\ \bibnamefont {Leroy}},
  \bibinfo {author} {\bibfnamefont {A.}~\bibnamefont {Le-Yaouanc}}, \bibinfo
  {author} {\bibfnamefont {J.}~\bibnamefont {Micheli}}, \bibinfo {author}
  {\bibfnamefont {O.}~\bibnamefont {Pene}},\ and\ \bibinfo {author}
  {\bibfnamefont {J.}~\bibnamefont {Rodr{\'i}guez-Quintero}},\ }\href@noop {}
  {\bibfield  {journal} {\bibinfo  {journal} {Few Body Syst.}\ }\textbf
  {\bibinfo {volume} {53}},\ \bibinfo {pages} {387} (\bibinfo {year}
  {2012})}\BibitemShut {NoStop}%
\bibitem [{\citenamefont {Pennington}\ and\ \citenamefont
  {Wilson}(2011)}]{Pennington:2011xs}%
  \BibitemOpen
  \bibfield  {author} {\bibinfo {author} {\bibfnamefont {M.~R.}\ \bibnamefont
  {Pennington}}\ and\ \bibinfo {author} {\bibfnamefont {D.~J.}\ \bibnamefont
  {Wilson}},\ }\href@noop {} {\bibfield  {journal} {\bibinfo  {journal} {Phys.
  Rev. D}\ }\textbf {\bibinfo {volume} {84}},\ \bibinfo {pages} {119901}
  (\bibinfo {year} {2011})}\BibitemShut {NoStop}%
\bibitem [{\citenamefont {Binosi}\ \emph {et~al.}(2012)\citenamefont {Binosi},
  \citenamefont {Ib{\'a}{\~n}ez},\ and\ \citenamefont
  {Papavassiliou}}]{Binosi:2012sj}%
  \BibitemOpen
  \bibfield  {author} {\bibinfo {author} {\bibfnamefont {D.}~\bibnamefont
  {Binosi}}, \bibinfo {author} {\bibfnamefont {D.}~\bibnamefont
  {Ib{\'a}{\~n}ez}},\ and\ \bibinfo {author} {\bibfnamefont {J.}~\bibnamefont
  {Papavassiliou}},\ }\href@noop {} {\bibfield  {journal} {\bibinfo  {journal}
  {Phys. Rev. D}\ }\textbf {\bibinfo {volume} {86}},\ \bibinfo {pages} {085033}
  (\bibinfo {year} {2012})}\BibitemShut {NoStop}%
\bibitem [{\citenamefont {Strauss}\ \emph {et~al.}(2012)\citenamefont
  {Strauss}, \citenamefont {Fischer},\ and\ \citenamefont
  {Kellermann}}]{Strauss:2012dg}%
  \BibitemOpen
  \bibfield  {author} {\bibinfo {author} {\bibfnamefont {S.}~\bibnamefont
  {Strauss}}, \bibinfo {author} {\bibfnamefont {C.~S.}\ \bibnamefont
  {Fischer}},\ and\ \bibinfo {author} {\bibfnamefont {C.}~\bibnamefont
  {Kellermann}},\ }\href@noop {} {\bibfield  {journal} {\bibinfo  {journal}
  {Phys. Rev. Lett.}\ }\textbf {\bibinfo {volume} {109}},\ \bibinfo {pages}
  {252001} (\bibinfo {year} {2012})}\BibitemShut {NoStop}%
\bibitem [{\citenamefont {Meyers}\ and\ \citenamefont
  {Swanson}(2014)}]{Meyers:2014iwa}%
  \BibitemOpen
  \bibfield  {author} {\bibinfo {author} {\bibfnamefont {J.}~\bibnamefont
  {Meyers}}\ and\ \bibinfo {author} {\bibfnamefont {E.~S.}\ \bibnamefont
  {Swanson}},\ }\href@noop {} {\bibfield  {journal} {\bibinfo  {journal} {Phys.
  Rev. D}\ }\textbf {\bibinfo {volume} {90}},\ \bibinfo {pages} {045037}
  (\bibinfo {year} {2014})}\BibitemShut {NoStop}%
\bibitem [{\citenamefont {Huber}(2020{\natexlab{b}})}]{Huber:2020keu}%
  \BibitemOpen
  \bibfield  {author} {\bibinfo {author} {\bibfnamefont {M.~Q.}\ \bibnamefont
  {Huber}},\ }\href@noop {} {\bibfield  {journal} {\bibinfo  {journal} {Phys.
  Rev. D}\ }\textbf {\bibinfo {volume} {101}},\ \bibinfo {pages} {114009}
  (\bibinfo {year} {2020}{\natexlab{b}})}\BibitemShut {NoStop}%
\bibitem [{\citenamefont {Binosi}\ \emph {et~al.}(2015)\citenamefont {Binosi},
  \citenamefont {Chang}, \citenamefont {Papavassiliou},\ and\ \citenamefont
  {Roberts}}]{Binosi:2014aea}%
  \BibitemOpen
  \bibfield  {author} {\bibinfo {author} {\bibfnamefont {D.}~\bibnamefont
  {Binosi}}, \bibinfo {author} {\bibfnamefont {L.}~\bibnamefont {Chang}},
  \bibinfo {author} {\bibfnamefont {J.}~\bibnamefont {Papavassiliou}},\ and\
  \bibinfo {author} {\bibfnamefont {C.~D.}\ \bibnamefont {Roberts}},\
  }\href@noop {} {\bibfield  {journal} {\bibinfo  {journal} {Phys. Lett. B}\
  }\textbf {\bibinfo {volume} {742}},\ \bibinfo {pages} {183} (\bibinfo {year}
  {2015})}\BibitemShut {NoStop}%
\bibitem [{\citenamefont {Pauli}(1957)}]{Pauli:1957ot}%
  \BibitemOpen
  \bibfield  {author} {\bibinfo {author} {\bibfnamefont {W.}~\bibnamefont
  {Pauli}},\ }\href@noop {} {\bibfield  {journal} {\bibinfo  {journal} {Nuovo
  Cim.}\ }\textbf {\bibinfo {volume} {6}},\ \bibinfo {pages} {204} (\bibinfo
  {year} {1957})}\BibitemShut {NoStop}%
\bibitem [{\citenamefont {G{\"u}rsey}(1958)}]{Guersey:1958ro}%
  \BibitemOpen
  \bibfield  {author} {\bibinfo {author} {\bibfnamefont {F.}~\bibnamefont
  {G{\"u}rsey}},\ }\href@noop {} {\bibfield  {journal} {\bibinfo  {journal}
  {Nuovo Cim.}\ }\textbf {\bibinfo {volume} {7}},\ \bibinfo {pages} {411}
  (\bibinfo {year} {1958})}\BibitemShut {NoStop}%
\bibitem [{\citenamefont {Reinhardt}(1990)}]{Reinhardt:1989rw}%
  \BibitemOpen
  \bibfield  {author} {\bibinfo {author} {\bibfnamefont {H.}~\bibnamefont
  {Reinhardt}},\ }\href@noop {} {\bibfield  {journal} {\bibinfo  {journal}
  {Phys. Lett. B}\ }\textbf {\bibinfo {volume} {244}},\ \bibinfo {pages} {316}
  (\bibinfo {year} {1990})}\BibitemShut {NoStop}%
\bibitem [{\citenamefont {Efimov}\ \emph {et~al.}(1990)\citenamefont {Efimov},
  \citenamefont {Ivanov},\ and\ \citenamefont {Lyubovitskij}}]{Efimov:1990uz}%
  \BibitemOpen
  \bibfield  {author} {\bibinfo {author} {\bibfnamefont {G.~V.}\ \bibnamefont
  {Efimov}}, \bibinfo {author} {\bibfnamefont {M.~A.}\ \bibnamefont {Ivanov}},\
  and\ \bibinfo {author} {\bibfnamefont {V.~E.}\ \bibnamefont {Lyubovitskij}},\
  }\href@noop {} {\bibfield  {journal} {\bibinfo  {journal} {Z. Phys. C}\
  }\textbf {\bibinfo {volume} {47}},\ \bibinfo {pages} {583} (\bibinfo {year}
  {1990})}\BibitemShut {NoStop}%
\bibitem [{\citenamefont {Oettel}\ \emph
  {et~al.}(2000{\natexlab{b}})\citenamefont {Oettel}, \citenamefont
  {Pichowsky},\ and\ \citenamefont {von Smekal}}]{Oettel:1999gc}%
  \BibitemOpen
  \bibfield  {author} {\bibinfo {author} {\bibfnamefont {M.}~\bibnamefont
  {Oettel}}, \bibinfo {author} {\bibfnamefont {M.}~\bibnamefont {Pichowsky}},\
  and\ \bibinfo {author} {\bibfnamefont {L.}~\bibnamefont {von Smekal}},\
  }\href@noop {} {\bibfield  {journal} {\bibinfo  {journal} {Eur. Phys. J. A}\
  }\textbf {\bibinfo {volume} {8}},\ \bibinfo {pages} {251} (\bibinfo {year}
  {2000}{\natexlab{b}})}\BibitemShut {NoStop}%
\bibitem [{\citenamefont {Hecht}\ \emph {et~al.}(2002)\citenamefont {Hecht},
  \citenamefont {Oettel}, \citenamefont {Roberts}, \citenamefont {Schmidt},
  \citenamefont {Tandy},\ and\ \citenamefont {Thomas}}]{Hecht:2002ej}%
  \BibitemOpen
  \bibfield  {author} {\bibinfo {author} {\bibfnamefont {M.~B.}\ \bibnamefont
  {Hecht}}, \bibinfo {author} {\bibfnamefont {M.}~\bibnamefont {Oettel}},
  \bibinfo {author} {\bibfnamefont {C.~D.}\ \bibnamefont {Roberts}}, \bibinfo
  {author} {\bibfnamefont {S.~M.}\ \bibnamefont {Schmidt}}, \bibinfo {author}
  {\bibfnamefont {P.~C.}\ \bibnamefont {Tandy}},\ and\ \bibinfo {author}
  {\bibfnamefont {A.~W.}\ \bibnamefont {Thomas}},\ }\href@noop {} {\bibfield
  {journal} {\bibinfo  {journal} {Phys. Rev. C}\ }\textbf {\bibinfo {volume}
  {65}},\ \bibinfo {pages} {055204} (\bibinfo {year} {2002})}\BibitemShut
  {NoStop}%
\bibitem [{\citenamefont {Segovia}\ \emph
  {et~al.}(2015{\natexlab{b}})\citenamefont {Segovia}, \citenamefont
  {El-Bennich}, \citenamefont {Rojas}, \citenamefont {Cloet}, \citenamefont
  {Roberts}, \citenamefont {Xu},\ and\ \citenamefont {Zong}}]{Segovia:2015hra}%
  \BibitemOpen
  \bibfield  {author} {\bibinfo {author} {\bibfnamefont {J.}~\bibnamefont
  {Segovia}}, \bibinfo {author} {\bibfnamefont {B.}~\bibnamefont {El-Bennich}},
  \bibinfo {author} {\bibfnamefont {E.}~\bibnamefont {Rojas}}, \bibinfo
  {author} {\bibfnamefont {I.~C.}\ \bibnamefont {Cloet}}, \bibinfo {author}
  {\bibfnamefont {C.~D.}\ \bibnamefont {Roberts}}, \bibinfo {author}
  {\bibfnamefont {S.-S.}\ \bibnamefont {Xu}},\ and\ \bibinfo {author}
  {\bibfnamefont {H.-S.}\ \bibnamefont {Zong}},\ }\href@noop {} {\bibfield
  {journal} {\bibinfo  {journal} {Phys. Rev. Lett.}\ }\textbf {\bibinfo
  {volume} {115}},\ \bibinfo {pages} {171801} (\bibinfo {year}
  {2015}{\natexlab{b}})}\BibitemShut {NoStop}%
\bibitem [{\citenamefont {Segovia}\ and\ \citenamefont
  {Roberts}(2016)}]{Segovia:2016zyc}%
  \BibitemOpen
  \bibfield  {author} {\bibinfo {author} {\bibfnamefont {J.}~\bibnamefont
  {Segovia}}\ and\ \bibinfo {author} {\bibfnamefont {C.~D.}\ \bibnamefont
  {Roberts}},\ }\href@noop {} {\bibfield  {journal} {\bibinfo  {journal} {Phys.
  Rev. C}\ }\textbf {\bibinfo {volume} {94}},\ \bibinfo {pages} {042201(R)}
  (\bibinfo {year} {2016})}\BibitemShut {NoStop}%
\bibitem [{\citenamefont {Maris}\ \emph {et~al.}(1998)\citenamefont {Maris},
  \citenamefont {Roberts},\ and\ \citenamefont {Tandy}}]{Maris:1997hd}%
  \BibitemOpen
  \bibfield  {author} {\bibinfo {author} {\bibfnamefont {P.}~\bibnamefont
  {Maris}}, \bibinfo {author} {\bibfnamefont {C.~D.}\ \bibnamefont {Roberts}},\
  and\ \bibinfo {author} {\bibfnamefont {P.~C.}\ \bibnamefont {Tandy}},\
  }\href@noop {} {\bibfield  {journal} {\bibinfo  {journal} {Phys. Lett. B}\
  }\textbf {\bibinfo {volume} {420}},\ \bibinfo {pages} {267} (\bibinfo {year}
  {1998})}\BibitemShut {NoStop}%
\bibitem [{\citenamefont {Cloet}\ \emph {et~al.}(2009)\citenamefont {Cloet},
  \citenamefont {Eichmann}, \citenamefont {El-Bennich}, \citenamefont
  {Kl{\"a}hn},\ and\ \citenamefont {Roberts}}]{Cloet:2008re}%
  \BibitemOpen
  \bibfield  {author} {\bibinfo {author} {\bibfnamefont {I.~C.}\ \bibnamefont
  {Cloet}}, \bibinfo {author} {\bibfnamefont {G.}~\bibnamefont {Eichmann}},
  \bibinfo {author} {\bibfnamefont {B.}~\bibnamefont {El-Bennich}}, \bibinfo
  {author} {\bibfnamefont {T.}~\bibnamefont {Kl{\"a}hn}},\ and\ \bibinfo
  {author} {\bibfnamefont {C.~D.}\ \bibnamefont {Roberts}},\ }\href@noop {}
  {\bibfield  {journal} {\bibinfo  {journal} {Few Body Syst.}\ }\textbf
  {\bibinfo {volume} {46}},\ \bibinfo {pages} {1} (\bibinfo {year}
  {2009})}\BibitemShut {NoStop}%
\bibitem [{\citenamefont {He}\ and\ \citenamefont {Ji}(1995)}]{He:1994gz}%
  \BibitemOpen
  \bibfield  {author} {\bibinfo {author} {\bibfnamefont {H.}~\bibnamefont
  {He}}\ and\ \bibinfo {author} {\bibfnamefont {X.}~\bibnamefont {Ji}},\
  }\href@noop {} {\bibfield  {journal} {\bibinfo  {journal} {Phys. Rev. D}\
  }\textbf {\bibinfo {volume} {52}},\ \bibinfo {pages} {2960} (\bibinfo {year}
  {1995})}\BibitemShut {NoStop}%
\bibitem [{\citenamefont {Bhattacharya}\ \emph {et~al.}(2016)\citenamefont
  {Bhattacharya}, \citenamefont {Cirigliano}, \citenamefont {Cohen},
  \citenamefont {Gupta}, \citenamefont {Lin},\ and\ \citenamefont
  {Yoon}}]{Bhattacharya:2016zcn}%
  \BibitemOpen
  \bibfield  {author} {\bibinfo {author} {\bibfnamefont {T.}~\bibnamefont
  {Bhattacharya}}, \bibinfo {author} {\bibfnamefont {V.}~\bibnamefont
  {Cirigliano}}, \bibinfo {author} {\bibfnamefont {S.}~\bibnamefont {Cohen}},
  \bibinfo {author} {\bibfnamefont {R.}~\bibnamefont {Gupta}}, \bibinfo
  {author} {\bibfnamefont {H.-W.}\ \bibnamefont {Lin}},\ and\ \bibinfo {author}
  {\bibfnamefont {B.}~\bibnamefont {Yoon}},\ }\href@noop {} {\bibfield
  {journal} {\bibinfo  {journal} {Phys. Rev. D}\ }\textbf {\bibinfo {volume}
  {94}},\ \bibinfo {pages} {054508} (\bibinfo {year} {2016})}\BibitemShut
  {NoStop}%
\bibitem [{\citenamefont {Alexandrou}\ \emph
  {et~al.}(2020{\natexlab{b}})\citenamefont {Alexandrou}, \citenamefont
  {Bacchio}, \citenamefont {Constantinou}, \citenamefont {Finkenrath},
  \citenamefont {Hadjiyiannakou}, \citenamefont {Jansen}, \citenamefont
  {Koutsou},\ and\ \citenamefont {Vaquero Aviles-Casco}}]{Alexandrou:2019brg}%
  \BibitemOpen
  \bibfield  {author} {\bibinfo {author} {\bibfnamefont {C.}~\bibnamefont
  {Alexandrou}}, \bibinfo {author} {\bibfnamefont {S.}~\bibnamefont {Bacchio}},
  \bibinfo {author} {\bibfnamefont {M.}~\bibnamefont {Constantinou}}, \bibinfo
  {author} {\bibfnamefont {J.}~\bibnamefont {Finkenrath}}, \bibinfo {author}
  {\bibfnamefont {K.}~\bibnamefont {Hadjiyiannakou}}, \bibinfo {author}
  {\bibfnamefont {K.}~\bibnamefont {Jansen}}, \bibinfo {author} {\bibfnamefont
  {G.}~\bibnamefont {Koutsou}},\ and\ \bibinfo {author} {\bibfnamefont
  {A.}~\bibnamefont {Vaquero Aviles-Casco}},\ }\href@noop {} {\bibfield
  {journal} {\bibinfo  {journal} {Phys. Rev. D}\ }\textbf {\bibinfo {volume}
  {102}},\ \bibinfo {pages} {054517} (\bibinfo {year}
  {2020}{\natexlab{b}})}\BibitemShut {NoStop}%
\bibitem [{\citenamefont {Cui}\ \emph {et~al.}(2021)\citenamefont {Cui},
  \citenamefont {Ding}, \citenamefont {Gao}, \citenamefont {Raya},
  \citenamefont {Binosi}, \citenamefont {Chang}, \citenamefont {Roberts},
  \citenamefont {Rodr\'{\i}guez-Quintero},\ and\ \citenamefont
  {Schmidt}}]{Cui:2020dlm}%
  \BibitemOpen
  \bibfield  {author} {\bibinfo {author} {\bibfnamefont {Z.-F.}\ \bibnamefont
  {Cui}}, \bibinfo {author} {\bibfnamefont {M.}~\bibnamefont {Ding}}, \bibinfo
  {author} {\bibfnamefont {F.}~\bibnamefont {Gao}}, \bibinfo {author}
  {\bibfnamefont {K.}~\bibnamefont {Raya}}, \bibinfo {author} {\bibfnamefont
  {D.}~\bibnamefont {Binosi}}, \bibinfo {author} {\bibfnamefont
  {L.}~\bibnamefont {Chang}}, \bibinfo {author} {\bibfnamefont {C.~D.}\
  \bibnamefont {Roberts}}, \bibinfo {author} {\bibfnamefont {J.}~\bibnamefont
  {Rodr\'{\i}guez-Quintero}},\ and\ \bibinfo {author} {\bibfnamefont {S.~M.}\
  \bibnamefont {Schmidt}},\ }\href@noop {} {\bibfield  {journal} {\bibinfo
  {journal} {Eur. Phys. J. A (Lett.)}\ }\textbf {\bibinfo {volume} {57}},\
  \bibinfo {pages} {5} (\bibinfo {year} {2021})}\BibitemShut {NoStop}%
\bibitem [{\citenamefont {Cui}\ \emph {et~al.}(2020{\natexlab{b}})\citenamefont
  {Cui}, \citenamefont {Ding}, \citenamefont {Gao}, \citenamefont {Raya},
  \citenamefont {Binosi}, \citenamefont {Chang}, \citenamefont {Roberts},
  \citenamefont {Rodr\'{\i}guez-Quintero},\ and\ \citenamefont
  {Schmidt}}]{Cui:2020tdf}%
  \BibitemOpen
  \bibfield  {author} {\bibinfo {author} {\bibfnamefont {Z.-F.}\ \bibnamefont
  {Cui}}, \bibinfo {author} {\bibfnamefont {M.}~\bibnamefont {Ding}}, \bibinfo
  {author} {\bibfnamefont {F.}~\bibnamefont {Gao}}, \bibinfo {author}
  {\bibfnamefont {K.}~\bibnamefont {Raya}}, \bibinfo {author} {\bibfnamefont
  {D.}~\bibnamefont {Binosi}}, \bibinfo {author} {\bibfnamefont
  {L.}~\bibnamefont {Chang}}, \bibinfo {author} {\bibfnamefont {C.~D.}\
  \bibnamefont {Roberts}}, \bibinfo {author} {\bibfnamefont {J.}~\bibnamefont
  {Rodr\'{\i}guez-Quintero}},\ and\ \bibinfo {author} {\bibfnamefont {S.~M.}\
  \bibnamefont {Schmidt}},\ }\href@noop {} {\bibfield  {journal} {\bibinfo
  {journal} {Eur. Phys. J. C}\ }\textbf {\bibinfo {volume} {80}},\ \bibinfo
  {pages} {1064} (\bibinfo {year} {2020}{\natexlab{b}})}\BibitemShut {NoStop}%
\bibitem [{\citenamefont {Baru}\ \emph {et~al.}(2011)\citenamefont {Baru},
  \citenamefont {Hanhart}, \citenamefont {Hoferichter}, \citenamefont {Kubis},
  \citenamefont {Nogga},\ and\ \citenamefont {Phillips}}]{Baru:2011bw}%
  \BibitemOpen
  \bibfield  {author} {\bibinfo {author} {\bibfnamefont {V.}~\bibnamefont
  {Baru}}, \bibinfo {author} {\bibfnamefont {C.}~\bibnamefont {Hanhart}},
  \bibinfo {author} {\bibfnamefont {M.}~\bibnamefont {Hoferichter}}, \bibinfo
  {author} {\bibfnamefont {B.}~\bibnamefont {Kubis}}, \bibinfo {author}
  {\bibfnamefont {A.}~\bibnamefont {Nogga}},\ and\ \bibinfo {author}
  {\bibfnamefont {D.~R.}\ \bibnamefont {Phillips}},\ }\href@noop {} {\bibfield
  {journal} {\bibinfo  {journal} {Nucl. Phys. A}\ }\textbf {\bibinfo {volume}
  {872}},\ \bibinfo {pages} {69} (\bibinfo {year} {2011})}\BibitemShut
  {NoStop}%
\bibitem [{\citenamefont {Navarro~P\'erez}\ \emph {et~al.}(2017)\citenamefont
  {Navarro~P\'erez}, \citenamefont {Amaro},\ and\ \citenamefont
  {Ruiz~Arriola}}]{NavarroPerez:2016eli}%
  \BibitemOpen
  \bibfield  {author} {\bibinfo {author} {\bibfnamefont {R.}~\bibnamefont
  {Navarro~P\'erez}}, \bibinfo {author} {\bibfnamefont {J.~E.}\ \bibnamefont
  {Amaro}},\ and\ \bibinfo {author} {\bibfnamefont {E.}~\bibnamefont
  {Ruiz~Arriola}},\ }\href@noop {} {\bibfield  {journal} {\bibinfo  {journal}
  {Phys. Rev. C}\ }\textbf {\bibinfo {volume} {95}},\ \bibinfo {pages} {064001}
  (\bibinfo {year} {2017})}\BibitemShut {NoStop}%
\bibitem [{\citenamefont {Reinert}\ \emph {et~al.}(2021)\citenamefont
  {Reinert}, \citenamefont {Krebs},\ and\ \citenamefont
  {Epelbaum}}]{Reinert:2020mcu}%
  \BibitemOpen
  \bibfield  {author} {\bibinfo {author} {\bibfnamefont {P.}~\bibnamefont
  {Reinert}}, \bibinfo {author} {\bibfnamefont {H.}~\bibnamefont {Krebs}},\
  and\ \bibinfo {author} {\bibfnamefont {E.}~\bibnamefont {Epelbaum}},\
  }\href@noop {} {\bibfield  {journal} {\bibinfo  {journal} {Phys. Rev. Lett.}\
  }\textbf {\bibinfo {volume} {126}},\ \bibinfo {pages} {092501} (\bibinfo
  {year} {2021})}\BibitemShut {NoStop}%
\bibitem [{\citenamefont {Nagy}\ and\ \citenamefont
  {Scadron}(2003)}]{Nagy:2004tp}%
  \BibitemOpen
  \bibfield  {author} {\bibinfo {author} {\bibfnamefont {M.}~\bibnamefont
  {Nagy}}\ and\ \bibinfo {author} {\bibfnamefont {M.~D.}\ \bibnamefont
  {Scadron}},\ }\href@noop {} {\bibfield  {journal} {\bibinfo  {journal} {Acta
  Phys. Slov.}\ }\textbf {\bibinfo {volume} {54}},\ \bibinfo {pages} {427}
  (\bibinfo {year} {2003})}\BibitemShut {NoStop}%
\bibitem [{\citenamefont {Kamano}\ \emph {et~al.}(2013)\citenamefont {Kamano},
  \citenamefont {Nakamura}, \citenamefont {Lee},\ and\ \citenamefont
  {Sato}}]{Kamano:2013iva}%
  \BibitemOpen
  \bibfield  {author} {\bibinfo {author} {\bibfnamefont {H.}~\bibnamefont
  {Kamano}}, \bibinfo {author} {\bibfnamefont {S.~X.}\ \bibnamefont
  {Nakamura}}, \bibinfo {author} {\bibfnamefont {T.~S.~H.}\ \bibnamefont
  {Lee}},\ and\ \bibinfo {author} {\bibfnamefont {T.}~\bibnamefont {Sato}},\
  }\href@noop {} {\bibfield  {journal} {\bibinfo  {journal} {Phys. Rev. C}\
  }\textbf {\bibinfo {volume} {88}},\ \bibinfo {pages} {035209} (\bibinfo
  {year} {2013})}\BibitemShut {NoStop}%
\bibitem [{\citenamefont {Ivanov}\ \emph {et~al.}(1999)\citenamefont {Ivanov},
  \citenamefont {Kalinovsky},\ and\ \citenamefont {Roberts}}]{Ivanov:1998ms}%
  \BibitemOpen
  \bibfield  {author} {\bibinfo {author} {\bibfnamefont {M.~A.}\ \bibnamefont
  {Ivanov}}, \bibinfo {author} {\bibfnamefont {{\mbox{Yu}}.~L.}\ \bibnamefont
  {Kalinovsky}},\ and\ \bibinfo {author} {\bibfnamefont {C.~D.}\ \bibnamefont
  {Roberts}},\ }\href@noop {} {\bibfield  {journal} {\bibinfo  {journal} {Phys.
  Rev. D}\ }\textbf {\bibinfo {volume} {60}},\ \bibinfo {pages} {034018}
  (\bibinfo {year} {1999})}\BibitemShut {NoStop}%
\bibitem [{\citenamefont {Hecht}\ \emph {et~al.}(2001)\citenamefont {Hecht},
  \citenamefont {Roberts},\ and\ \citenamefont {Schmidt}}]{Hecht:2000xa}%
  \BibitemOpen
  \bibfield  {author} {\bibinfo {author} {\bibfnamefont {M.~B.}\ \bibnamefont
  {Hecht}}, \bibinfo {author} {\bibfnamefont {C.~D.}\ \bibnamefont {Roberts}},\
  and\ \bibinfo {author} {\bibfnamefont {S.~M.}\ \bibnamefont {Schmidt}},\
  }\href@noop {} {\bibfield  {journal} {\bibinfo  {journal} {Phys. Rev. C}\
  }\textbf {\bibinfo {volume} {63}},\ \bibinfo {pages} {025213} (\bibinfo
  {year} {2001})}\BibitemShut {NoStop}%
\bibitem [{\citenamefont {Alkofer}\ \emph {et~al.}(2005)\citenamefont
  {Alkofer}, \citenamefont {H{\"o}ll}, \citenamefont {Kloker}, \citenamefont
  {Krassnigg},\ and\ \citenamefont {Roberts}}]{Alkofer:2004yf}%
  \BibitemOpen
  \bibfield  {author} {\bibinfo {author} {\bibfnamefont {R.}~\bibnamefont
  {Alkofer}}, \bibinfo {author} {\bibfnamefont {A.}~\bibnamefont {H{\"o}ll}},
  \bibinfo {author} {\bibfnamefont {M.}~\bibnamefont {Kloker}}, \bibinfo
  {author} {\bibfnamefont {A.}~\bibnamefont {Krassnigg}},\ and\ \bibinfo
  {author} {\bibfnamefont {C.~D.}\ \bibnamefont {Roberts}},\ }\href@noop {}
  {\bibfield  {journal} {\bibinfo  {journal} {Few Body Syst.}\ }\textbf
  {\bibinfo {volume} {37}},\ \bibinfo {pages} {1} (\bibinfo {year}
  {2005})}\BibitemShut {NoStop}%
\bibitem [{\citenamefont {Lane}(1974)}]{Lane:1974he}%
  \BibitemOpen
  \bibfield  {author} {\bibinfo {author} {\bibfnamefont {K.~D.}\ \bibnamefont
  {Lane}},\ }\href@noop {} {\bibfield  {journal} {\bibinfo  {journal} {Phys.
  Rev. D}\ }\textbf {\bibinfo {volume} {10}},\ \bibinfo {pages} {2605}
  (\bibinfo {year} {1974})}\BibitemShut {NoStop}%
\bibitem [{\citenamefont {Politzer}(1976)}]{Politzer:1976tv}%
  \BibitemOpen
  \bibfield  {author} {\bibinfo {author} {\bibfnamefont {H.~D.}\ \bibnamefont
  {Politzer}},\ }\href@noop {} {\bibfield  {journal} {\bibinfo  {journal}
  {Nucl. Phys. B}\ }\textbf {\bibinfo {volume} {117}},\ \bibinfo {pages} {397}
  (\bibinfo {year} {1976})}\BibitemShut {NoStop}%
\bibitem [{\citenamefont {Bhagwat}\ and\ \citenamefont
  {Tandy}(2006)}]{Bhagwat:2006tu}%
  \BibitemOpen
  \bibfield  {author} {\bibinfo {author} {\bibfnamefont {M.~S.}\ \bibnamefont
  {Bhagwat}}\ and\ \bibinfo {author} {\bibfnamefont {P.~C.}\ \bibnamefont
  {Tandy}},\ }\href@noop {} {\bibfield  {journal} {\bibinfo  {journal} {AIP
  Conf. Proc.}\ }\textbf {\bibinfo {volume} {842}},\ \bibinfo {pages} {225}
  (\bibinfo {year} {2006})}\BibitemShut {NoStop}%
\bibitem [{\citenamefont {Binosi}\ \emph {et~al.}(2017)\citenamefont {Binosi},
  \citenamefont {Chang}, \citenamefont {Papavassiliou}, \citenamefont {Qin},\
  and\ \citenamefont {Roberts}}]{Binosi:2016wcx}%
  \BibitemOpen
  \bibfield  {author} {\bibinfo {author} {\bibfnamefont {D.}~\bibnamefont
  {Binosi}}, \bibinfo {author} {\bibfnamefont {L.}~\bibnamefont {Chang}},
  \bibinfo {author} {\bibfnamefont {J.}~\bibnamefont {Papavassiliou}}, \bibinfo
  {author} {\bibfnamefont {S.-X.}\ \bibnamefont {Qin}},\ and\ \bibinfo {author}
  {\bibfnamefont {C.~D.}\ \bibnamefont {Roberts}},\ }\href@noop {} {\bibfield
  {journal} {\bibinfo  {journal} {Phys. Rev. D}\ }\textbf {\bibinfo {volume}
  {95}},\ \bibinfo {pages} {031501(R)} (\bibinfo {year} {2017})}\BibitemShut
  {NoStop}%
\bibitem [{\citenamefont {Roberts}(2020)}]{Roberts:2020hiw}%
  \BibitemOpen
  \bibfield  {author} {\bibinfo {author} {\bibfnamefont {C.~D.}\ \bibnamefont
  {Roberts}},\ }\href@noop {} {\bibfield  {journal} {\bibinfo  {journal}
  {Symmetry}\ }\textbf {\bibinfo {volume} {12}},\ \bibinfo {pages} {1468}
  (\bibinfo {year} {2020})}\BibitemShut {NoStop}%
\bibitem [{\citenamefont {Chang}\ \emph
  {et~al.}(2013{\natexlab{a}})\citenamefont {Chang}, \citenamefont {Cloet},
  \citenamefont {Cobos-Martinez}, \citenamefont {Roberts}, \citenamefont
  {Schmidt},\ and\ \citenamefont {Tandy}}]{Chang:2013pq}%
  \BibitemOpen
  \bibfield  {author} {\bibinfo {author} {\bibfnamefont {L.}~\bibnamefont
  {Chang}}, \bibinfo {author} {\bibfnamefont {I.~C.}\ \bibnamefont {Cloet}},
  \bibinfo {author} {\bibfnamefont {J.~J.}\ \bibnamefont {Cobos-Martinez}},
  \bibinfo {author} {\bibfnamefont {C.~D.}\ \bibnamefont {Roberts}}, \bibinfo
  {author} {\bibfnamefont {S.~M.}\ \bibnamefont {Schmidt}},\ and\ \bibinfo
  {author} {\bibfnamefont {P.~C.}\ \bibnamefont {Tandy}},\ }\href@noop {}
  {\bibfield  {journal} {\bibinfo  {journal} {Phys. Rev. Lett.}\ }\textbf
  {\bibinfo {volume} {110}},\ \bibinfo {pages} {132001} (\bibinfo {year}
  {2013}{\natexlab{a}})}\BibitemShut {NoStop}%
\bibitem [{\citenamefont {Chang}\ \emph
  {et~al.}(2013{\natexlab{b}})\citenamefont {Chang}, \citenamefont {Roberts},\
  and\ \citenamefont {Schmidt}}]{Chang:2013epa}%
  \BibitemOpen
  \bibfield  {author} {\bibinfo {author} {\bibfnamefont {L.}~\bibnamefont
  {Chang}}, \bibinfo {author} {\bibfnamefont {C.~D.}\ \bibnamefont {Roberts}},\
  and\ \bibinfo {author} {\bibfnamefont {S.~M.}\ \bibnamefont {Schmidt}},\
  }\href@noop {} {\bibfield  {journal} {\bibinfo  {journal} {Phys. Lett. B}\
  }\textbf {\bibinfo {volume} {727}},\ \bibinfo {pages} {255} (\bibinfo {year}
  {2013}{\natexlab{b}})}\BibitemShut {NoStop}%
\bibitem [{\citenamefont {Burden}\ \emph {et~al.}(1996)\citenamefont {Burden},
  \citenamefont {Roberts},\ and\ \citenamefont {Thomson}}]{Burden:1995ve}%
  \BibitemOpen
  \bibfield  {author} {\bibinfo {author} {\bibfnamefont {C.~J.}\ \bibnamefont
  {Burden}}, \bibinfo {author} {\bibfnamefont {C.~D.}\ \bibnamefont
  {Roberts}},\ and\ \bibinfo {author} {\bibfnamefont {M.~J.}\ \bibnamefont
  {Thomson}},\ }\href@noop {} {\bibfield  {journal} {\bibinfo  {journal} {Phys.
  Lett. B}\ }\textbf {\bibinfo {volume} {371}},\ \bibinfo {pages} {163}
  (\bibinfo {year} {1996})}\BibitemShut {NoStop}%
\bibitem [{\citenamefont {Horn}\ and\ \citenamefont
  {Roberts}(2016)}]{Horn:2016rip}%
  \BibitemOpen
  \bibfield  {author} {\bibinfo {author} {\bibfnamefont {T.}~\bibnamefont
  {Horn}}\ and\ \bibinfo {author} {\bibfnamefont {C.~D.}\ \bibnamefont
  {Roberts}},\ }\href@noop {} {\bibfield  {journal} {\bibinfo  {journal} {J.
  Phys. G.}\ }\textbf {\bibinfo {volume} {43}},\ \bibinfo {pages} {073001}
  (\bibinfo {year} {2016})}\BibitemShut {NoStop}%
\bibitem [{\citenamefont {Ishii}(2001)}]{Ishii:2000zy}%
  \BibitemOpen
  \bibfield  {author} {\bibinfo {author} {\bibfnamefont {N.}~\bibnamefont
  {Ishii}},\ }\href@noop {} {\bibfield  {journal} {\bibinfo  {journal} {Nucl.
  Phys. A}\ }\textbf {\bibinfo {volume} {689}},\ \bibinfo {pages} {793}
  (\bibinfo {year} {2001})}\BibitemShut {NoStop}%
\end{thebibliography}

%

\end{document}